\IfFileExists{patch-revtex4-2.tex}{\input{patch-revtex4-2.tex}}{}		

\documentclass[letterpaper,11pt,superscriptaddress,nofootinbib]{revtex4-2}
\pdfoutput=1
\usepackage{xcolor}
\usepackage{amsmath,amsthm,amsfonts,amssymb,braket}
\usepackage[pdftex,colorlinks=true,linkcolor=blue,citecolor=darkred,urlcolor=darkblue]{hyperref}
\usepackage[all]{xy}
\usepackage{enumitem}	
\usepackage{fullpage}
\usepackage{mathtools}

\usepackage{tikz}   
\usepackage{rotating}  
\usetikzlibrary{patterns}
\usepackage{placeins} 

\usepackage{multirow}

\newtheorem{proposition-definition}[lemma]{Proposition-Definition}
\theoremstyle{definition}

\definecolor{darkblue}{rgb}{0.,0.,0.4}
\definecolor{darkred}{rgb}{0.5,0.,0.}
\definecolor{darkpurple}{rgb}{0.5,0.,0.5}
\definecolor{ltgreen}{rgb}{0.1,.59,.43}
\definecolor{orange}{rgb}{1.0, 0.5, 0.0}
\newcommand{\dave}[1]{{\textcolor{ltgreen}{DA: #1}}}
\newcommand{\linnea}[1]{{\textcolor{darkred}{LGS: #1}}}
\newcommand{\parsa}[1]{{\textcolor{cyan}{PB: #1}}}

\renewcommand{\dave}[1]{}
\renewcommand{\linnea}[1]{}
\renewcommand{\parsa}[1]{}

\makeatletter
\def\l@subsubsection#1#2{}
\makeatother
\newcommand{\nocontentsline}[3]{}
\newcommand{\tocless}[2]{\bgroup\let\addcontentsline=\nocontentsline#1{#2}\egroup}

\newcommand*{\MyDef}{\mathrm{def}}
\newcommand*{\MyEqdefU}{\ensuremath{\mathrel{\overset{\MyDef}{=}}}}
\newcommand*{\eqdef}{\mathrel{\overset{\MyDef}{\resizebox{\widthof{\kern1.25pt\MyEqdefU}}{\heightof{$=$}}{$=$}}}}

\begin{document}
\title{A fault-tolerant pairwise measurement-based code on eight qubits}
\author{Linnea Grans-Samuelsson}
\affiliation{Microsoft Station Q, Santa Barbara, California 93106-6105 USA}
\author{David Aasen}
\affiliation{Microsoft Station Q, Santa Barbara, California 93106-6105 USA}
\author{Parsa Bonderson}
\affiliation{Microsoft Station Q, Santa Barbara, California 93106-6105 USA}

\begin{abstract}
We construct a pairwise measurement-based code on eight qubits that is error correcting for circuit noise, with fault distance 3. 
The code can be implemented on a subset of a rectangular array of qubits with nearest neighbor connectivity of pairwise Pauli measurements, with a syndrome extraction circuit of depth 28. 
We describe fault-tolerant logical operations on patches of this eight-qubit code that generate the full Clifford group.
We estimate the performance under circuit noise both during logical idle and during a logical two-qubit measurement. 
We estimate the pseudo-threshold to be between  $10^{-5}$ and $2\times 10^{-4}$, depending on the amount of noise on idle physical qubits.
The use of post-selection in addition to error correction (correcting all degree one faults and rejecting a subset of the higher degree faults) can improve the pseudo-threshold by up to an order of magnitude.
\end{abstract}

\maketitle


\section{Introduction}

In the quest to build utility-scale quantum computers, quantum error correcting codes are a vital tool for combating noise.
In parallel with the development of codes for scalable systems, it is interesting to consider codes requiring only a small number of physical qubits.
Such codes offer the possibility to explore quantum error correction and fault-tolerant operations at relatively early stages of hardware development.
Lessons from small-scale experiments can then inform the development of fault-tolerant quantum computing in larger systems.
During the last few years there have been several notable small-scale experiments, demonstrating key features of quantum error correction such as suppression of error rates with increased code distance~\cite{Acharya2022SuppressingQE,Bluvstein:2023zmt} and improvement over physical error rates during repeated error correction~\cite{dasilva2024demonstration}.

Motivated by recently developed Floquet codes~\cite{Hastings2021, Paetznick2023} as well as Majorana-based quantum computing hardware~\cite{Karzig2017,aghaee2024interferometric}, the goal of our paper is to construct an error correcting code based on one- and two-qubit Pauli measurements using the smallest number of physical qubits possible. 
We introduce a new fault-tolerant pairwise measurement-based code on eight qubits, which is the smallest such code to our knowledge. 
For comparison, fault-tolerant error correction in pairwise measurement-based code realizations is achieved using 17 qubits for the windmill realization of the surface code~\cite{Chao2020}, 25 qubits for the Bacon-Shor subsystem code~\cite{Knapp2018modelingnoiseerror}, and 25 qubits for the surface code realization of Ref.~\onlinecite{granssamuelsson2023improved}.
As a precursor to producing this code, we first construct a pairwise measurement-based code on seven qubits. This smaller code can correct all single qubit errors, but not all single faults in the more general circuit noise model, where faults can cause correlated (multi-qubit and measurement readout) errors \cite{Chao2020}.
Fault-tolerance is attained by adding a qubit and modifying the circuit appropriately to enable detection and correction of any single fault in the circuit noise model.
These codes may be viewed as pairwise measurement-based implementations of the of the perfect code $[[5,1,3]]$, so we refer to them as the 5+3 code and the 5+2 codes, respectively.
Another nice feature of the 5+3 code is that it can be implemented within a rectangular array of qubits with only nearest neighbor connectivity of pairwise measurements, which makes this code suitable for potential operation in measurement-based Majorana hardware.

As the circuit noise model is more realistic than independent identically distributed Pauli noise, the main focus of this paper will be on the 5+3 code. 
The 5+3 code can be compared to the fault-tolerant CNOT-based implementation of the $[[5,1,3]]$ code in Ref.~\onlinecite{Chao_2018}, which requires only seven qubits.
As with the flag scheme of Ref.~\onlinecite{Chao_2018}, the syndrome extraction circuit of the present paper is not unique to the $[[5,1,3]]$ code; a similar construction was used by the present authors in a measurement-based implementation of the surface code \cite{granssamuelsson2023improved}, and we provide an example for how to modify it to implement the $[[8,3,2]]$ color code with pairwise measurements in Appendix~\ref{ColorCode}.  
Simulating the performance of the 5+3 code under circuit noise, we find that the estimated pseudo-threshold is approximately $2\times 10^{-4}$ under the assumption of no idling errors.
However, by combining error correction with post-selection, the pseudo-threshold can be increased by up to an order of magnitude, exemplifying how post-selection can have a strong impact even in an odd-distance code.
The  performance is significantly decreased in the presence of idle noise, indicating that ``temporally sparse codes,'' such as the present code, are best suited for hardware with good idle fidelity.


The paper is structured as follows.
In Section \ref{Circuits}, we define the measurement circuits for the new codes.
We note that the 5+2 code cannot correct degree one faults in the circuit noise model, and show how the addition of an extra qubit together with repeated measurements makes the 5+3 code error correcting for circuit noise. 
In Section \ref{Square}, we show how the 5+3 code can be implemented within a rectangular qubit array. 
In Section \ref{Logical}, we describe how logical operations can be performed on logical qubits encoded in the 5+3 code.
In particular, we show that square patches of the code, each encoding one logical qubit, can be placed on a square grid such that the full Clifford group is obtained from only nearest neighbor measurements of the physical qubits. 
Finally, in Section \ref{Performance}, we show Monte Carlo estimates of the logical failure rate $p_{logical}$ for the 5+3 code as a function of the physical failure rate $p_{physical}$ in the circuit noise model. 
This performance estimate is done both during a logical idle  and during a pairwise $XX$ measurement of two logical qubits. 
We also show how adding post-selection on top of error correction boosts the logical fidelity.

\section{The measurement circuits}\label{Circuits}

The notion of \emph{fault distance}  will be important for what follows, and we briefly introduce it here. We consider the \emph{circuit noise model} of Ref.~\onlinecite{Chao2020}, which is discussed in more detail in Section \ref{Performance}. In the circuit noise model, gates fail with a probability $p_{physical}$, such that the errors on qubits that are involved in a multi-qubit gate are correlated. The model also includes readout errors. We say that a fault is of degree $t$ in a given noise model if it occurs with a probability proportional to $(p_{physical})^t$, and refer to the minimum degree of a fault in the circuit noise model that can cause an undetectable logical error as the \emph{fault distance} $d_{\rm f}$ of a code \cite{bombin_logical_2023}.
We say that a code is fault-tolerantly error correcting if it has $d_{\rm f} \geq 3$.

We also introduce the notion of \emph{detectors}, as discussed in Ref.~\onlinecite{Delfosse2023}. (See also Ref.~\onlinecite{McEwen2023}.) 
In a sequence of one- and two-qubit measurements, a detector is a set of measurements for which the sum of the measurement outcomes mod 2 \parsa{I usually think of Pauli measurements as having $\pm1$ outcomes, not 0 and 1, so products and parities rather than sum mod 2}\linnea{I usually see 0 and 1 in the context of qec, and it's also what we use in the surface code paper} takes a fixed value in the absence of errors. 
The \emph{syndrome} of an error is the set of detectors for which the sum of the measurement outcomes mod 2 differs from its value in the absence of errors. 
Faults up to degree $t$ can be detected if each such fault has a nontrivial syndrome, and can be corrected if it is possible to distinguish each such fault from inequivalent faults based on its syndrome.\footnote{If a fault does not affect the logical information, we do not require it to be detectable or correctable.} 
A crucial difference between the 5+2 code and the 5+3 code presented below is that the latter contains more detectors, by construction, making it possible to distinguish between the equivalence classes of all degree 1 faults in the circuit noise model and, thus, achieve fault distance 3. 
We identify the detectors using the fault propagation as described in Ref.~\onlinecite{Delfosse2023} [see Algorithm 1 within this reference].
The detectors for the 5+2 code and the 5+3 code are shown in Appendix~\ref{Detectors}.


The stabilizer group of the perfect code [[5,1,3]] is generated by the weight four stabilizers $XZZXI, IXZZX, XIXZZ,ZXIXZ$. 
In terms of number of qubits it is the smallest distance 3 code, but it has two practical drawbacks: it is generally not feasible to perform a four-qubit measurement in hardware, and it does not have \emph{fault distance} 3.\footnote{
If four-qubit measurements were feasible, the circuit noise model would include correlated errors on up to four qubits occurring with probability $p_{physical}$. 
The [[5,1,3]] code can only handle single-qubit errors.} 
With the help of auxiliary qubits, a four qubit measurement can be decomposed into different sequences of one- and two-qubit measurements, and the 5+2 code and 5+3 code presented in this section are two such decompositions. (A convenient tool for making such decompositions is ZX-calculus. We do not give an overview of ZX-calculus here, referring instead the interested reader to introductions such as Ref.~\onlinecite{Wetering2020}.)

For the sake of compact notation, we represent one- and two-qubit measurements as
\begin{equation}
\vcenter{\hbox{
\begin{tikzpicture}
   \newcommand{\scaling}{0.6}
   \node at (-2*\scaling, 0.0000*\scaling)  [anchor=center] () {};
   \draw[line width = 0.6mm] (-1.5*\scaling, 0.0000*\scaling)   -- (1.5000*\scaling, 0.0000*\scaling);
   \draw[dotted] (-1.300000*\scaling, 0.000000*\scaling +1*\scaling)  -- (-1.300000*\scaling, 0.000000*\scaling -1*\scaling);
   \draw[dotted] (1.300000*\scaling, 0.000000*\scaling +1*\scaling)  -- (1.300000*\scaling, 0.000000*\scaling -1*\scaling);
   \node at (0.000000*\scaling, 0.000000*\scaling)  [rectangle,fill=white,draw, minimum width = 12pt, minimum height = 12pt,text width=, anchor=center] () {\tiny $\rm P$};
\end{tikzpicture}
}} \text{and}
\vcenter{\hbox{
\begin{tikzpicture}
   \newcommand{\scaling}{0.6}
   \node at (-2*\scaling, 0.0000*\scaling)  [anchor=center] () {};
   \draw[line width = 0.6mm] (-1.5*\scaling, 0.0000*\scaling)   -- (1.5000*\scaling, 0.0000*\scaling);
   \node at (-2*\scaling, -1.0000*\scaling)  [anchor=center] () {};
   \draw[line width = 0.6mm] (-1.5*\scaling, -1.0000*\scaling)   -- (1.5000*\scaling, -1.0000*\scaling);
   \draw[dotted] (-1.300000*\scaling, 0.000000*\scaling +1*\scaling)  -- (-1.300000*\scaling, -1.000000*\scaling -1*\scaling);
   \draw[dotted] (1.300000*\scaling, 0.000000*\scaling +1*\scaling)  -- (1.300000*\scaling, -1.000000*\scaling -1*\scaling);
   \draw[line width=1mm, white] (0.0000*\scaling,  -1.000000*\scaling)   -- (0.0000*\scaling, 0.000000*\scaling);
   \draw (0.0000*\scaling,  -1.000000*\scaling)   -- (0.0000*\scaling, 0.000000*\scaling);
   \node at (0.0000*\scaling, -1.000000*\scaling)  [rectangle,fill=white,draw, minimum width = 12pt, minimum height = 12pt,text width=, anchor=center] () {\tiny $\rm P_1$};
   \node at (0.0000*\scaling, 0.000000*\scaling)  [rectangle,fill=white,draw, minimum width = 12pt, minimum height = 12pt,text width=, anchor=center] () {\tiny $\rm P_2$};
\end{tikzpicture}
}},
\end{equation}
respectively, where the boxes are labeled by the basis $ P, P_1,P_2 \in \{ X,Y,Z \}$ in which the measurement is performed. We emphasize that two boxes connected by a solid vertical line denotes a pairwise measurement, rather than two single-qubit measurements. The dashed vertical lines separate different time steps.

In the 5+2 code, an XZZXI measurement is performed through the following circuit of one- and two-qubit measurements:
\begin{equation}
\vcenter{\hbox{
\begin{tikzpicture}[scale=0.8]
\foreach \y in {1,2,3,4,5,6,7}{
  \draw (-0.4,\y*0.2)--(2.4,\y*0.2);
}
\filldraw[fill=gray!10!white, rounded corners] (0,0) rectangle (2,7*0.2 + 0.2);
\node[align=center] at (1,4*0.2) {XZZXI};
\end{tikzpicture}}} \quad  = \quad 
\vcenter{\hbox{
\begin{tikzpicture}
   \newcommand{\scaling}{0.6}
   \node at (-2*\scaling, 0.0000*\scaling)  [anchor=center] () {1};
   \draw[line width = 0.6mm, gray] (-1.5*\scaling, 0.0000*\scaling)   -- (11.9000*\scaling, 0.0000*\scaling);
   \node at (-2*\scaling, -1.0000*\scaling)  [anchor=center] () {2};
   \draw[line width = 0.6mm, gray] (-1.5*\scaling, -1.0000*\scaling)   -- (11.9000*\scaling, -1.0000*\scaling);
   \node at (-2*\scaling, -2.0000*\scaling)  [anchor=center] () {3};
   \draw[line width = 0.6mm, gray] (-1.5*\scaling, -2.0000*\scaling)   -- (11.9000*\scaling, -2.0000*\scaling);
   \node at (-2*\scaling, -3.0000*\scaling)  [anchor=center] () {4};
   \draw[line width = 0.6mm, gray] (-1.5*\scaling, -3.0000*\scaling)   -- (11.9000*\scaling, -3.0000*\scaling);
   \node at (-2*\scaling, -4.0000*\scaling)  [anchor=center] () {5};
   \draw[line width = 0.6mm, gray] (-1.5*\scaling, -4.0000*\scaling)   -- (11.9000*\scaling, -4.0000*\scaling);
   \node at (-2*\scaling, -5.0000*\scaling)  [anchor=center] () {A};
   \draw[line width = 0.6mm] (-1.5*\scaling, -5.0000*\scaling)   -- (11.9000*\scaling, -5.0000*\scaling);
   \node at (-2*\scaling, -6.0000*\scaling)  [anchor=center] () {B};
   \draw[line width = 0.6mm] (-1.5*\scaling, -6.0000*\scaling)   -- (11.9000*\scaling, -6.0000*\scaling);
   \draw[dotted] (-1.300000*\scaling, 0.000000*\scaling +1*\scaling)  -- (-1.300000*\scaling, -6.000000*\scaling -1*\scaling);
   \draw[dotted] (1.300000*\scaling, 0.000000*\scaling +1*\scaling)  -- (1.300000*\scaling, -6.000000*\scaling -1*\scaling);
   \draw[dotted] (3.900000*\scaling, 0.000000*\scaling +1*\scaling)  -- (3.900000*\scaling, -6.000000*\scaling -1*\scaling);
   \draw[dotted] (6.500000*\scaling, 0.000000*\scaling +1*\scaling)  -- (6.500000*\scaling, -6.000000*\scaling -1*\scaling);
   \draw[dotted] (9.100000*\scaling, 0.000000*\scaling +1*\scaling)  -- (9.100000*\scaling, -6.000000*\scaling -1*\scaling);
   \draw[dotted] (11.700000*\scaling, 0.000000*\scaling +1*\scaling)  -- (11.700000*\scaling, -6.000000*\scaling -1*\scaling);
   \node at (0.0000*\scaling, -6.0000*\scaling -1*\scaling) {1};
   \node at (2.6000*\scaling, -6.0000*\scaling -1*\scaling) {2};
   \node at (5.2000*\scaling, -6.0000*\scaling -1*\scaling) {3};
   \node at (7.8000*\scaling, -6.0000*\scaling -1*\scaling) {4};
   \node at (10.4000*\scaling, -6.0000*\scaling -1*\scaling) {5};
   \node at (0.000000*\scaling, -5.000000*\scaling)  [rectangle,fill=white,draw, minimum width = 12pt, minimum height = 12pt,text width=, anchor=center] () {\tiny Z};
   \node at (0.000000*\scaling, -6.000000*\scaling)  [rectangle,fill=white,draw, minimum width = 12pt, minimum height = 12pt,text width=, anchor=center] () {\tiny Z};
   \draw[line width=1mm, white] (2.1580*\scaling,  -5.000000*\scaling)   -- (2.1580*\scaling, 0.000000*\scaling);
   \draw (2.1580*\scaling,  -5.000000*\scaling)   -- (2.1580*\scaling, 0.000000*\scaling);
   \node at (2.1580*\scaling, -5.000000*\scaling)  [rectangle,fill=white,draw, minimum width = 12pt, minimum height = 12pt,text width=, anchor=center] () {\tiny X};
   \node at (2.1580*\scaling, 0.000000*\scaling)  [rectangle,fill=white,draw, minimum width = 12pt, minimum height = 12pt,text width=, anchor=center] () {\tiny X};
   \draw[line width=1mm, white] (3.0420*\scaling,  -6.000000*\scaling)   -- (3.0420*\scaling, -3.000000*\scaling);
   \draw (3.0420*\scaling,  -6.000000*\scaling)   -- (3.0420*\scaling, -3.000000*\scaling);
   \node at (3.0420*\scaling, -6.000000*\scaling)  [rectangle,fill=white,draw, minimum width = 12pt, minimum height = 12pt,text width=, anchor=center] () {\tiny X};
   \node at (3.0420*\scaling, -3.000000*\scaling)  [rectangle,fill=white,draw, minimum width = 12pt, minimum height = 12pt,text width=, anchor=center] () {\tiny X};
   \draw[line width=1mm, white] (5.2000*\scaling,  -6.000000*\scaling)   -- (5.2000*\scaling, -5.000000*\scaling);
   \draw (5.2000*\scaling,  -6.000000*\scaling)   -- (5.2000*\scaling, -5.000000*\scaling);
   \node at (5.2000*\scaling, -6.000000*\scaling)  [rectangle,fill=white,draw, minimum width = 12pt, minimum height = 12pt,text width=, anchor=center] () {\tiny Z};
   \node at (5.2000*\scaling, -5.000000*\scaling)  [rectangle,fill=white,draw, minimum width = 12pt, minimum height = 12pt,text width=, anchor=center] () {\tiny Z};
   \draw[line width=1mm, white] (7.3580*\scaling,  -5.000000*\scaling)   -- (7.3580*\scaling, -1.000000*\scaling);
   \draw (7.3580*\scaling,  -5.000000*\scaling)   -- (7.3580*\scaling, -1.000000*\scaling);
   \node at (7.3580*\scaling, -5.000000*\scaling)  [rectangle,fill=white,draw, minimum width = 12pt, minimum height = 12pt,text width=, anchor=center] () {\tiny X};
   \node at (7.3580*\scaling, -1.000000*\scaling)  [rectangle,fill=white,draw, minimum width = 12pt, minimum height = 12pt,text width=, anchor=center] () {\tiny Z};
   \draw[line width=1mm, white] (8.2420*\scaling,  -6.000000*\scaling)   -- (8.2420*\scaling, -2.000000*\scaling);
   \draw (8.2420*\scaling,  -6.000000*\scaling)   -- (8.2420*\scaling, -2.000000*\scaling);
   \node at (8.2420*\scaling, -6.000000*\scaling)  [rectangle,fill=white,draw, minimum width = 12pt, minimum height = 12pt,text width=, anchor=center] () {\tiny X};
   \node at (8.2420*\scaling, -2.000000*\scaling)  [rectangle,fill=white,draw, minimum width = 12pt, minimum height = 12pt,text width=, anchor=center] () {\tiny Z};
   \node at (10.400000*\scaling, -5.000000*\scaling)  [rectangle,fill=white,draw, minimum width = 12pt, minimum height = 12pt,text width=, anchor=center] () {\tiny Z};
   \node at (10.400000*\scaling, -6.000000*\scaling)  [rectangle,fill=white,draw, minimum width = 12pt, minimum height = 12pt,text width=, anchor=center] () {\tiny Z};
\end{tikzpicture}
}}
\end{equation}
The upper five lines represent the data qubits, which are labeled 1-5, while the lower two lines represent the auxiliary qubits, which are labeled $A,B$. 
Step $5$ in the measurement sequence can serve as both the final measurement step of one stabilizer's measurement circuit and the initial measurement step of the subsequent circuit.
Operating in this way, each stabilizer measurements have an effective period of 4 steps.
This circuit is very similar to the pairwise measurement-based circuit introduced in Ref.~\onlinecite{Gidney2023pairmeasurement} for measuring surface code stabilizers; the difference is the use of two mixed Pauli ($XZ$) measurements. 
(We note that an equivalent circuit using slight different measurements can be obtained by conjugating the two auxiliary qubits by Hadamard operators, making the replacement $X \leftrightarrow Z$ for qubits $A$ and $B$.
The $SH$ automorphism of the $[[5,1,3]]$ code also allows us to modify the data qubit operators entering the stabilizer measurements, by choosing a different set of generators of the stabilizer group to measure.) 
The circuits for the other generators of the stabilizer group follow from cyclic permutation of the data qubits. 
To generate the full stabilizer group of the five qubit code, four independent generators must be measured, resulting in a circuit depth of $16r+1$ for $r$ rounds of syndrome extraction:

\begin{equation}
\vcenter{\hbox{
\begin{tikzpicture}[scale=0.8]
\foreach \y in {1,2,3,4,5,6,7}{
  \draw (-0.4,\y*0.2)--(2.5,\y*0.2);
}
\filldraw[fill=gray!10!white, rounded corners] (0,0) rectangle (2,7*0.2 + 0.2);
\node[align=center] at (1,4*0.2) {XZZXI};
\foreach \y in {1,2,3,4,5,6,7}{
  \draw (-0.5+3,\y*0.2)--(2.5+3,\y*0.2);
}
\filldraw[fill=gray!10!white, rounded corners] (0+ 2.5,0) rectangle (2+ 2.5,7*0.2 + 0.2);
\node[align=center] at (1+ 2.5,4*0.2) {IXZZX};
\foreach \y in {1,2,3,4,5,6,7}{
  \draw (-0.5+6,\y*0.2)--(2.5+6,\y*0.2);
}
\filldraw[fill=gray!10!white, rounded corners] (0+5,0) rectangle (2+5,7*0.2 + 0.2);
\node[align=center] at (1+5,4*0.2) {XIXZZ};
\foreach \y in {1,2,3,4,5,6,7}{
  \draw (-0.5+7.5,\y*0.2)--(2.4+7.5,\y*0.2);
}
\filldraw[fill=gray!10!white, rounded corners] (0+7.5,0) rectangle (2+7.5,7*0.2 + 0.2);
\node[align=center] at (1+7.5,4*0.2) {ZXIXZ};
\end{tikzpicture}}}
\end{equation}

The 5+2 code can correct any degree one Pauli fault, but it will fail for some degree one faults in the circuit noise model. 
In particular, readout errors on the single and pairwise auxiliary qubit measurements are equivalent to degree two data qubit errors, as are certain correlated (two-qubit) Pauli errors at the pairwise auxiliary qubit measurements. 
More specifically, a readout error on a single qubit $Z$ measurement is equivalent to two Pauli $X$ errors on that qubit -- one just before the measurement, and one after.
Each of these Pauli $X$ errors is then equivalent to a Pauli error on a (different) data qubit, yielding a degree two data qubit error.
A readout error on a pairwise auxiliary qubit $ZZ$ measurement is similarly equivalent to two Pauli $X$ errors on either one of the two auxiliary qubits, and thus a degree two data qubit error ($X_1 Z_2 \sim Z_3 X_4$ for this circuit).
Meanwhile, the following equivalence holds for two auxiliary qubit Pauli $X$ errors, which can result from a single circuit fault on the pairwise auxiliary qubit $ZZ$ measurements in step 3 (left), and two data qubit Pauli ($Z_2 Z_3 \sim X_1 X_4$ for this circuit) errors (right):
\begin{equation}
\vcenter{\hbox{
\begin{tikzpicture}
   \newcommand{\scaling}{0.6}
   \node at (-2*\scaling, 0.0000*\scaling)  [anchor=center] () {1};
   \draw[line width = 0.6mm, gray] (-1.5*\scaling, 0.0000*\scaling)   -- (9.3000*\scaling, 0.0000*\scaling);
   \node at (-2*\scaling, -1.0000*\scaling)  [anchor=center] () {2};
   \draw[line width = 0.6mm, gray] (-1.5*\scaling, -1.0000*\scaling)   -- (9.3000*\scaling, -1.0000*\scaling);
   \node at (-2*\scaling, -2.0000*\scaling)  [anchor=center] () {3};
   \draw[line width = 0.6mm, gray] (-1.5*\scaling, -2.0000*\scaling)   -- (9.3000*\scaling, -2.0000*\scaling);
   \node at (-2*\scaling, -3.0000*\scaling)  [anchor=center] () {4};
   \draw[line width = 0.6mm, gray] (-1.5*\scaling, -3.0000*\scaling)   -- (9.3000*\scaling, -3.0000*\scaling);
   \node at (-2*\scaling, -4.0000*\scaling)  [anchor=center] () {5};
   \draw[line width = 0.6mm, gray] (-1.5*\scaling, -4.0000*\scaling)   -- (9.3000*\scaling, -4.0000*\scaling);
   \node at (-2*\scaling, -5.0000*\scaling)  [anchor=center] () {A};
   \draw[line width = 0.6mm] (-1.5*\scaling, -5.0000*\scaling)   -- (9.3000*\scaling, -5.0000*\scaling);
   \node at (-2*\scaling, -6.0000*\scaling)  [anchor=center] () {B};
   \draw[line width = 0.6mm] (-1.5*\scaling, -6.0000*\scaling)   -- (9.3000*\scaling, -6.0000*\scaling);
   \draw[dotted] (-1.300000*\scaling, 0.000000*\scaling +1*\scaling)  -- (-1.300000*\scaling, -6.000000*\scaling -1*\scaling);
   \draw[dotted] (1.300000*\scaling, 0.000000*\scaling +1*\scaling)  -- (1.300000*\scaling, -6.000000*\scaling -1*\scaling);
   \draw[dotted] (3.900000*\scaling, 0.000000*\scaling +1*\scaling)  -- (3.900000*\scaling, -6.000000*\scaling -1*\scaling);
   \draw[dotted] (6.500000*\scaling, 0.000000*\scaling +1*\scaling)  -- (6.500000*\scaling, -6.000000*\scaling -1*\scaling);
   \draw[dotted] (9.100000*\scaling, 0.000000*\scaling +1*\scaling)  -- (9.100000*\scaling, -6.000000*\scaling -1*\scaling);
   \node at (0.0000*\scaling, -6.0000*\scaling -1*\scaling) {1};
   \node at (2.6000*\scaling, -6.0000*\scaling -1*\scaling) {2};
   \node at (5.2000*\scaling, -6.0000*\scaling -1*\scaling) {3};
   \node at (7.8000*\scaling, -6.0000*\scaling -1*\scaling) {4};
   \node at (0.000000*\scaling, -5.000000*\scaling)  [rectangle,fill=white,draw, minimum width = 12pt, minimum height = 12pt,text width=, anchor=center] () {\tiny Z};
   \node at (0.000000*\scaling, -6.000000*\scaling)  [rectangle,fill=white,draw, minimum width = 12pt, minimum height = 12pt,text width=, anchor=center] () {\tiny Z};
   \draw[line width=1mm, white] (2.1580*\scaling,  -5.000000*\scaling)   -- (2.1580*\scaling, 0.000000*\scaling);
   \draw (2.1580*\scaling,  -5.000000*\scaling)   -- (2.1580*\scaling, 0.000000*\scaling);
   \node at (2.1580*\scaling, -5.000000*\scaling)  [rectangle,fill=white,draw, minimum width = 12pt, minimum height = 12pt,text width=, anchor=center] () {\tiny X};
   \node at (2.1580*\scaling, 0.000000*\scaling)  [rectangle,fill=white,draw, minimum width = 12pt, minimum height = 12pt,text width=, anchor=center] () {\tiny X};
   \draw[line width=1mm, white] (3.0420*\scaling,  -6.000000*\scaling)   -- (3.0420*\scaling, -3.000000*\scaling);
   \draw (3.0420*\scaling,  -6.000000*\scaling)   -- (3.0420*\scaling, -3.000000*\scaling);
   \node at (3.0420*\scaling, -6.000000*\scaling)  [rectangle,fill=white,draw, minimum width = 12pt, minimum height = 12pt,text width=, anchor=center] () {\tiny X};
   \node at (3.0420*\scaling, -3.000000*\scaling)  [rectangle,fill=white,draw, minimum width = 12pt, minimum height = 12pt,text width=, anchor=center] () {\tiny X};
   \draw[line width=1mm, white] (5.2000*\scaling,  -6.000000*\scaling)   -- (5.2000*\scaling, -5.000000*\scaling);
   \draw (5.2000*\scaling,  -6.000000*\scaling)   -- (5.2000*\scaling, -5.000000*\scaling);
   \node at (5.2000*\scaling, -6.000000*\scaling)  [rectangle,fill=white,draw, minimum width = 12pt, minimum height = 12pt,text width=, anchor=center] () {\tiny Z};
   \node at (5.2000*\scaling, -5.000000*\scaling)  [rectangle,fill=white,draw, minimum width = 12pt, minimum height = 12pt,text width=, anchor=center] () {\tiny Z};
   \draw[line width=1mm, white] (7.3580*\scaling,  -5.000000*\scaling)   -- (7.3580*\scaling, -1.000000*\scaling);
   \draw (7.3580*\scaling,  -5.000000*\scaling)   -- (7.3580*\scaling, -1.000000*\scaling);
   \node at (7.3580*\scaling, -5.000000*\scaling)  [rectangle,fill=white,draw, minimum width = 12pt, minimum height = 12pt,text width=, anchor=center] () {\tiny X};
   \node at (7.3580*\scaling, -1.000000*\scaling)  [rectangle,fill=white,draw, minimum width = 12pt, minimum height = 12pt,text width=, anchor=center] () {\tiny Z};
   \draw[line width=1mm, white] (8.2420*\scaling,  -6.000000*\scaling)   -- (8.2420*\scaling, -2.000000*\scaling);
   \draw (8.2420*\scaling,  -6.000000*\scaling)   -- (8.2420*\scaling, -2.000000*\scaling);
   \node at (8.2420*\scaling, -6.000000*\scaling)  [rectangle,fill=white,draw, minimum width = 12pt, minimum height = 12pt,text width=, anchor=center] () {\tiny X};
   \node at (8.2420*\scaling, -2.000000*\scaling)  [rectangle,fill=white,draw, minimum width = 12pt, minimum height = 12pt,text width=, anchor=center] () {\tiny Z};
\end{tikzpicture}
\put(-55,36){\scalebox{0.7}{\begin{tikzpicture}
\draw[fill = yellow] (0,0) circle(0.2);
\node at (0,0) {\color{red} \bf X};
\end{tikzpicture}}}
\put(-55,19){\scalebox{0.7}{\begin{tikzpicture}
\draw[fill = yellow] (0,0) circle(0.2);
\node at (0,0) {\color{red} \bf X};
\end{tikzpicture}}}
}} \quad  \sim  \quad 
\vcenter{\hbox{
\begin{tikzpicture}
   \newcommand{\scaling}{0.6}
   \node at (-2*\scaling, 0.0000*\scaling)  [anchor=center] () {1};
   \draw[line width = 0.6mm, gray] (-1.5*\scaling, 0.0000*\scaling)   -- (9.3000*\scaling, 0.0000*\scaling);
   \node at (-2*\scaling, -1.0000*\scaling)  [anchor=center] () {2};
   \draw[line width = 0.6mm, gray] (-1.5*\scaling, -1.0000*\scaling)   -- (9.3000*\scaling, -1.0000*\scaling);
   \node at (-2*\scaling, -2.0000*\scaling)  [anchor=center] () {3};
   \draw[line width = 0.6mm, gray] (-1.5*\scaling, -2.0000*\scaling)   -- (9.3000*\scaling, -2.0000*\scaling);
   \node at (-2*\scaling, -3.0000*\scaling)  [anchor=center] () {4};
   \draw[line width = 0.6mm, gray] (-1.5*\scaling, -3.0000*\scaling)   -- (9.3000*\scaling, -3.0000*\scaling);
   \node at (-2*\scaling, -4.0000*\scaling)  [anchor=center] () {5};
   \draw[line width = 0.6mm, gray] (-1.5*\scaling, -4.0000*\scaling)   -- (9.3000*\scaling, -4.0000*\scaling);
   \node at (-2*\scaling, -5.0000*\scaling)  [anchor=center] () {A};
   \draw[line width = 0.6mm] (-1.5*\scaling, -5.0000*\scaling)   -- (9.3000*\scaling, -5.0000*\scaling);
   \node at (-2*\scaling, -6.0000*\scaling)  [anchor=center] () {B};
   \draw[line width = 0.6mm] (-1.5*\scaling, -6.0000*\scaling)   -- (9.3000*\scaling, -6.0000*\scaling);
   \draw[dotted] (-1.300000*\scaling, 0.000000*\scaling +1*\scaling)  -- (-1.300000*\scaling, -6.000000*\scaling -1*\scaling);
   \draw[dotted] (1.300000*\scaling, 0.000000*\scaling +1*\scaling)  -- (1.300000*\scaling, -6.000000*\scaling -1*\scaling);
   \draw[dotted] (3.900000*\scaling, 0.000000*\scaling +1*\scaling)  -- (3.900000*\scaling, -6.000000*\scaling -1*\scaling);
   \draw[dotted] (6.500000*\scaling, 0.000000*\scaling +1*\scaling)  -- (6.500000*\scaling, -6.000000*\scaling -1*\scaling);
   \draw[dotted] (9.100000*\scaling, 0.000000*\scaling +1*\scaling)  -- (9.100000*\scaling, -6.000000*\scaling -1*\scaling);
   \node at (0.0000*\scaling, -6.0000*\scaling -1*\scaling) {1};
   \node at (2.6000*\scaling, -6.0000*\scaling -1*\scaling) {2};
   \node at (5.2000*\scaling, -6.0000*\scaling -1*\scaling) {3};
   \node at (7.8000*\scaling, -6.0000*\scaling -1*\scaling) {4};
   \node at (0.000000*\scaling, -5.000000*\scaling)  [rectangle,fill=white,draw, minimum width = 12pt, minimum height = 12pt,text width=, anchor=center] () {\tiny Z};
   \node at (0.000000*\scaling, -6.000000*\scaling)  [rectangle,fill=white,draw, minimum width = 12pt, minimum height = 12pt,text width=, anchor=center] () {\tiny Z};
   \draw[line width=1mm, white] (2.1580*\scaling,  -5.000000*\scaling)   -- (2.1580*\scaling, 0.000000*\scaling);
   \draw (2.1580*\scaling,  -5.000000*\scaling)   -- (2.1580*\scaling, 0.000000*\scaling);
   \node at (2.1580*\scaling, -5.000000*\scaling)  [rectangle,fill=white,draw, minimum width = 12pt, minimum height = 12pt,text width=, anchor=center] () {\tiny X};
   \node at (2.1580*\scaling, 0.000000*\scaling)  [rectangle,fill=white,draw, minimum width = 12pt, minimum height = 12pt,text width=, anchor=center] () {\tiny X};
   \draw[line width=1mm, white] (3.0420*\scaling,  -6.000000*\scaling)   -- (3.0420*\scaling, -3.000000*\scaling);
   \draw (3.0420*\scaling,  -6.000000*\scaling)   -- (3.0420*\scaling, -3.000000*\scaling);
   \node at (3.0420*\scaling, -6.000000*\scaling)  [rectangle,fill=white,draw, minimum width = 12pt, minimum height = 12pt,text width=, anchor=center] () {\tiny X};
   \node at (3.0420*\scaling, -3.000000*\scaling)  [rectangle,fill=white,draw, minimum width = 12pt, minimum height = 12pt,text width=, anchor=center] () {\tiny X};
   \draw[line width=1mm, white] (5.2000*\scaling,  -6.000000*\scaling)   -- (5.2000*\scaling, -5.000000*\scaling);
   \draw (5.2000*\scaling,  -6.000000*\scaling)   -- (5.2000*\scaling, -5.000000*\scaling);
   \node at (5.2000*\scaling, -6.000000*\scaling)  [rectangle,fill=white,draw, minimum width = 12pt, minimum height = 12pt,text width=, anchor=center] () {\tiny Z};
   \node at (5.2000*\scaling, -5.000000*\scaling)  [rectangle,fill=white,draw, minimum width = 12pt, minimum height = 12pt,text width=, anchor=center] () {\tiny Z};
   \draw[line width=1mm, white] (7.3580*\scaling,  -5.000000*\scaling)   -- (7.3580*\scaling, -1.000000*\scaling);
   \draw (7.3580*\scaling,  -5.000000*\scaling)   -- (7.3580*\scaling, -1.000000*\scaling);
   \node at (7.3580*\scaling, -5.000000*\scaling)  [rectangle,fill=white,draw, minimum width = 12pt, minimum height = 12pt,text width=, anchor=center] () {\tiny X};
   \node at (7.3580*\scaling, -1.000000*\scaling)  [rectangle,fill=white,draw, minimum width = 12pt, minimum height = 12pt,text width=, anchor=center] () {\tiny Z};
   \draw[line width=1mm, white] (8.2420*\scaling,  -6.000000*\scaling)   -- (8.2420*\scaling, -2.000000*\scaling);
   \draw (8.2420*\scaling,  -6.000000*\scaling)   -- (8.2420*\scaling, -2.000000*\scaling);
   \node at (8.2420*\scaling, -6.000000*\scaling)  [rectangle,fill=white,draw, minimum width = 12pt, minimum height = 12pt,text width=, anchor=center] () {\tiny X};
   \node at (8.2420*\scaling, -2.000000*\scaling)  [rectangle,fill=white,draw, minimum width = 12pt, minimum height = 12pt,text width=, anchor=center] () {\tiny Z};
\end{tikzpicture}
\put(-54,105){\scalebox{0.7}{\begin{tikzpicture}
\draw[fill = yellow] (0,0) circle(0.2);
\node at (0,0) {\color{red} \bf Z};
\end{tikzpicture}}}
\put(-54,88){\scalebox{0.7}{\begin{tikzpicture}
\draw[fill = yellow] (0,0) circle(0.2);
\node at (0,0) {\color{red} \bf Z};
\end{tikzpicture}}}
\put(-52.5,20){\scalebox{0.65}{\begin{tikzpicture}
\draw[fill = gray!50!white] (0,0) circle(0.2);
\end{tikzpicture}}}
\put(-52.5,37){\scalebox{0.65}{\begin{tikzpicture}
\draw[fill = gray!50!white] (0,0) circle(0.2);
\end{tikzpicture}}}
\put(-45.5,23.75){\scalebox{0.65}{\begin{tikzpicture}
\draw[-latex, red, ultra thick] plot [smooth, tension=0] coordinates {(0.2,0.85)  (1.6,0.85)  (1.6,4.55)  (0.2,4.55)};
\end{tikzpicture}}}
\put(-45.5,40.75){\scalebox{0.65}{\begin{tikzpicture}
\draw[white, ultra thick] plot [smooth, tension=0] coordinates { (0.85,2.85)  (0.85,4.15)};
\draw[white, ultra thick] plot [smooth, tension=0] coordinates { (0.75,2.85)  (0.75,4.15)};
\draw[-latex, red, ultra thick] plot [smooth, tension=0] coordinates {(0.2,0.85)  (0.8,0.85)  (0.8,4.55)  (0.2,4.55)};
\end{tikzpicture}}}
}}
\end{equation}
Here, errors are marked in yellow, and the red arrows indicate the path through which the Pauli operators are ``pushed'' through the circuit using equivalences.
We refer to a degree one circuit fault as a \emph{hook error} if it acts on the logical operators as if it were a data qubit Pauli fault of degree two rather than one.

Readout errors on the single auxiliary qubit measurements can easily be dealt with by repeating those measurements, creating new detectors for the corresponding errors.
Similarly, readout errors on the pairwise auxiliary qubit measurements can be dealt with by repeating those measurements.
However, this strategy leaves the hook errors associated with correlated Pauli faults on the pairwise auxiliary qubit measurements, which require a more involved alteration of the circuit.
For the correlated Pauli faults, a na\"ive solution would be to add two flag qubits, one for each auxiliary qubit.
This solution works, but we can produce a different solution that requires only one additional qubit.
In this method, the additional qubit is used to mediate the pairwise auxiliary qubit measurement in a way that adds detectors that make the code robust against the circuit faults.

The shortest sequence to implement a mediated measurement would be of the form
\begin{equation}
\vcenter{\hbox{
\begin{tikzpicture}
   \newcommand{\scaling}{0.6}
   \node at (-2*\scaling, 0.0000*\scaling)  [anchor=center] () {A};
   \draw[line width = 0.6mm] (-1.5*\scaling, 0.0000*\scaling)   -- (1.5000*\scaling, 0.0000*\scaling);
   \node at (-2*\scaling, -1.0000*\scaling)  [anchor=center] () {B};
   \draw[line width = 0.6mm] (-1.5*\scaling, -1.0000*\scaling)   -- (1.5000*\scaling, -1.0000*\scaling);
   \draw[dotted] (-1.300000*\scaling, 0.000000*\scaling +1*\scaling)  -- (-1.300000*\scaling, -1.000000*\scaling -1*\scaling);
   \draw[dotted] (1.300000*\scaling, 0.000000*\scaling +1*\scaling)  -- (1.300000*\scaling, -1.000000*\scaling -1*\scaling);
   \draw[line width=1mm, white] (0.0000*\scaling,  -1.000000*\scaling)   -- (0.0000*\scaling, 0.000000*\scaling);
   \draw (0.0000*\scaling,  -1.000000*\scaling)   -- (0.0000*\scaling, 0.000000*\scaling);
   \node at (0.0000*\scaling, -1.000000*\scaling)  [rectangle,fill=white,draw, minimum width = 12pt, minimum height = 12pt,text width=, anchor=center] () {\tiny Z};
   \node at (0.0000*\scaling, 0.000000*\scaling)  [rectangle,fill=white,draw, minimum width = 12pt, minimum height = 12pt,text width=, anchor=center] () {\tiny Z};
\end{tikzpicture}
}} \quad  \rightarrow \quad 
\vcenter{\hbox{
\begin{tikzpicture}
   \newcommand{\scaling}{0.6}
   \node at (-2*\scaling, 0.0000*\scaling)  [anchor=center] () {A};
   \draw[line width = 0.6mm] (-1.5*\scaling, 0.0000*\scaling)   -- (9.3000*\scaling, 0.0000*\scaling);
   \node at (-2*\scaling, -1.0000*\scaling)  [anchor=center] () {B};
   \draw[line width = 0.6mm] (-1.5*\scaling, -1.0000*\scaling)   -- (9.3000*\scaling, -1.0000*\scaling);
   \node at (-2*\scaling, -2.0000*\scaling)  [anchor=center] () {C};
   \draw[line width = 0.6mm] (-1.5*\scaling, -2.0000*\scaling)   -- (9.3000*\scaling, -2.0000*\scaling);
   \draw[dotted] (-1.300000*\scaling, 0.000000*\scaling +1*\scaling)  -- (-1.300000*\scaling, -2.000000*\scaling -1*\scaling);
   \draw[dotted] (1.300000*\scaling, 0.000000*\scaling +1*\scaling)  -- (1.300000*\scaling, -2.000000*\scaling -1*\scaling);
   \draw[dotted] (3.900000*\scaling, 0.000000*\scaling +1*\scaling)  -- (3.900000*\scaling, -2.000000*\scaling -1*\scaling);
   \draw[dotted] (6.500000*\scaling, 0.000000*\scaling +1*\scaling)  -- (6.500000*\scaling, -2.000000*\scaling -1*\scaling);
   \draw[dotted] (9.100000*\scaling, 0.000000*\scaling +1*\scaling)  -- (9.100000*\scaling, -2.000000*\scaling -1*\scaling);
   \node at (0.000000*\scaling, -2.000000*\scaling)  [rectangle,fill=white,draw, minimum width = 12pt, minimum height = 12pt,text width=, anchor=center] () {\tiny X};
   \draw[line width=1mm, white] (2.6000*\scaling,  -2.000000*\scaling)   -- (2.6000*\scaling, 0.000000*\scaling);
   \draw (2.6000*\scaling,  -2.000000*\scaling)   -- (2.6000*\scaling, 0.000000*\scaling);
   \node at (2.6000*\scaling, -2.000000*\scaling)  [rectangle,fill=white,draw, minimum width = 12pt, minimum height = 12pt,text width=, anchor=center] () {\tiny Z};
   \node at (2.6000*\scaling, 0.000000*\scaling)  [rectangle,fill=white,draw, minimum width = 12pt, minimum height = 12pt,text width=, anchor=center] () {\tiny Z};
   \draw[line width=1mm, white] (5.2000*\scaling,  -2.000000*\scaling)   -- (5.2000*\scaling, -1.000000*\scaling);
   \draw (5.2000*\scaling,  -2.000000*\scaling)   -- (5.2000*\scaling, -1.000000*\scaling);
   \node at (5.2000*\scaling, -2.000000*\scaling)  [rectangle,fill=white,draw, minimum width = 12pt, minimum height = 12pt,text width=, anchor=center] () {\tiny Z};
   \node at (5.2000*\scaling, -1.000000*\scaling)  [rectangle,fill=white,draw, minimum width = 12pt, minimum height = 12pt,text width=, anchor=center] () {\tiny Z};
   \node at (7.800000*\scaling, -2.000000*\scaling)  [rectangle,fill=white,draw, minimum width = 12pt, minimum height = 12pt,text width=, anchor=center] () {\tiny X};
\end{tikzpicture}
}}.
\end{equation}
However, while this modification dispels the hook error from correlated two qubit Pauli faults on the pairwise auxiliary qubit measurements, it leaves in place the hook errors associated with readout errors and introduces another hook error associated with a single Pauli fault on the new (mediating) auxiliary qubit. 
As illustrated below, a single Pauli $X$ error on auxiliary qubit $C$ is equivalent to two Pauli $X$ errors on auxiliary qubit $B$ (or equivalently on auxiliary qubit $A$), which would further propagate onto two data qubit Pauli errors through pairwise measurements of $B$ and the data qubits.\footnote{For an example of propagation to data qubits of such an error, see Figure 5 of Ref. \onlinecite{granssamuelsson2023improved}.}
\begin{equation}
\vcenter{\hbox{
\begin{tikzpicture}
   \newcommand{\scaling}{0.6}
   \node at (-2*\scaling, 0.0000*\scaling)  [anchor=center] () {A};
   \draw[line width = 0.6mm] (-1.5*\scaling, 0.0000*\scaling)   -- (9.3000*\scaling, 0.0000*\scaling);
   \node at (-2*\scaling, -1.0000*\scaling)  [anchor=center] () {B};
   \draw[line width = 0.6mm] (-1.5*\scaling, -1.0000*\scaling)   -- (9.3000*\scaling, -1.0000*\scaling);
   \node at (-2*\scaling, -2.0000*\scaling)  [anchor=center] () {C};
   \draw[line width = 0.6mm] (-1.5*\scaling, -2.0000*\scaling)   -- (9.3000*\scaling, -2.0000*\scaling);
   \draw[dotted] (-1.300000*\scaling, 0.000000*\scaling +1*\scaling)  -- (-1.300000*\scaling, -2.000000*\scaling -1*\scaling);
   \draw[dotted] (1.300000*\scaling, 0.000000*\scaling +1*\scaling)  -- (1.300000*\scaling, -2.000000*\scaling -1*\scaling);
   \draw[dotted] (3.900000*\scaling, 0.000000*\scaling +1*\scaling)  -- (3.900000*\scaling, -2.000000*\scaling -1*\scaling);
   \draw[dotted] (6.500000*\scaling, 0.000000*\scaling +1*\scaling)  -- (6.500000*\scaling, -2.000000*\scaling -1*\scaling);
   \draw[dotted] (9.100000*\scaling, 0.000000*\scaling +1*\scaling)  -- (9.100000*\scaling, -2.000000*\scaling -1*\scaling);
   \node at (0.000000*\scaling, -2.000000*\scaling)  [rectangle,fill=white,draw, minimum width = 12pt, minimum height = 12pt,text width=, anchor=center] () {\tiny X};
   \draw[line width=1mm, white] (2.6000*\scaling,  -2.000000*\scaling)   -- (2.6000*\scaling, 0.000000*\scaling);
   \draw (2.6000*\scaling,  -2.000000*\scaling)   -- (2.6000*\scaling, 0.000000*\scaling);
   \node at (2.6000*\scaling, -2.000000*\scaling)  [rectangle,fill=white,draw, minimum width = 12pt, minimum height = 12pt,text width=, anchor=center] () {\tiny Z};
   \node at (2.6000*\scaling, 0.000000*\scaling)  [rectangle,fill=white,draw, minimum width = 12pt, minimum height = 12pt,text width=, anchor=center] () {\tiny Z};
   \draw[line width=1mm, white] (5.2000*\scaling,  -2.000000*\scaling)   -- (5.2000*\scaling, -1.000000*\scaling);
   \draw (5.2000*\scaling,  -2.000000*\scaling)   -- (5.2000*\scaling, -1.000000*\scaling);
   \node at (5.2000*\scaling, -2.000000*\scaling)  [rectangle,fill=white,draw, minimum width = 12pt, minimum height = 12pt,text width=, anchor=center] () {\tiny Z};
   \node at (5.2000*\scaling, -1.000000*\scaling)  [rectangle,fill=white,draw, minimum width = 12pt, minimum height = 12pt,text width=, anchor=center] () {\tiny Z};
   \node at (7.800000*\scaling, -2.000000*\scaling)  [rectangle,fill=white,draw, minimum width = 12pt, minimum height = 12pt,text width=, anchor=center] () {\tiny X};
\end{tikzpicture}
\put(-99,12){\scalebox{0.7}{\begin{tikzpicture}
\draw[fill = yellow] (0,0) circle(0.2);
\node at (0,0) {\color{red} \bf X};
\end{tikzpicture}}}
}} \quad \sim
\vcenter{\hbox{
\begin{tikzpicture}
   \newcommand{\scaling}{0.6}
   \node at (-2*\scaling, 0.0000*\scaling)  [anchor=center] () {A};
   \draw[line width = 0.6mm] (-1.5*\scaling, 0.0000*\scaling)   -- (9.3000*\scaling, 0.0000*\scaling);
   \node at (-2*\scaling, -1.0000*\scaling)  [anchor=center] () {B};
   \draw[line width = 0.6mm] (-1.5*\scaling, -1.0000*\scaling)   -- (9.3000*\scaling, -1.0000*\scaling);
   \node at (-2*\scaling, -2.0000*\scaling)  [anchor=center] () {C};
   \draw[line width = 0.6mm] (-1.5*\scaling, -2.0000*\scaling)   -- (9.3000*\scaling, -2.0000*\scaling);
   \draw[dotted] (-1.300000*\scaling, 0.000000*\scaling +1*\scaling)  -- (-1.300000*\scaling, -2.000000*\scaling -1*\scaling);
   \draw[dotted] (1.300000*\scaling, 0.000000*\scaling +1*\scaling)  -- (1.300000*\scaling, -2.000000*\scaling -1*\scaling);
   \draw[dotted] (3.900000*\scaling, 0.000000*\scaling +1*\scaling)  -- (3.900000*\scaling, -2.000000*\scaling -1*\scaling);
   \draw[dotted] (6.500000*\scaling, 0.000000*\scaling +1*\scaling)  -- (6.500000*\scaling, -2.000000*\scaling -1*\scaling);
   \draw[dotted] (9.100000*\scaling, 0.000000*\scaling +1*\scaling)  -- (9.100000*\scaling, -2.000000*\scaling -1*\scaling);
   \node at (0.000000*\scaling, -2.000000*\scaling)  [rectangle,fill=white,draw, minimum width = 12pt, minimum height = 12pt,text width=, anchor=center] () {\tiny X};
   \draw[line width=1mm, white] (2.6000*\scaling,  -2.000000*\scaling)   -- (2.6000*\scaling, 0.000000*\scaling);
   \draw (2.6000*\scaling,  -2.000000*\scaling)   -- (2.6000*\scaling, 0.000000*\scaling);
   \node at (2.6000*\scaling, -2.000000*\scaling)  [rectangle,fill=white,draw, minimum width = 12pt, minimum height = 12pt,text width=, anchor=center] () {\tiny Z};
   \node at (2.6000*\scaling, 0.000000*\scaling)  [rectangle,fill=white,draw, minimum width = 12pt, minimum height = 12pt,text width=, anchor=center] () {\tiny Z};
   \draw[line width=1mm, white] (5.2000*\scaling,  -2.000000*\scaling)   -- (5.2000*\scaling, -1.000000*\scaling);
   \draw (5.2000*\scaling,  -2.000000*\scaling)   -- (5.2000*\scaling, -1.000000*\scaling);
   \node at (5.2000*\scaling, -2.000000*\scaling)  [rectangle,fill=white,draw, minimum width = 12pt, minimum height = 12pt,text width=, anchor=center] () {\tiny Z};
   \node at (5.2000*\scaling, -1.000000*\scaling)  [rectangle,fill=white,draw, minimum width = 12pt, minimum height = 12pt,text width=, anchor=center] () {\tiny Z};
   \node at (7.800000*\scaling, -2.000000*\scaling)  [rectangle,fill=white,draw, minimum width = 12pt, minimum height = 12pt,text width=, anchor=center] () {\tiny X};
\end{tikzpicture}
\put(-99,30){\scalebox{0.7}{\begin{tikzpicture}
\draw[fill = yellow] (0,0) circle(0.2);
\node at (0,0) {\color{red} \bf X};
\end{tikzpicture}}}
\put(-55,30){\scalebox{0.7}{\begin{tikzpicture}
\draw[fill = yellow] (0,0) circle(0.2);
\node at (0,0) {\color{red} \bf X};
\end{tikzpicture}}}
\put(-97,13){\scalebox{0.65}{\begin{tikzpicture}
\draw[fill = gray!50!white] (0,0) circle(0.2);
\end{tikzpicture}}}
\put(-90,16.75){\scalebox{0.65}{\begin{tikzpicture}
\draw[-latex, red, ultra thick] plot [smooth, tension=0] coordinates {(0.2,0.85)  (1.2,0.85)  (1.2,1.77)  (0.2,1.77)};
\draw[-latex, red, ultra thick] plot [smooth, tension=0] coordinates {(0.2,0.85)  (1.2,0.85)  (1.2,1.77)  (2.2,1.77)};
\draw[-latex, red, ultra thick] plot [smooth, tension=0] coordinates {(0.2,0.85)  (3.6,0.85)};
\end{tikzpicture}}}
}}.
\end{equation}
Fortunately, all the remaining hook errors in this three auxiliary qubit circuit can be deal with by repeating auxiliary qubit measurements.
For the pairwise auxiliary qubit measurements, they should be repeated in an alternating fashion, which generates two new detectors (shown in Appendix~\ref{Detectors}, circuit diagram~\eqref{hook_detectors}) that combine to ensure that no single circuit fault is equivalent to a degree two data qubit Pauli error. 
Putting it all together to yield the 5+3 code, an XZZXI stabilizer measurement is performed through the following depth 8 circuit: 
\begin{equation}
\vcenter{\hbox{
\begin{tikzpicture}
   \newcommand{\scaling}{0.6}
   \node at (-2*\scaling, 0.0000*\scaling)  [anchor=center] () {1};
   \draw[line width = 0.6mm, gray] (-1.5*\scaling, 0.0000*\scaling)   -- (19.7000*\scaling, 0.0000*\scaling);
   \node at (-2*\scaling, -1.0000*\scaling)  [anchor=center] () {2};
   \draw[line width = 0.6mm, gray] (-1.5*\scaling, -1.0000*\scaling)   -- (19.7000*\scaling, -1.0000*\scaling);
   \node at (-2*\scaling, -2.0000*\scaling)  [anchor=center] () {3};
   \draw[line width = 0.6mm, gray] (-1.5*\scaling, -2.0000*\scaling)   -- (19.7000*\scaling, -2.0000*\scaling);
   \node at (-2*\scaling, -3.0000*\scaling)  [anchor=center] () {4};
   \draw[line width = 0.6mm, gray] (-1.5*\scaling, -3.0000*\scaling)   -- (19.7000*\scaling, -3.0000*\scaling);
   \node at (-2*\scaling, -4.0000*\scaling)  [anchor=center] () {5};
   \draw[line width = 0.6mm, gray] (-1.5*\scaling, -4.0000*\scaling)   -- (19.7000*\scaling, -4.0000*\scaling);
   \node at (-2*\scaling, -5.0000*\scaling)  [anchor=center] () {A};
   \draw[line width = 0.6mm] (-1.5*\scaling, -5.0000*\scaling)   -- (19.7000*\scaling, -5.0000*\scaling);
   \node at (-2*\scaling, -6.0000*\scaling)  [anchor=center] () {B};
   \draw[line width = 0.6mm] (-1.5*\scaling, -6.0000*\scaling)   -- (19.7000*\scaling, -6.0000*\scaling);
   \node at (-2*\scaling, -7.0000*\scaling)  [anchor=center] () {C};
   \draw[line width = 0.6mm] (-1.5*\scaling, -7.0000*\scaling)   -- (19.7000*\scaling, -7.0000*\scaling);
   \draw[dotted] (-1.300000*\scaling, 0.000000*\scaling +1*\scaling)  -- (-1.300000*\scaling, -7.000000*\scaling -1*\scaling);
   \draw[dotted] (1.300000*\scaling, 0.000000*\scaling +1*\scaling)  -- (1.300000*\scaling, -7.000000*\scaling -1*\scaling);
   \draw[dotted] (3.900000*\scaling, 0.000000*\scaling +1*\scaling)  -- (3.900000*\scaling, -7.000000*\scaling -1*\scaling);
   \draw[dotted] (6.500000*\scaling, 0.000000*\scaling +1*\scaling)  -- (6.500000*\scaling, -7.000000*\scaling -1*\scaling);
   \draw[dotted] (9.100000*\scaling, 0.000000*\scaling +1*\scaling)  -- (9.100000*\scaling, -7.000000*\scaling -1*\scaling);
   \draw[dotted] (11.700000*\scaling, 0.000000*\scaling +1*\scaling)  -- (11.700000*\scaling, -7.000000*\scaling -1*\scaling);
   \draw[dotted] (14.300000*\scaling, 0.000000*\scaling +1*\scaling)  -- (14.300000*\scaling, -7.000000*\scaling -1*\scaling);
   \draw[dotted] (16.900000*\scaling, 0.000000*\scaling +1*\scaling)  -- (16.900000*\scaling, -7.000000*\scaling -1*\scaling);
   \draw[dotted] (19.500000*\scaling, 0.000000*\scaling +1*\scaling)  -- (19.500000*\scaling, -7.000000*\scaling -1*\scaling);
   \node at (0.0000*\scaling, -7.0000*\scaling -1*\scaling) {1};
   \node at (2.6000*\scaling, -7.0000*\scaling -1*\scaling) {2};
   \node at (5.2000*\scaling, -7.0000*\scaling -1*\scaling) {3};
   \node at (7.8000*\scaling, -7.0000*\scaling -1*\scaling) {4};
   \node at (10.4000*\scaling, -7.0000*\scaling -1*\scaling) {5};
   \node at (13.0000*\scaling, -7.0000*\scaling -1*\scaling) {6};
   \node at (15.6000*\scaling, -7.0000*\scaling -1*\scaling) {7};
   \node at (18.2000*\scaling, -7.0000*\scaling -1*\scaling) {8};
   \node at (0.000000*\scaling, -5.000000*\scaling)  [rectangle,fill=white,draw, minimum width = 12pt, minimum height = 12pt,text width=, anchor=center] () {\tiny Z};
   \node at (0.000000*\scaling, -6.000000*\scaling)  [rectangle,fill=white,draw, minimum width = 12pt, minimum height = 12pt,text width=, anchor=center] () {\tiny Z};
   \node at (0.000000*\scaling, -7.000000*\scaling)  [rectangle,fill=white,draw, minimum width = 12pt, minimum height = 12pt,text width=, anchor=center] () {\tiny X};
   \draw[line width=1mm, white] (2.1580*\scaling,  -5.000000*\scaling)   -- (2.1580*\scaling, 0.000000*\scaling);
   \draw (2.1580*\scaling,  -5.000000*\scaling)   -- (2.1580*\scaling, 0.000000*\scaling);
   \node at (2.1580*\scaling, -5.000000*\scaling)  [rectangle,fill=white,draw, minimum width = 12pt, minimum height = 12pt,text width=, anchor=center] () {\tiny X};
   \node at (2.1580*\scaling, 0.000000*\scaling)  [rectangle,fill=white,draw, minimum width = 12pt, minimum height = 12pt,text width=, anchor=center] () {\tiny X};
   \draw[line width=1mm, white] (3.0420*\scaling,  -6.000000*\scaling)   -- (3.0420*\scaling, -3.000000*\scaling);
   \draw (3.0420*\scaling,  -6.000000*\scaling)   -- (3.0420*\scaling, -3.000000*\scaling);
   \node at (3.0420*\scaling, -6.000000*\scaling)  [rectangle,fill=white,draw, minimum width = 12pt, minimum height = 12pt,text width=, anchor=center] () {\tiny X};
   \node at (3.0420*\scaling, -3.000000*\scaling)  [rectangle,fill=white,draw, minimum width = 12pt, minimum height = 12pt,text width=, anchor=center] () {\tiny X};
   \draw[line width=1mm, white] (5.2000*\scaling,  -7.000000*\scaling)   -- (5.2000*\scaling, -5.000000*\scaling);
   \draw (5.2000*\scaling,  -7.000000*\scaling)   -- (5.2000*\scaling, -5.000000*\scaling);
   \node at (5.2000*\scaling, -7.000000*\scaling)  [rectangle,fill=white,draw, minimum width = 12pt, minimum height = 12pt,text width=, anchor=center] () {\tiny Z};
   \node at (5.2000*\scaling, -5.000000*\scaling)  [rectangle,fill=white,draw, minimum width = 12pt, minimum height = 12pt,text width=, anchor=center] () {\tiny Z};
   \draw[line width=1mm, white] (7.8000*\scaling,  -7.000000*\scaling)   -- (7.8000*\scaling, -6.000000*\scaling);
   \draw (7.8000*\scaling,  -7.000000*\scaling)   -- (7.8000*\scaling, -6.000000*\scaling);
   \node at (7.8000*\scaling, -7.000000*\scaling)  [rectangle,fill=white,draw, minimum width = 12pt, minimum height = 12pt,text width=, anchor=center] () {\tiny Z};
   \node at (7.8000*\scaling, -6.000000*\scaling)  [rectangle,fill=white,draw, minimum width = 12pt, minimum height = 12pt,text width=, anchor=center] () {\tiny Z};
   \draw[line width=1mm, white] (10.4000*\scaling,  -7.000000*\scaling)   -- (10.4000*\scaling, -5.000000*\scaling);
   \draw (10.4000*\scaling,  -7.000000*\scaling)   -- (10.4000*\scaling, -5.000000*\scaling);
   \node at (10.4000*\scaling, -7.000000*\scaling)  [rectangle,fill=white,draw, minimum width = 12pt, minimum height = 12pt,text width=, anchor=center] () {\tiny Z};
   \node at (10.4000*\scaling, -5.000000*\scaling)  [rectangle,fill=white,draw, minimum width = 12pt, minimum height = 12pt,text width=, anchor=center] () {\tiny Z};
   \draw[line width=1mm, white] (13.0000*\scaling,  -7.000000*\scaling)   -- (13.0000*\scaling, -6.000000*\scaling);
   \draw (13.0000*\scaling,  -7.000000*\scaling)   -- (13.0000*\scaling, -6.000000*\scaling);
   \node at (13.0000*\scaling, -7.000000*\scaling)  [rectangle,fill=white,draw, minimum width = 12pt, minimum height = 12pt,text width=, anchor=center] () {\tiny Z};
   \node at (13.0000*\scaling, -6.000000*\scaling)  [rectangle,fill=white,draw, minimum width = 12pt, minimum height = 12pt,text width=, anchor=center] () {\tiny Z};
   \draw[line width=1mm, white] (15.1580*\scaling,  -5.000000*\scaling)   -- (15.1580*\scaling, -1.000000*\scaling);
   \draw (15.1580*\scaling,  -5.000000*\scaling)   -- (15.1580*\scaling, -1.000000*\scaling);
   \node at (15.1580*\scaling, -5.000000*\scaling)  [rectangle,fill=white,draw, minimum width = 12pt, minimum height = 12pt,text width=, anchor=center] () {\tiny X};
   \node at (15.1580*\scaling, -1.000000*\scaling)  [rectangle,fill=white,draw, minimum width = 12pt, minimum height = 12pt,text width=, anchor=center] () {\tiny Z};
   \draw[line width=1mm, white] (16.0420*\scaling,  -6.000000*\scaling)   -- (16.0420*\scaling, -2.000000*\scaling);
   \draw (16.0420*\scaling,  -6.000000*\scaling)   -- (16.0420*\scaling, -2.000000*\scaling);
   \node at (16.0420*\scaling, -6.000000*\scaling)  [rectangle,fill=white,draw, minimum width = 12pt, minimum height = 12pt,text width=, anchor=center] () {\tiny X};
   \node at (16.0420*\scaling, -2.000000*\scaling)  [rectangle,fill=white,draw, minimum width = 12pt, minimum height = 12pt,text width=, anchor=center] () {\tiny Z};
   \node at (18.200000*\scaling, -5.000000*\scaling)  [rectangle,fill=white,draw, minimum width = 12pt, minimum height = 12pt,text width=, anchor=center] () {\tiny Z};
   \node at (18.200000*\scaling, -6.000000*\scaling)  [rectangle,fill=white,draw, minimum width = 12pt, minimum height = 12pt,text width=, anchor=center] () {\tiny Z};
   \node at (18.200000*\scaling, -7.000000*\scaling)  [rectangle,fill=white,draw, minimum width = 12pt, minimum height = 12pt,text width=, anchor=center] () {\tiny X};
\end{tikzpicture}
}}
,
\end{equation}
which is free of hook errors.
The data qubits are again labeled 1-5, and the auxiliary qubits are labeled $A,B,C$. The circuits for the other generators of the stabilizer group follow from cyclic permutation of the data qubits. 

The measurements can be shifted in time such that the final (step 8) measurement of one stabilizer can be performed at the same time as the initial (step 1) measurement of the next, so that sequences of stabilizer measurement circuits effectively have a period of 7 steps.
This reduces the total circuit depth of $r$ rounds of syndrome extraction (measuring four different stabilizers) to $28r+1$.
The following circuit shows the more compact syndrome extraction for the XZZXI stabilizer:
\begin{equation}
\vcenter{\hbox{
\begin{tikzpicture}
   \newcommand{\scaling}{0.6}
   \node at (-2*\scaling, 0.0000*\scaling)  [anchor=center] () {1};
   \draw[line width = 0.6mm, gray] (-1.5*\scaling, 0.0000*\scaling)   -- (19.7000*\scaling, 0.0000*\scaling);
   \node at (-2*\scaling, -1.0000*\scaling)  [anchor=center] () {2};
   \draw[line width = 0.6mm, gray] (-1.5*\scaling, -1.0000*\scaling)   -- (19.7000*\scaling, -1.0000*\scaling);
   \node at (-2*\scaling, -2.0000*\scaling)  [anchor=center] () {3};
   \draw[line width = 0.6mm, gray] (-1.5*\scaling, -2.0000*\scaling)   -- (19.7000*\scaling, -2.0000*\scaling);
   \node at (-2*\scaling, -3.0000*\scaling)  [anchor=center] () {4};
   \draw[line width = 0.6mm, gray] (-1.5*\scaling, -3.0000*\scaling)   -- (19.7000*\scaling, -3.0000*\scaling);
   \node at (-2*\scaling, -4.0000*\scaling)  [anchor=center] () {5};
   \draw[line width = 0.6mm, gray] (-1.5*\scaling, -4.0000*\scaling)   -- (19.7000*\scaling, -4.0000*\scaling);
   \node at (-2*\scaling, -5.0000*\scaling)  [anchor=center] () {A};
   \draw[line width = 0.6mm] (-1.5*\scaling, -5.0000*\scaling)   -- (19.7000*\scaling, -5.0000*\scaling);
   \node at (-2*\scaling, -6.0000*\scaling)  [anchor=center] () {B};
   \draw[line width = 0.6mm] (-1.5*\scaling, -6.0000*\scaling)   -- (19.7000*\scaling, -6.0000*\scaling);
   \node at (-2*\scaling, -7.0000*\scaling)  [anchor=center] () {C};
   \draw[line width = 0.6mm] (-1.5*\scaling, -7.0000*\scaling)   -- (19.7000*\scaling, -7.0000*\scaling);
   \draw[dotted] (-1.300000*\scaling, 0.000000*\scaling +1*\scaling)  -- (-1.300000*\scaling, -7.000000*\scaling -1*\scaling);
   \draw[dotted] (1.300000*\scaling, 0.000000*\scaling +1*\scaling)  -- (1.300000*\scaling, -7.000000*\scaling -1*\scaling);
   \draw[dotted] (3.900000*\scaling, 0.000000*\scaling +1*\scaling)  -- (3.900000*\scaling, -7.000000*\scaling -1*\scaling);
   \draw[dotted] (6.500000*\scaling, 0.000000*\scaling +1*\scaling)  -- (6.500000*\scaling, -7.000000*\scaling -1*\scaling);
   \draw[dotted] (9.100000*\scaling, 0.000000*\scaling +1*\scaling)  -- (9.100000*\scaling, -7.000000*\scaling -1*\scaling);
   \draw[dotted] (11.700000*\scaling, 0.000000*\scaling +1*\scaling)  -- (11.700000*\scaling, -7.000000*\scaling -1*\scaling);
   \draw[dotted] (14.300000*\scaling, 0.000000*\scaling +1*\scaling)  -- (14.300000*\scaling, -7.000000*\scaling -1*\scaling);
   \draw[dotted] (16.900000*\scaling, 0.000000*\scaling +1*\scaling)  -- (16.900000*\scaling, -7.000000*\scaling -1*\scaling);
   \draw[dotted] (19.500000*\scaling, 0.000000*\scaling +1*\scaling)  -- (19.500000*\scaling, -7.000000*\scaling -1*\scaling);
   \node at (0.0000*\scaling, -7.0000*\scaling -1*\scaling) {1};
   \node at (2.6000*\scaling, -7.0000*\scaling -1*\scaling) {2};
   \node at (5.2000*\scaling, -7.0000*\scaling -1*\scaling) {3};
   \node at (7.8000*\scaling, -7.0000*\scaling -1*\scaling) {4};
   \node at (10.4000*\scaling, -7.0000*\scaling -1*\scaling) {5};
   \node at (13.0000*\scaling, -7.0000*\scaling -1*\scaling) {6};
   \node at (15.6000*\scaling, -7.0000*\scaling -1*\scaling) {7};
   \node at (18.2000*\scaling, -7.0000*\scaling -1*\scaling) {8};
   \node at (0.000000*\scaling, -5.000000*\scaling)  [rectangle,fill=white,draw, minimum width = 12pt, minimum height = 12pt,text width=, anchor=center] () {\tiny Z};
   \node at (2.600000*\scaling, -7.000000*\scaling)  [rectangle,fill=white,draw, minimum width = 12pt, minimum height = 12pt,text width=, anchor=center] () {\tiny X};
   \node at (2.600000*\scaling, -6.000000*\scaling)  [rectangle,fill=white,draw, minimum width = 12pt, minimum height = 12pt,text width=, anchor=center] () {\tiny Z};
   \draw[line width=1mm, white] (2.6000*\scaling,  -5.000000*\scaling)   -- (2.6000*\scaling, 0.000000*\scaling);
   \draw (2.6000*\scaling,  -5.000000*\scaling)   -- (2.6000*\scaling, 0.000000*\scaling);
   \node at (2.6000*\scaling, -5.000000*\scaling)  [rectangle,fill=white,draw, minimum width = 12pt, minimum height = 12pt,text width=, anchor=center] () {\tiny X};
   \node at (2.6000*\scaling, 0.000000*\scaling)  [rectangle,fill=white,draw, minimum width = 12pt, minimum height = 12pt,text width=, anchor=center] () {\tiny X};
   \draw[line width=1mm, white] (4.7580*\scaling,  -6.000000*\scaling)   -- (4.7580*\scaling, -3.000000*\scaling);
   \draw (4.7580*\scaling,  -6.000000*\scaling)   -- (4.7580*\scaling, -3.000000*\scaling);
   \node at (4.7580*\scaling, -6.000000*\scaling)  [rectangle,fill=white,draw, minimum width = 12pt, minimum height = 12pt,text width=, anchor=center] () {\tiny X};
   \node at (4.7580*\scaling, -3.000000*\scaling)  [rectangle,fill=white,draw, minimum width = 12pt, minimum height = 12pt,text width=, anchor=center] () {\tiny X};
   \draw[line width=1mm, white] (5.6420*\scaling,  -7.000000*\scaling)   -- (5.6420*\scaling, -5.000000*\scaling);
   \draw (5.6420*\scaling,  -7.000000*\scaling)   -- (5.6420*\scaling, -5.000000*\scaling);
   \node at (5.6420*\scaling, -7.000000*\scaling)  [rectangle,fill=white,draw, minimum width = 12pt, minimum height = 12pt,text width=, anchor=center] () {\tiny Z};
   \node at (5.6420*\scaling, -5.000000*\scaling)  [rectangle,fill=white,draw, minimum width = 12pt, minimum height = 12pt,text width=, anchor=center] () {\tiny Z};
   \draw[line width=1mm, white] (7.8000*\scaling,  -7.000000*\scaling)   -- (7.8000*\scaling, -6.000000*\scaling);
   \draw (7.8000*\scaling,  -7.000000*\scaling)   -- (7.8000*\scaling, -6.000000*\scaling);
   \node at (7.8000*\scaling, -7.000000*\scaling)  [rectangle,fill=white,draw, minimum width = 12pt, minimum height = 12pt,text width=, anchor=center] () {\tiny Z};
   \node at (7.8000*\scaling, -6.000000*\scaling)  [rectangle,fill=white,draw, minimum width = 12pt, minimum height = 12pt,text width=, anchor=center] () {\tiny Z};
   \draw[line width=1mm, white] (10.4000*\scaling,  -7.000000*\scaling)   -- (10.4000*\scaling, -5.000000*\scaling);
   \draw (10.4000*\scaling,  -7.000000*\scaling)   -- (10.4000*\scaling, -5.000000*\scaling);
   \node at (10.4000*\scaling, -7.000000*\scaling)  [rectangle,fill=white,draw, minimum width = 12pt, minimum height = 12pt,text width=, anchor=center] () {\tiny Z};
   \node at (10.4000*\scaling, -5.000000*\scaling)  [rectangle,fill=white,draw, minimum width = 12pt, minimum height = 12pt,text width=, anchor=center] () {\tiny Z};
   \draw[line width=1mm, white] (13.0000*\scaling,  -7.000000*\scaling)   -- (13.0000*\scaling, -6.000000*\scaling);
   \draw (13.0000*\scaling,  -7.000000*\scaling)   -- (13.0000*\scaling, -6.000000*\scaling);
   \node at (13.0000*\scaling, -7.000000*\scaling)  [rectangle,fill=white,draw, minimum width = 12pt, minimum height = 12pt,text width=, anchor=center] () {\tiny Z};
   \node at (13.0000*\scaling, -6.000000*\scaling)  [rectangle,fill=white,draw, minimum width = 12pt, minimum height = 12pt,text width=, anchor=center] () {\tiny Z};
   \draw[line width=1mm, white] (13.0000*\scaling,  -5.000000*\scaling)   -- (13.0000*\scaling, -1.000000*\scaling);
   \draw (13.0000*\scaling,  -5.000000*\scaling)   -- (13.0000*\scaling, -1.000000*\scaling);
   \node at (13.0000*\scaling, -5.000000*\scaling)  [rectangle,fill=white,draw, minimum width = 12pt, minimum height = 12pt,text width=, anchor=center] () {\tiny X};
   \node at (13.0000*\scaling, -1.000000*\scaling)  [rectangle,fill=white,draw, minimum width = 12pt, minimum height = 12pt,text width=, anchor=center] () {\tiny Z};
   \draw[line width=1mm, white] (15.1580*\scaling,  -6.000000*\scaling)   -- (15.1580*\scaling, -2.000000*\scaling);
   \draw (15.1580*\scaling,  -6.000000*\scaling)   -- (15.1580*\scaling, -2.000000*\scaling);
   \node at (15.1580*\scaling, -6.000000*\scaling)  [rectangle,fill=white,draw, minimum width = 12pt, minimum height = 12pt,text width=, anchor=center] () {\tiny X};
   \node at (15.1580*\scaling, -2.000000*\scaling)  [rectangle,fill=white,draw, minimum width = 12pt, minimum height = 12pt,text width=, anchor=center] () {\tiny Z};
   \node at (16.042000*\scaling, -5.000000*\scaling)  [rectangle,fill=white,draw, minimum width = 12pt, minimum height = 12pt,text width=, anchor=center] () {\tiny Z};
   \node at (16.042000*\scaling, -7.000000*\scaling)  [rectangle,fill=white,draw, minimum width = 12pt, minimum height = 12pt,text width=, anchor=center] () {\tiny X};
   \node at (18.200000*\scaling, -6.000000*\scaling)  [rectangle,fill=white,draw, minimum width = 12pt, minimum height = 12pt,text width=, anchor=center] () {\tiny Z};
\end{tikzpicture}
}}
\end{equation}

\FloatBarrier

\section{Square lattice layout}\label{Square}

In this Section, we describe how the 5+3 code can be implemented with a square lattice connectivity, such that pairwise measurements are only performed between horizontal and vertical nearest neighbors. (A square lattice connectivity can also be obtained through similar means for the 5+2 code, but it is then to the price of a slightly deeper syndrome extraction circuit.) 
In the following, we denote data qubits using open dots and auxiliary qubits using solid dots. 
With slight modifications of the 5+3 code syndrome extraction (that do not increase the depth), we can implement the 5+3 code using the following connectivity:
\begin{equation}
\vcenter{\hbox{
\begin{tikzpicture}[scale = 0.5]
\newcommand{\circlesize}{2}
 \draw (1, 0) -- (1, 1) ;
 \draw (2, 0) -- (2, 1) ;
 \draw (0,0) -- (0,1);
  \draw (0,0) -- (2,0);
 \draw (0, 0) -- (0, -1) ;
 \draw (2, 0) -- (2, -1) ;
  \draw (0, 0) node[circle, draw, fill=black, minimum size = \circlesize mm, inner sep = 0]{};
  \draw (1, 0) node[circle, draw, fill=black, minimum size = \circlesize mm, inner sep = 0]{};
  \draw (2, 0) node[circle, draw, fill=black, minimum size = \circlesize mm, inner sep = 0]{};
  %
  \draw (0, 1) node[draw,fill=white, circle, minimum size = \circlesize mm, inner sep = 0]{};
   \draw (1, 1) node[draw,fill=white, circle, minimum size = \circlesize mm, inner sep = 0]{};
  \draw (2, 1) node[draw,fill=white, circle, minimum size = \circlesize mm, inner sep = 0]{};
  \draw (0, -1) node[draw,fill=white, circle, minimum size = \circlesize mm, inner sep = 0]{};
   \draw (2, -1) node[draw,fill=white, circle, minimum size = \circlesize mm, inner sep = 0]{};
\end{tikzpicture}}}
\end{equation}
More specifically, one of the stabilizer measurements is performed in the usual fashion, with the layout 
\begin{equation}
\vcenter{\hbox{
\begin{tikzpicture}[scale = 0.5]
\newcommand{\circlesize}{2}
 \draw (2, 0) -- (2, 1) ;
 \draw (0,0) -- (0,1);
  \draw (0,0) -- (2,0);
 \draw (0, 0) -- (0, -1) ;
 \draw (2, 0) -- (2, -1) ;
  \draw (0, 0) node[circle, draw, fill=black, minimum size = \circlesize mm, inner sep = 0]{};
  \draw (1, 0) node[circle, draw, fill=black, minimum size = \circlesize mm, inner sep = 0]{};
  \draw (2, 0) node[circle, draw, fill=black, minimum size = \circlesize mm, inner sep = 0]{};
  %
  \draw (0, 1) node[draw,fill=white, circle, minimum size = \circlesize mm, inner sep = 0]{};
  \draw (2, 1) node[draw,fill=white, circle, minimum size = \circlesize mm, inner sep = 0]{};
  \draw (0, -1) node[draw,fill=white, circle, minimum size = \circlesize mm, inner sep = 0]{};
   \draw (2, -1) node[draw,fill=white, circle, minimum size = \circlesize mm, inner sep = 0]{};
\end{tikzpicture}}}
\end{equation}
while the other three stabilizers will involve the upper middle data qubit, with layout of the form
\begin{equation}
\vcenter{\hbox{
\begin{tikzpicture}[scale = 0.5]
\newcommand{\circlesize}{2}
 \draw (1, 0) -- (1, 1) ;
 \draw (2, 0) -- (2, 1) ;
  \draw (0,0) -- (2,0);
 \draw (0, 0) -- (0, -1) ;
 \draw (2, 0) -- (2, -1) ;
  \draw (0, 0) node[circle, draw, fill=black, minimum size = \circlesize mm, inner sep = 0]{};
  \draw (1, 0) node[circle, draw, fill=black, minimum size = \circlesize mm, inner sep = 0]{};
  \draw (2, 0) node[circle, draw, fill=black, minimum size = \circlesize mm, inner sep = 0]{};
  %
   \draw (1, 1) node[draw,fill=white, circle, minimum size = \circlesize mm, inner sep = 0]{};
  \draw (2, 1) node[draw,fill=white, circle, minimum size = \circlesize mm, inner sep = 0]{};
  \draw (0, -1) node[draw,fill=white, circle, minimum size = \circlesize mm, inner sep = 0]{};
   \draw (2, -1) node[draw,fill=white, circle, minimum size = \circlesize mm, inner sep = 0]{};
\end{tikzpicture}}}
.
\end{equation}
The measurement sequence for the stabilizers involving the upper middle qubit must then be modified to exchange the roles of the middle auxiliary qubit with one of the outer auxiliary qubits at the end of the sequence. 
The corresponding circuit is shown in Appendix~\ref{SquareCircuit}. 
Shown in terms of the qubit layout, the measurement sequences are:

\vspace{2mm}

\begin{minipage}{0.15\linewidth}
${XZZXI}$
\end{minipage}
\begin{minipage}{0.8\linewidth}
\scalebox{0.71}{
\input{XZZXI_connections_time}}
\end{minipage}

\vspace{3mm}

\begin{minipage}{0.15\linewidth}
${IXZZX}$
\end{minipage}
\begin{minipage}{0.8\linewidth}
\scalebox{0.71}{
\input{IXZZX_connections_time}}
\end{minipage}

\vspace{3mm}

\begin{minipage}{0.15\linewidth}
${ZXIXZ}$
\end{minipage}
\begin{minipage}{0.8\linewidth}
\scalebox{0.71}{
\input{ZXIXZ_connections_time}}
\end{minipage}

\vspace{3mm}

\begin{minipage}{0.15\linewidth}
${XIXZZ}$
\end{minipage}
\begin{minipage}{0.8\linewidth}
\scalebox{0.71}{
\input{XIXZZ_connections_time}}
\end{minipage}

\vspace{2mm}

\noindent 
The ordering $XZZXI$, $IXZZX$, $ZXIXZ$, $XIXZZ$ ensures that  the measurement in step $8$ of each stabilizer measurement can be done at the same time as the measurement in step $1$ in the subsequent stabilizer measurement. 
We make all performance estimates (Section~\ref{Performance}) using the square lattice layout circuit with period 28 steps.

\section{Logical operations}\label{Logical}

In this section, we describe logical operations that can be performed fault-tolerantly, focusing on logical two-qubit measurements and the logical SH gate. A logical two-qubit measurement can in particular be implemented on the square lattice layout. Together with the use of ``standards'' (here in the form of auxiliary logical qubits), as described in Ref.~\onlinecite{PhysRevB.95.235305}, logical one-qubit and two-qubit measurements alone generate the full logical Clifford group, and if supplemented with an approximate magic state, logical one-qubit measurements and logical two-qubit measurements between nearest neighbors are sufficient for universal logical quantum computations.

A logical qubit can be initialized to the $\pm 1$ eigenstate of any basis $P \in \{ X,Y,Z \}$ by initializing each data qubit in that state, followed by measurements of the stabilizer generators to project onto the code space, followed by measurements of the corresponding logical operator using at least three different representatives out of $P^{\otimes 5}\times XZZXI$ + cyclic permutations. The logical measurement is performed through a modified syndrome extraction circuit, which measures the weight 3 logical operator rather than a weight 4 stabilizer. This circuit is shown in Appendix \ref{SquareCircuit}. 
A non-destructive logical one-qubit measurement can be performed by measuring the logical operator using at least three different representatives, and taking a majority vote.

The discussion of logical operations applies to the 5+2 code as well, although without a straightforward square lattice layout for logical two-qubit measurements.  We note that, being based on the $[[5,1,3]]$ perfect code, the 5+3 and 5+2 codes can be used for magic state distillation.\cite{Bravyi:2004isx}

\subsection{Logical two-qubit measurements}\label{LogicalTwoQubit}

We write representatives of logical operators $\bar{X},\bar{Y},\bar{Z}$ such that they have support on the three data qubits that share an edge. Orienting two of 5+3 code ``patches'' so that the edges with three data qubits face each other,
\begin{equation}\label{twopatches}
\vcenter{\hbox{
\rotatebox{90}{
\begin{tikzpicture}[scale = 0.5]
\newcommand{\circlesize}{2}
%
 \draw (1+3, 0+3) -- (1+3, -1+3) ;
 \draw (2+3, 0+3) -- (2+3, 1+3) ;
 \draw (0+3,0+3) -- (0+3,1+3);
  \draw (0+3,0+3) -- (2+3,0+3);
 \draw (0+3, 0+3) -- (0+3, -1+3) ;
 \draw (2+3, 0+3) -- (2+3, -1+3) ;
  \draw (0+3, 0+3) node[circle, draw, fill=black, minimum size = \circlesize mm, inner sep = 0]{};
  \draw (1+3, 0+3) node[circle, draw, fill=black, minimum size = \circlesize mm, inner sep = 0]{};
  \draw (2+3, 0+3) node[circle, draw, fill=black, minimum size = \circlesize mm, inner sep = 0]{};
 \draw (1+3, -1+3) node[fill=white, draw, circle, minimum size = \circlesize mm, inner sep = 0]{};
  \draw (0+3, 1+3) node[fill=white, draw, circle, minimum size = \circlesize mm, inner sep = 0]{};
  \draw (2+3, 1+3) node[fill=white, draw, circle, minimum size = \circlesize mm, inner sep = 0]{};
  \draw (0+3, -1+3) node[fill=white, draw, circle, minimum size = \circlesize mm, inner sep = 0]{};
  \draw (2+3, -1+3) node[fill=white, draw, circle, minimum size = \circlesize mm, inner sep = 0]{};
%
%
 \draw (1+3, 0) -- (1+3, 1) ;
 \draw (2+3, 0) -- (2+3, 1) ;
 \draw (0+3,0) -- (0+3,1);
  \draw (0+3,0) -- (2+3,0);
 \draw (0+3, 0) -- (0+3, -1) ;
 \draw (2+3, 0) -- (2+3, -1) ;
  \draw (0+3, 0) node[circle, draw, fill=black, minimum size = \circlesize mm, inner sep = 0]{};
  \draw (1+3, 0) node[circle, draw, fill=black, minimum size = \circlesize mm, inner sep = 0]{};
  \draw (2+3, 0) node[circle, draw, fill=black, minimum size = \circlesize mm, inner sep = 0]{};
  %
  \draw (0+3, 1) node[draw, fill=white, circle, minimum size = \circlesize mm, inner sep = 0]{};
   \draw (1+3, 1) node[draw, fill=white, circle, minimum size = \circlesize mm, inner sep = 0]{};
  \draw (2+3, 1) node[draw, fill=white, circle, minimum size = \circlesize mm, inner sep = 0]{};
  \draw (0+3, -1) node[draw, fill=white, circle, minimum size = \circlesize mm, inner sep = 0]{};
   \draw (2+3, -1) node[draw, fill=white, circle, minimum size = \circlesize mm, inner sep = 0]{};
%
 \end{tikzpicture}
 }
}}
\end{equation}
we perform two-qubit measurements along the ``seam'' in accordance with the logical operators to be measured. Including the measurements along the seam (marked with thicker, red edges below), the connectivity is:
\begin{equation}
\vcenter{\hbox{
\rotatebox{90}{
\begin{tikzpicture}[scale = 0.5]
\newcommand{\circlesize}{2}
 \draw [thick, red] (3, 1) -- (3, -1+3) ; 
 \draw [thick, red] (1+3, 1) -- (1+3, -1+3) ;
 \draw [thick, red] (2+3, 1) -- (2+3, -1+3) ; 
%
%
 \draw (1+3, 0+3) -- (1+3, -1+3) ;
 \draw (2+3, 0+3) -- (2+3, 1+3) ;
 \draw (0+3,0+3) -- (0+3,1+3);
  \draw (0+3,0+3) -- (2+3,0+3);
 \draw (0+3, 0+3) -- (0+3, -1+3) ;
 \draw (2+3, 0+3) -- (2+3, -1+3) ;
  \draw (0+3, 0+3) node[circle, draw, fill=black, minimum size = \circlesize mm, inner sep = 0]{};
  \draw (1+3, 0+3) node[circle, draw, fill=black, minimum size = \circlesize mm, inner sep = 0]{};
  \draw (2+3, 0+3) node[circle, draw, fill=black, minimum size = \circlesize mm, inner sep = 0]{};
 \draw (1+3, -1+3) node[draw, fill=white, circle, minimum size = \circlesize mm, inner sep = 0]{};
  \draw (0+3, 1+3) node[draw, fill=white, circle, minimum size = \circlesize mm, inner sep = 0]{};
  \draw (2+3, 1+3) node[draw, fill=white, circle, minimum size = \circlesize mm, inner sep = 0]{};
  \draw (0+3, -1+3) node[draw, fill=white, circle, minimum size = \circlesize mm, inner sep = 0]{};
  \draw (2+3, -1+3) node[draw, fill=white, circle, minimum size = \circlesize mm, inner sep = 0]{};
%
%
 \draw (1+3, 0) -- (1+3, 1) ;
 \draw (2+3, 0) -- (2+3, 1) ;
 \draw (0+3,0) -- (0+3,1);
  \draw (0+3,0) -- (2+3,0);
 \draw (0+3, 0) -- (0+3, -1) ;
 \draw (2+3, 0) -- (2+3, -1) ;
  \draw (0+3, 0) node[circle, draw, fill=black, minimum size = \circlesize mm, inner sep = 0]{};
  \draw (1+3, 0) node[circle, draw, fill=black, minimum size = \circlesize mm, inner sep = 0]{};
  \draw (2+3, 0) node[circle, draw, fill=black, minimum size = \circlesize mm, inner sep = 0]{};
  %
  \draw (0+3, 1) node[draw, fill=white, circle, minimum size = \circlesize mm, inner sep = 0]{};
   \draw (1+3, 1) node[draw, fill=white, circle, minimum size = \circlesize mm, inner sep = 0]{};
  \draw (2+3, 1) node[draw, fill=white, circle, minimum size = \circlesize mm, inner sep = 0]{};
  \draw (0+3, -1) node[draw, fill=white, circle, minimum size = \circlesize mm, inner sep = 0]{};
   \draw (2+3, -1) node[draw, fill=white, circle, minimum size = \circlesize mm, inner sep = 0]{};
%
 \end{tikzpicture}
 }
}}
\end{equation}
As an example, a logical $\bar{Z}\bar{Z}$ measurement is performed through pairwise $YY$ and $ZZ$ measurements between the patches:
\begin{equation}
\vcenter{\hbox{
\rotatebox{90}{
\begin{tikzpicture}[scale = 0.5]
\newcommand{\circlesize}{2}
 \draw [thick, red] (3, 1) -- (3, -1+3) ; 
 \draw [thick, red] (1+3, 1) -- (1+3, -1+3) ;
 \draw [thick, red] (2+3, 1) -- (2+3, -1+3) ; 
%
%
 \draw (1+3, 0+3) -- (1+3, -1+3) ;
 \draw (2+3, 0+3) -- (2+3, 1+3) ;
 \draw (0+3,0+3) -- (0+3,1+3);
  \draw (0+3,0+3) -- (2+3,0+3);
 \draw (0+3, 0+3) -- (0+3, -1+3) ;
 \draw (2+3, 0+3) -- (2+3, -1+3) ;
  \draw (0+3, 0+3) node[circle, draw, fill=black, minimum size = \circlesize mm, inner sep = 0]{};
  \draw (1+3, 0+3) node[circle, draw, fill=black, minimum size = \circlesize mm, inner sep = 0]{};
  \draw (2+3, 0+3) node[circle, draw, fill=black, minimum size = \circlesize mm, inner sep = 0]{};
 \draw (1+3, -1+3) node[fill, circle, draw , fill=white, minimum size = \circlesize mm, inner sep = 0]{};
  \draw (0+3, 1+3) node[fill, circle, draw , fill=white, minimum size = \circlesize mm, inner sep = 0]{};
  \draw (2+3, 1+3) node[fill, circle, draw , fill=white, minimum size = \circlesize mm, inner sep = 0]{};
  \draw (0+3, -1+3) node[fill, circle, draw , fill=white, minimum size = \circlesize mm, inner sep = 0]{};
  \draw (2+3, -1+3) node[fill, circle, draw , fill=white, minimum size = \circlesize mm, inner sep = 0]{};
%
%
 \draw (1+3, 0) -- (1+3, 1) ;
 \draw (2+3, 0) -- (2+3, 1) ;
 \draw (0+3,0) -- (0+3,1);
  \draw (0+3,0) -- (2+3,0);
 \draw (0+3, 0) -- (0+3, -1) ;
 \draw (2+3, 0) -- (2+3, -1) ;
  \draw (0+3, 0) node[circle, draw, fill=black, minimum size = \circlesize mm, inner sep = 0]{};
  \draw (1+3, 0) node[circle, draw, fill=black, minimum size = \circlesize mm, inner sep = 0]{};
  \draw (2+3, 0) node[circle, draw, fill=black, minimum size = \circlesize mm, inner sep = 0]{};
  %
  \draw (0+3, 1) node[fill, circle, draw , fill=white, minimum size = \circlesize mm, inner sep = 0]{};
   \draw (1+3, 1) node[fill, circle, draw , fill=white, minimum size = \circlesize mm, inner sep = 0]{};
  \draw (2+3, 1) node[fill, circle, draw , fill=white, minimum size = \circlesize mm, inner sep = 0]{};
  \draw (0+3, -1) node[fill, circle, draw , fill=white, minimum size = \circlesize mm, inner sep = 0]{};
   \draw (2+3, -1) node[fill, circle, draw , fill=white, minimum size = \circlesize mm, inner sep = 0]{};
%
\node [red] at (3.3,1.5) {\rotatebox{-90}{\scriptsize YY}};
\node [red] at (4.3,1.5) {\rotatebox{-90}{\scriptsize ZZ}};
\node [red] at (5.3,1.5) {\rotatebox{-90}{\scriptsize YY}};
 \end{tikzpicture}
 }
 }}
\end{equation}

In order to perform this logical two-qubit measurement in such a way that any single circuit fault during the operation can be corrected, the inter-patch measurements need to be repeated three times, with a round of syndrome extraction on each patch in between two of the repetitions and another round at the very end. The three repetitions of the inter-patch measurements allow for correction of readout faults on these measurements, and the rounds of syndrome extraction allow for correction of Pauli faults during the procedure.

In order to connect with other neighboring logical qubits, we can use teleportation to change which edge has three data qubits. This requires a ninth qubit to make the last edge available:

\begin{equation}
\vcenter{\hbox{
\begin{tikzpicture}
\newcommand{\circlesize}{2}
 \draw[-latex] (1, 0.75) -- (1, 0.25) ;
 \draw[-latex] (1.25, 0) -- (1.75, 0) ;
 \draw[-latex] (0.75, 0) -- (0.25, 0) ;
 \draw[-latex] (1, -0.25) -- (1, -0.75) ;
  \draw (0, 0) node[circle, draw, fill=black, minimum size = \circlesize mm, inner sep = 0]{};
  \draw (1, 0) node[circle, draw, fill=black, minimum size = \circlesize mm, inner sep = 0]{};
  \draw (2, 0) node[circle, draw, fill=black, minimum size = \circlesize mm, inner sep = 0]{};
  \draw (1, -1) node[circle, draw, fill=black, minimum size = \circlesize mm, inner sep = 0]{};
  \draw (0, 1) node[draw, fill=white, circle, minimum size = \circlesize mm, inner sep = 0]{};
   \draw (1, 1) node[draw, fill=white, circle, minimum size = \circlesize mm, inner sep = 0]{};
  \draw (2, 1) node[draw, fill=white, circle, minimum size = \circlesize mm, inner sep = 0]{};
  \draw (0, -1) node[draw, fill=white, circle, minimum size = \circlesize mm, inner sep = 0]{};
   \draw (2, -1) node[draw, fill=white, circle, minimum size = \circlesize mm, inner sep = 0]{};
\end{tikzpicture}   
}}
\end{equation}

The teleportation is mediated by the center auxiliary qubit $C$, as described by the following circuit, where the upper line at the beginning represents the data qubit $D$ whose state we wish to teleport, the middle line represents the center auxiliary qubit and the lower line represents the target auxiliary qubit $A$: 
\begin{equation}
\vcenter{\hbox{
\begin{tikzpicture}
   \newcommand{\scaling}{0.6}
   \node at (-2*\scaling, 0.0000*\scaling)  [anchor=center] () {D};
    \node at (9.8*\scaling, 0.0000*\scaling)  [anchor=center] () {A};
   \draw[line width = 0.6mm, gray] (-1.5*\scaling, 0.0000*\scaling)   -- (9.3000*\scaling, 0.0000*\scaling);
   \node at (-2*\scaling, -1.0000*\scaling)  [anchor=center] () {C};
    \node at (9.8*\scaling, -1.0000*\scaling)  [anchor=center] () {C};
   \draw[line width = 0.6mm] (-1.5*\scaling, -1.0000*\scaling)   -- (9.3000*\scaling, -1.0000*\scaling);
   \node at (-2*\scaling, -2.0000*\scaling)  [anchor=center] () {A};
    \node at (9.8*\scaling, -2.0000*\scaling)  [anchor=center] () {D};
   \draw[line width = 0.6mm, gray] (-1.5*\scaling, -2.0000*\scaling)   -- (9.3000*\scaling, -2.0000*\scaling);
    \draw[line width = 0.6mm] (-1.5*\scaling, -2.0000*\scaling)   -- (5.3000*\scaling, -2.0000*\scaling);
   \draw[line width = 0.6mm] (2.3*\scaling, 0.0000*\scaling)   -- (9.3000*\scaling, 0.0000*\scaling);
   \draw[line width = 0.6mm, gray] (2.3*\scaling, -1.0000*\scaling)   -- (5.3000*\scaling, -1.0000*\scaling);
   \draw[dotted] (-1.300000*\scaling, 0.000000*\scaling +1*\scaling)  -- (-1.300000*\scaling, -2.000000*\scaling -1*\scaling);
   \draw[dotted] (1.300000*\scaling, 0.000000*\scaling +1*\scaling)  -- (1.300000*\scaling, -2.000000*\scaling -1*\scaling);
   \draw[dotted] (3.900000*\scaling, 0.000000*\scaling +1*\scaling)  -- (3.900000*\scaling, -2.000000*\scaling -1*\scaling);
   \draw[dotted] (6.500000*\scaling, 0.000000*\scaling +1*\scaling)  -- (6.500000*\scaling, -2.000000*\scaling -1*\scaling);
   \draw[dotted] (9.100000*\scaling, 0.000000*\scaling +1*\scaling)  -- (9.100000*\scaling, -2.000000*\scaling -1*\scaling);
   \node at (0.000000*\scaling, -2.000000*\scaling)  [rectangle,fill=white,draw, minimum width = 12pt, minimum height = 12pt,text width=, anchor=center] () {\tiny Z};
   \node at (0.000000*\scaling, -1.000000*\scaling)  [rectangle,fill=white,draw, minimum width = 12pt, minimum height = 12pt,text width=, anchor=center] () {\tiny Z};
   \draw[line width=1mm, white] (2.6000*\scaling,  -1.000000*\scaling)   -- (2.6000*\scaling, 0.000000*\scaling);
   \draw (2.6000*\scaling,  -1.000000*\scaling)   -- (2.6000*\scaling, 0.000000*\scaling);
   \node at (2.6000*\scaling, -1.000000*\scaling)  [rectangle,fill=white,draw, minimum width = 12pt, minimum height = 12pt,text width=, anchor=center] () {\tiny X};
   \node at (2.6000*\scaling, 0.000000*\scaling)  [rectangle,fill=white,draw, minimum width = 12pt, minimum height = 12pt,text width=, anchor=center] () {\tiny X};
   \draw[line width=1mm, white] (5.2000*\scaling,  -2.000000*\scaling)   -- (5.2000*\scaling, -1.000000*\scaling);
   \draw (5.2000*\scaling,  -2.000000*\scaling)   -- (5.2000*\scaling, -1.000000*\scaling);
   \node at (5.2000*\scaling, -2.000000*\scaling)  [rectangle,fill=white,draw, minimum width = 12pt, minimum height = 12pt,text width=, anchor=center] () {\tiny X};
   \node at (5.2000*\scaling, -1.000000*\scaling)  [rectangle,fill=white,draw, minimum width = 12pt, minimum height = 12pt,text width=, anchor=center] () {\tiny X};
   \node at (7.800000*\scaling, 0.000000*\scaling)  [rectangle,fill=white,draw, minimum width = 12pt, minimum height = 12pt,text width=, anchor=center] () {\tiny Z};
   \node at (7.800000*\scaling, -1.000000*\scaling)  [rectangle,fill=white,draw, minimum width = 12pt, minimum height = 12pt,text width=, anchor=center] () {\tiny Z};
\end{tikzpicture}
}}
\end{equation}
At the end of the circuit, the lower line represents the data qubit $D$ and the upper line is denoted by $A$ for auxiliary qubit. Performing a round of syndrome extraction before and after this teleportation, any circuit fault can be corrected. 
In this fashion, we can place logical qubits on a square grid, and perform logical two-qubit measurements between any horizontal and vertical neighbors:

\begin{equation}
\vcenter{\hbox{
\begin{tikzpicture}[scale = 0.5]
\newcommand{\circlesize}{2}
 \draw [thick, red] (3, 1) -- (3, -1+3) ; 
 \draw [thick, red] (1+3, 1) -- (1+3, -1+3) ;
 \draw [thick, red] (2+3, 1) -- (2+3, -1+3) ; 
%
 \draw (1, 0) -- (1, 1) ;
 \draw (2, 0) -- (2, 1) ;
 \draw (0,0) -- (0,1);
  \draw (0,0) -- (2,0);
 \draw (0, 0) -- (0, -1) ;
 \draw (2, 0) -- (2, -1) ;
  \draw (0, 0) node[circle, draw, fill=black, minimum size = \circlesize mm, inner sep = 0]{};
  \draw (1, 0) node[circle, draw, fill=black, minimum size = \circlesize mm, inner sep = 0]{};
  \draw (2, 0) node[circle, draw, fill=black, minimum size = \circlesize mm, inner sep = 0]{};
  \draw (1, -1) node[circle, draw, fill=black, minimum size = \circlesize mm, inner sep = 0]{};
  \draw (0, 1) node[draw, fill=white, circle, minimum size = \circlesize mm, inner sep = 0]{};
   \draw (1, 1) node[draw, fill=white, circle, minimum size = \circlesize mm, inner sep = 0]{};
  \draw (2, 1) node[draw, fill=white, circle, minimum size = \circlesize mm, inner sep = 0]{};
  \draw (0, -1) node[draw, fill=white, circle, minimum size = \circlesize mm, inner sep = 0]{};
   \draw (2, -1) node[draw, fill=white, circle, minimum size = \circlesize mm, inner sep = 0]{};
%
%
 \draw (1+3, 0+3) -- (1+3, -1+3) ;
 \draw (2+3, 0+3) -- (2+3, 1+3) ;
 \draw (0+3,0+3) -- (0+3,1+3);
  \draw (0+3,0+3) -- (2+3,0+3);
 \draw (0+3, 0+3) -- (0+3, -1+3) ;
 \draw (2+3, 0+3) -- (2+3, -1+3) ;
  \draw (0+3, 0+3) node[circle, draw, fill=black, minimum size = \circlesize mm, inner sep = 0]{};
  \draw (1+3, 0+3) node[circle, draw, fill=black, minimum size = \circlesize mm, inner sep = 0]{};
  \draw (2+3, 0+3) node[circle, draw, fill=black, minimum size = \circlesize mm, inner sep = 0]{};
 \draw (1+3, -1+3) node[draw, fill=white, circle, minimum size = \circlesize mm, inner sep = 0]{};
  \draw (0+3, 1+3) node[draw, fill=white, circle, minimum size = \circlesize mm, inner sep = 0]{};
   \draw (1+3, 1+3) node[circle, draw, fill=black, minimum size = \circlesize mm, inner sep = 0]{};
  \draw (2+3, 1+3) node[draw, fill=white, circle, minimum size = \circlesize mm, inner sep = 0]{};
  \draw (0+3, -1+3) node[draw, fill=white, circle, minimum size = \circlesize mm, inner sep = 0]{};
  \draw (2+3, -1+3) node[draw, fill=white, circle, minimum size = \circlesize mm, inner sep = 0]{};
%
%
 \draw (1+3, 0) -- (1+3, 1) ;
 \draw (2+3, 0) -- (2+3, 1) ;
 \draw (0+3,0) -- (0+3,1);
  \draw (0+3,0) -- (2+3,0);
 \draw (0+3, 0) -- (0+3, -1) ;
 \draw (2+3, 0) -- (2+3, -1) ;
  \draw (0+3, 0) node[circle, draw, fill=black, minimum size = \circlesize mm, inner sep = 0]{};
  \draw (1+3, 0) node[circle, draw, fill=black, minimum size = \circlesize mm, inner sep = 0]{};
  \draw (2+3, 0) node[circle, draw, fill=black, minimum size = \circlesize mm, inner sep = 0]{};
  \draw (1+3, -1) node[circle, draw, fill=black, minimum size = \circlesize mm, inner sep = 0]{};  
  \draw (0+3, 1) node[draw, fill=white, circle, minimum size = \circlesize mm, inner sep = 0]{};
   \draw (1+3, 1) node[draw, fill=white, circle, minimum size = \circlesize mm, inner sep = 0]{};
  \draw (2+3, 1) node[draw, fill=white, circle, minimum size = \circlesize mm, inner sep = 0]{};
  \draw (0+3, -1) node[draw, fill=white, circle, minimum size = \circlesize mm, inner sep = 0]{};
   \draw (2+3, -1) node[draw, fill=white, circle, minimum size = \circlesize mm, inner sep = 0]{};
%
 \draw (1, 0+3) -- (1, 1+3) ;
 \draw (2, 0+3) -- (2, 1+3) ;
 \draw (0,0+3) -- (0,1+3);
  \draw (0,0+3) -- (2,0+3);
 \draw (0, 0+3) -- (0, -1+3) ;
 \draw (2, 0+3) -- (2, -1+3) ;
  \draw (0, 0+3) node[circle, draw, fill=black, minimum size = \circlesize mm, inner sep = 0]{};
  \draw (1, 0+3) node[circle, draw, fill=black, minimum size = \circlesize mm, inner sep = 0]{};
  \draw (2, 0+3) node[circle, draw, fill=black, minimum size = \circlesize mm, inner sep = 0]{};
  \draw (1, -1+3) node[circle, draw, fill=black, minimum size = \circlesize mm, inner sep = 0]{};  
  \draw (0, 1+3) node[draw, fill=white, circle, minimum size = \circlesize mm, inner sep = 0]{};
   \draw (1, 1+3) node[draw, fill=white, circle, minimum size = \circlesize mm, inner sep = 0]{};
  \draw (2, 1+3) node[draw, fill=white, circle, minimum size = \circlesize mm, inner sep = 0]{};
  \draw (0, -1+3) node[draw, fill=white, circle, minimum size = \circlesize mm, inner sep = 0]{};
   \draw (2, -1+3) node[draw, fill=white, circle, minimum size = \circlesize mm, inner sep = 0]{};
%
 \end{tikzpicture}
}}
\end{equation}

\subsection{Logical SH gate through code automorphisms}

While logical two-qubit measurements suffice to implement the full Clifford group, we note that the SH gate is transversal on the perfect code $[[5,1,3]]$, and a logical SH gate can be implemented in the 5+2 and 5+3 codes by applying an SH gate on each of the five data qubits. In this section, we describe how this can be done for the 5+3 code during syndrom extraction, which allows for the correction of any circuit fault that occurs during the protocol. A similar scheme can be implemented for the 5+2 code. In contrast to the logical two-qubit measurements described above, the following circuits do not have the connectivity of a square grid. We consider the less compact syndrome extraction circuit when implementing the SH gate, for simplicity of presentation.

An SH gate on a data qubit $D$ can be implemented through one- and two-qubit measurements with the help of one auxiliary qubit $A$, using the following circuit:
\begin{equation}
\vcenter{\hbox{
\begin{tikzpicture}
   \newcommand{\scaling}{0.6}
   \node at (-2*\scaling, 0.0000*\scaling)  [anchor=center] () {D};
   \draw[line width = 0.6mm, gray] (-1.5*\scaling, 0.0000*\scaling)   -- (14.5000*\scaling, 0.0000*\scaling);
   \node at (-2*\scaling, -1.0000*\scaling)  [anchor=center] () {A};
   \draw[line width = 0.6mm] (-1.5*\scaling, -1.0000*\scaling)   -- (14.5000*\scaling, -1.0000*\scaling);
   \draw[dotted] (-1.300000*\scaling, 0.000000*\scaling +1*\scaling)  -- (-1.300000*\scaling, -1.000000*\scaling -1*\scaling);
   \draw[dotted] (1.300000*\scaling, 0.000000*\scaling +1*\scaling)  -- (1.300000*\scaling, -1.000000*\scaling -1*\scaling);
   \draw[dotted] (3.900000*\scaling, 0.000000*\scaling +1*\scaling)  -- (3.900000*\scaling, -1.000000*\scaling -1*\scaling);
   \draw[dotted] (6.500000*\scaling, 0.000000*\scaling +1*\scaling)  -- (6.500000*\scaling, -1.000000*\scaling -1*\scaling);
   \draw[dotted] (9.100000*\scaling, 0.000000*\scaling +1*\scaling)  -- (9.100000*\scaling, -1.000000*\scaling -1*\scaling);
   \draw[dotted] (11.700000*\scaling, 0.000000*\scaling +1*\scaling)  -- (11.700000*\scaling, -1.000000*\scaling -1*\scaling);
   \draw[dotted] (14.300000*\scaling, 0.000000*\scaling +1*\scaling)  -- (14.300000*\scaling, -1.000000*\scaling -1*\scaling);
   \node at (0.000000*\scaling, -1.000000*\scaling)  [rectangle,fill=white,draw, minimum width = 12pt, minimum height = 12pt,text width=, anchor=center] () {\tiny Z};
   \draw[line width=1mm, white] (2.6000*\scaling,  -1.000000*\scaling)   -- (2.6000*\scaling, 0.000000*\scaling);
   \draw (2.6000*\scaling,  -1.000000*\scaling)   -- (2.6000*\scaling, 0.000000*\scaling);
   \node at (2.6000*\scaling, -1.000000*\scaling)  [rectangle,fill=white,draw, minimum width = 12pt, minimum height = 12pt,text width=, anchor=center] () {\tiny X};
   \node at (2.6000*\scaling, 0.000000*\scaling)  [rectangle,fill=white,draw, minimum width = 12pt, minimum height = 12pt,text width=, anchor=center] () {\tiny X};
   \node at (5.200000*\scaling, 0.000000*\scaling)  [rectangle,fill=white,draw, minimum width = 12pt, minimum height = 12pt,text width=, anchor=center] () {\tiny Z};
   \node at (7.800000*\scaling, 0.000000*\scaling)  [rectangle,fill=white,draw, minimum width = 12pt, minimum height = 12pt,text width=, anchor=center] () {\tiny Y};
   \draw[line width=1mm, white] (10.4000*\scaling,  0.000000*\scaling)   -- (10.4000*\scaling, -1.000000*\scaling);
   \draw (10.4000*\scaling,  0.000000*\scaling)   -- (10.4000*\scaling, -1.000000*\scaling);
   \node at (10.4000*\scaling, 0.000000*\scaling)  [rectangle,fill=white,draw, minimum width = 12pt, minimum height = 12pt,text width=, anchor=center] () {\tiny Z};
   \node at (10.4000*\scaling, -1.000000*\scaling)  [rectangle,fill=white,draw, minimum width = 12pt, minimum height = 12pt,text width=, anchor=center] () {\tiny X};
   \node at (13.000000*\scaling, -1.000000*\scaling)  [rectangle,fill=white,draw, minimum width = 12pt, minimum height = 12pt,text width=, anchor=center] () {\tiny Z};
\end{tikzpicture}
}}
\end{equation}
This circuit implements two teleportations between the data qubit and the auxiliary qubit, with a basis change for the data qubit measurements in the second teleportation determined by action of the SH gate: $X\to Z, Z\to Y, Y\to X$.

We can modify the syndrome extraction circuit to implement teleportations, by changing which qubits play the role of auxiliary qubits at the end. In the following example, data qubits 1 and 4 exchange role with auxiliary qubits A and B in a modified $XZZXI$ measurement, that we denote by $XZZXI(1 \leftrightarrow A, 4 \leftrightarrow B)$:
\begin{equation}
\vcenter{\hbox{
\begin{tikzpicture}
   \newcommand{\scaling}{0.6}
   \node at (-2*\scaling, 0.0000*\scaling)  [anchor=center] () {1};
   \draw[line width = 0.6mm, gray] (-1.5*\scaling, 0.0000*\scaling)   -- (19.7000*\scaling, 0.0000*\scaling);
   \node at (-2*\scaling, -1.0000*\scaling)  [anchor=center] () {2};
   \draw[line width = 0.6mm, gray] (-1.5*\scaling, -1.0000*\scaling)   -- (19.7000*\scaling, -1.0000*\scaling);
   \node at (-2*\scaling, -2.0000*\scaling)  [anchor=center] () {3};
   \draw[line width = 0.6mm, gray] (-1.5*\scaling, -2.0000*\scaling)   -- (19.7000*\scaling, -2.0000*\scaling);
   \node at (-2*\scaling, -3.0000*\scaling)  [anchor=center] () {4};
   \draw[line width = 0.6mm, gray] (-1.5*\scaling, -3.0000*\scaling)   -- (19.7000*\scaling, -3.0000*\scaling);
   \node at (-2*\scaling, -4.0000*\scaling)  [anchor=center] () {5};
   \draw[line width = 0.6mm, gray] (-1.5*\scaling, -4.0000*\scaling)   -- (19.7000*\scaling, -4.0000*\scaling);
   \node at (-2*\scaling, -5.0000*\scaling)  [anchor=center] () {A};
   \draw[line width = 0.6mm] (-1.5*\scaling, -5.0000*\scaling)   -- (19.7000*\scaling, -5.0000*\scaling);
   \node at (-2*\scaling, -6.0000*\scaling)  [anchor=center] () {B};
   \draw[line width = 0.6mm] (-1.5*\scaling, -6.0000*\scaling)   -- (19.7000*\scaling, -6.0000*\scaling);
   \node at (-2*\scaling, -7.0000*\scaling)  [anchor=center] () {C};
   \draw[line width = 0.6mm] (-1.5*\scaling, -7.0000*\scaling)   -- (19.7000*\scaling, -7.0000*\scaling);
\node at (19.7*\scaling + 0.5*\scaling, -0.0000*\scaling)  [anchor=center] () {A};
\node at (19.7*\scaling + 0.5*\scaling, -1.0000*\scaling)  [anchor=center] () {2};
\node at (19.7*\scaling + 0.5*\scaling, -2.0000*\scaling)  [anchor=center] () {3};
\node at (19.7*\scaling + 0.5*\scaling, -3.0000*\scaling)  [anchor=center] () {B};
\node at (19.7*\scaling + 0.5*\scaling, -4.0000*\scaling)  [anchor=center] () {5};
\node at (19.7*\scaling + 0.5*\scaling, -5.0000*\scaling)  [anchor=center] () {1};
\node at (19.7*\scaling + 0.5*\scaling, -6.0000*\scaling)  [anchor=center] () {4};
\node at (19.7*\scaling + 0.5*\scaling, -7.0000*\scaling)  [anchor=center] () {C};
\draw[line width = 0.6mm, gray] (16.3*\scaling, -6.0000*\scaling)   -- (19.7000*\scaling, -6.0000*\scaling);
\draw[line width = 0.6mm] (16.3*\scaling, -3.0000*\scaling)   -- (19.7000*\scaling, -3.0000*\scaling);
\draw[line width = 0.6mm, gray] (15.3*\scaling, -5.0000*\scaling)   -- (19.7000*\scaling, -5.0000*\scaling);
\draw[line width = 0.6mm] (15.3*\scaling, -0.0000*\scaling)   -- (19.7000*\scaling, -0.0000*\scaling);
   \draw[dotted] (-1.300000*\scaling, 0.000000*\scaling +1*\scaling)  -- (-1.300000*\scaling, -7.000000*\scaling -1*\scaling);
   \draw[dotted] (1.300000*\scaling, 0.000000*\scaling +1*\scaling)  -- (1.300000*\scaling, -7.000000*\scaling -1*\scaling);
   \draw[dotted] (3.900000*\scaling, 0.000000*\scaling +1*\scaling)  -- (3.900000*\scaling, -7.000000*\scaling -1*\scaling);
   \draw[dotted] (6.500000*\scaling, 0.000000*\scaling +1*\scaling)  -- (6.500000*\scaling, -7.000000*\scaling -1*\scaling);
   \draw[dotted] (9.100000*\scaling, 0.000000*\scaling +1*\scaling)  -- (9.100000*\scaling, -7.000000*\scaling -1*\scaling);
   \draw[dotted] (11.700000*\scaling, 0.000000*\scaling +1*\scaling)  -- (11.700000*\scaling, -7.000000*\scaling -1*\scaling);
   \draw[dotted] (14.300000*\scaling, 0.000000*\scaling +1*\scaling)  -- (14.300000*\scaling, -7.000000*\scaling -1*\scaling);
   \draw[dotted] (16.900000*\scaling, 0.000000*\scaling +1*\scaling)  -- (16.900000*\scaling, -7.000000*\scaling -1*\scaling);
   \draw[dotted] (19.500000*\scaling, 0.000000*\scaling +1*\scaling)  -- (19.500000*\scaling, -7.000000*\scaling -1*\scaling);
   \node at (0.0000*\scaling, -7.0000*\scaling -1*\scaling) {1};
   \node at (2.6000*\scaling, -7.0000*\scaling -1*\scaling) {2};
   \node at (5.2000*\scaling, -7.0000*\scaling -1*\scaling) {3};
   \node at (7.8000*\scaling, -7.0000*\scaling -1*\scaling) {4};
   \node at (10.4000*\scaling, -7.0000*\scaling -1*\scaling) {5};
   \node at (13.0000*\scaling, -7.0000*\scaling -1*\scaling) {6};
   \node at (15.6000*\scaling, -7.0000*\scaling -1*\scaling) {7};
   \node at (18.2000*\scaling, -7.0000*\scaling -1*\scaling) {8};
   \node at (0.000000*\scaling, -5.000000*\scaling)  [rectangle,fill=white,draw, minimum width = 12pt, minimum height = 12pt,text width=, anchor=center] () {\tiny Z};
   \node at (0.000000*\scaling, -6.000000*\scaling)  [rectangle,fill=white,draw, minimum width = 12pt, minimum height = 12pt,text width=, anchor=center] () {\tiny Z};
   \node at (0.000000*\scaling, -7.000000*\scaling)  [rectangle,fill=white,draw, minimum width = 12pt, minimum height = 12pt,text width=, anchor=center] () {\tiny X};
   \draw[line width=1mm, white] (2.1580*\scaling,  -5.000000*\scaling)   -- (2.1580*\scaling, -1.000000*\scaling);
   \draw (2.1580*\scaling,  -5.000000*\scaling)   -- (2.1580*\scaling, -1.000000*\scaling);
   \node at (2.1580*\scaling, -5.000000*\scaling)  [rectangle,fill=white,draw, minimum width = 12pt, minimum height = 12pt,text width=, anchor=center] () {\tiny X};
   \node at (2.1580*\scaling, -1.000000*\scaling)  [rectangle,fill=white,draw, minimum width = 12pt, minimum height = 12pt,text width=, anchor=center] () {\tiny Z};
   \draw[line width=1mm, white] (3.0420*\scaling,  -6.000000*\scaling)   -- (3.0420*\scaling, -2.000000*\scaling);
   \draw (3.0420*\scaling,  -6.000000*\scaling)   -- (3.0420*\scaling, -2.000000*\scaling);
   \node at (3.0420*\scaling, -6.000000*\scaling)  [rectangle,fill=white,draw, minimum width = 12pt, minimum height = 12pt,text width=, anchor=center] () {\tiny X};
   \node at (3.0420*\scaling, -2.000000*\scaling)  [rectangle,fill=white,draw, minimum width = 12pt, minimum height = 12pt,text width=, anchor=center] () {\tiny Z};
   \draw[line width=1mm, white] (5.2000*\scaling,  -7.000000*\scaling)   -- (5.2000*\scaling, -5.000000*\scaling);
   \draw (5.2000*\scaling,  -7.000000*\scaling)   -- (5.2000*\scaling, -5.000000*\scaling);
   \node at (5.2000*\scaling, -7.000000*\scaling)  [rectangle,fill=white,draw, minimum width = 12pt, minimum height = 12pt,text width=, anchor=center] () {\tiny Z};
   \node at (5.2000*\scaling, -5.000000*\scaling)  [rectangle,fill=white,draw, minimum width = 12pt, minimum height = 12pt,text width=, anchor=center] () {\tiny Z};
   \draw[line width=1mm, white] (7.8000*\scaling,  -7.000000*\scaling)   -- (7.8000*\scaling, -6.000000*\scaling);
   \draw (7.8000*\scaling,  -7.000000*\scaling)   -- (7.8000*\scaling, -6.000000*\scaling);
   \node at (7.8000*\scaling, -7.000000*\scaling)  [rectangle,fill=white,draw, minimum width = 12pt, minimum height = 12pt,text width=, anchor=center] () {\tiny Z};
   \node at (7.8000*\scaling, -6.000000*\scaling)  [rectangle,fill=white,draw, minimum width = 12pt, minimum height = 12pt,text width=, anchor=center] () {\tiny Z};
   \draw[line width=1mm, white] (10.4000*\scaling,  -7.000000*\scaling)   -- (10.4000*\scaling, -5.000000*\scaling);
   \draw (10.4000*\scaling,  -7.000000*\scaling)   -- (10.4000*\scaling, -5.000000*\scaling);
   \node at (10.4000*\scaling, -7.000000*\scaling)  [rectangle,fill=white,draw, minimum width = 12pt, minimum height = 12pt,text width=, anchor=center] () {\tiny Z};
   \node at (10.4000*\scaling, -5.000000*\scaling)  [rectangle,fill=white,draw, minimum width = 12pt, minimum height = 12pt,text width=, anchor=center] () {\tiny Z};
   \draw[line width=1mm, white] (13.0000*\scaling,  -7.000000*\scaling)   -- (13.0000*\scaling, -6.000000*\scaling);
   \draw (13.0000*\scaling,  -7.000000*\scaling)   -- (13.0000*\scaling, -6.000000*\scaling);
   \node at (13.0000*\scaling, -7.000000*\scaling)  [rectangle,fill=white,draw, minimum width = 12pt, minimum height = 12pt,text width=, anchor=center] () {\tiny Z};
   \node at (13.0000*\scaling, -6.000000*\scaling)  [rectangle,fill=white,draw, minimum width = 12pt, minimum height = 12pt,text width=, anchor=center] () {\tiny Z};
   \draw[line width=1mm, white] (15.1580*\scaling,  -5.000000*\scaling)   -- (15.1580*\scaling, 0.000000*\scaling);
   \draw (15.1580*\scaling,  -5.000000*\scaling)   -- (15.1580*\scaling, 0.000000*\scaling);
   \node at (15.1580*\scaling, -5.000000*\scaling)  [rectangle,fill=white,draw, minimum width = 12pt, minimum height = 12pt,text width=, anchor=center] () {\tiny X};
   \node at (15.1580*\scaling, 0.000000*\scaling)  [rectangle,fill=white,draw, minimum width = 12pt, minimum height = 12pt,text width=, anchor=center] () {\tiny X};
   \draw[line width=1mm, white] (16.0420*\scaling,  -6.000000*\scaling)   -- (16.0420*\scaling, -3.000000*\scaling);
   \draw (16.0420*\scaling,  -6.000000*\scaling)   -- (16.0420*\scaling, -3.000000*\scaling);
   \node at (16.0420*\scaling, -6.000000*\scaling)  [rectangle,fill=white,draw, minimum width = 12pt, minimum height = 12pt,text width=, anchor=center] () {\tiny X};
   \node at (16.0420*\scaling, -3.000000*\scaling)  [rectangle,fill=white,draw, minimum width = 12pt, minimum height = 12pt,text width=, anchor=center] () {\tiny X};
   \node at (18.200000*\scaling, 0.000000*\scaling)  [rectangle,fill=white,draw, minimum width = 12pt, minimum height = 12pt,text width=, anchor=center] () {\tiny Z};
   \node at (18.200000*\scaling, -3.000000*\scaling)  [rectangle,fill=white,draw, minimum width = 12pt, minimum height = 12pt,text width=, anchor=center] () {\tiny Z};
   \node at (18.200000*\scaling, -7.000000*\scaling)  [rectangle,fill=white,draw, minimum width = 12pt, minimum height = 12pt,text width=, anchor=center] () {\tiny X};
\end{tikzpicture}
}}
\end{equation}

One protocol for performing an SH gate on each data qubit would be to perform several twice-repeated stabilizer measurements, each swapping swapping one or two data qubits with one or two of the auxiliary qubits $A,B$ and back, with the above basis change performed during the second swap.  
However, a more efficient protocol takes advantage of the [[5,1,3]] code's invariance under cyclic permutations. Thanks to this invariance, the data qubits need not be returned to their original positions, and we can instead perform the following sequence of swaps with auxiliary qubits that permutes the data qubits from 12345 to 51234: 
\begin{equation}
\vcenter{\hbox{
\begin{tikzpicture}[xscale=1.4,yscale=1.4]
\foreach \y in {1,2,3,4,5,6,7,8}{
  \draw (-0.2,\y*0.2)--(2.5,\y*0.2);
}
\node at (-0.4, 1.6) {\tiny 1};
\node at (-0.4, 1.4) {\tiny 2};
\node at (-0.4, 1.2) {\tiny 3};
\node at (-0.4,1) {\tiny 4};
\node at (-0.4, 0.8) {\tiny 5};
\node at (-0.4, 0.6) {\tiny A};
\node at (-0.4, 0.4) {\tiny B};
\node at (-0.4, 0.2) {\tiny C};
\filldraw[fill=gray!10!white, rounded corners] (0,0) rectangle (2,8*0.2 + 0.2);
\node[align=center] at (1,4.5*0.2) {XZZXI\\ $1 \leftrightarrow A$ \\ $4 \leftrightarrow B$};
\node[fill=white] at (2.25, 1.6) {\phantom{\tiny .}};
\node at (2.25, 1.6) {\tiny A};
\node[fill=white] at (2.25, 1.4) {\phantom{\tiny .}};
\node at (2.25, 1.4) {\tiny 2};
\node[fill=white] at (2.25, 1.2) {\phantom{\tiny .}};
\node at (2.25, 1.2) {\tiny 3};
\node[fill=white] at (2.25, 1.0) {\phantom{\tiny .}};
\node at (2.25, 1.0) {\tiny B};
\node[fill=white] at (2.25, 0.8) {\phantom{\tiny .}};
\node at (2.25, 0.8) {\tiny 5};
\node[fill=white] at (2.25, 0.6) {\phantom{\tiny .}};
\node at (2.25, 0.6) {\tiny 1};
\node[fill=white] at (2.25, 0.4) {\phantom{\tiny .}};
\node at (2.25, 0.4) {\tiny 4};
\node[fill=white] at (2.25, 0.2) {\phantom{\tiny .}};
\node at (2.25, 0.2) {\tiny C};
\foreach \y in {1,2,3,4,5,6,7,8}{
  \draw (-0.5+3,\y*0.2)--(2.5+3,\y*0.2);
}
\filldraw[fill=gray!10!white, rounded corners] (0+ 2.5,0) rectangle (2+ 2.5,8*0.2 + 0.2);
\node[align=center] at (1+ 2.5,4.5*0.2) {IXZZX\\ $2 \leftrightarrow 1$ \\ $5 \leftrightarrow 4$};
\node[fill=white] at (4.75, 1.6) {\phantom{\tiny .}};
\node at (4.75, 1.6) {\tiny A};
\node[fill=white] at (4.75, 1.4) {\phantom{\tiny .}};
\node at (4.75, 1.4) {\tiny 1};
\node[fill=white] at (4.75, 1.2) {\phantom{\tiny .}};
\node at (4.75, 1.2) {\tiny 3};
\node[fill=white] at (4.75, 1.0) {\phantom{\tiny .}};
\node at (4.75, 1.0) {\tiny B};
\node[fill=white] at (4.75, 0.8) {\phantom{\tiny .}};
\node at (4.75, 0.8) {\tiny 4};
\node[fill=white] at (4.75, 0.6) {\phantom{\tiny .}};
\node at (4.75, 0.6) {\tiny 2};
\node[fill=white] at (4.75, 0.4) {\phantom{\tiny .}};
\node at (4.75, 0.4) {\tiny 5};
\node[fill=white] at (4.75, 0.2) {\phantom{\tiny .}};
\node at (4.75, 0.2) {\tiny C};
\foreach \y in {1,2,3,4,5,6,7,8}{
  \draw (-0.5+6,\y*0.2)--(2.5+6,\y*0.2);
}
\filldraw[fill=gray!10!white, rounded corners] (0+5,0) rectangle (2+5,8*0.2 + 0.2);
\node[align=center] at (1+5,4.5*0.2) {XIXZZ\\ $A \leftrightarrow 5$ \\ $3 \leftrightarrow 2$};
\foreach \y in {1,2,3,4,5,6,7,8}{
  \draw (-0.5+7.5,\y*0.2)--(2.2+7.5,\y*0.2);
}
\node[fill=white] at (7.25, 1.6) {\phantom{\tiny .}};
\node at (7.25, 1.6) {\tiny 5};
\node[fill=white] at (7.25, 1.4) {\phantom{\tiny .}};
\node at (7.25, 1.4) {\tiny 1};
\node[fill=white] at (7.25, 1.2) {\phantom{\tiny .}};
\node at (7.25, 1.2) {\tiny 2};
\node[fill=white] at (7.25, 1.0) {\phantom{\tiny .}};
\node at (7.25, 1.0) {\tiny B};
\node[fill=white] at (7.25, 0.8) {\phantom{\tiny .}};
\node at (7.25, 0.8) {\tiny 4};
\node[fill=white] at (7.25, 0.6) {\phantom{\tiny .}};
\node at (7.25, 0.6) {\tiny 3};
\node[fill=white] at (7.25, 0.4) {\phantom{\tiny .}};
\node at (7.25, 0.4) {\tiny A};
\node[fill=white] at (7.25, 0.2) {\phantom{\tiny .}};
\node at (7.25, 0.2) {\tiny C};
\filldraw[fill=gray!10!white, rounded corners] (0+7.5,0) rectangle (2+7.5,8*0.2 + 0.2);
\node[align=center] at (1+7.5,4.5*0.2) {ZXIXZ\\ $B \leftrightarrow 3$\\ };
\node at (2.4+7.5, 1.6) {\tiny 5};
\node at (2.4+7.5, 1.4) {\tiny 1};
\node at (2.4+7.5, 1.2) {\tiny 2};
\node at (2.4+7.5,1) {\tiny 3};
\node at (2.4+7.5, 0.8) {\tiny 4};
\node at (2.4+7.5, 0.6) {\tiny B};
\node at (2.4+7.5, 0.4) {\tiny A};
\node at (2.4+7.5, 0.2) {\tiny C};
\end{tikzpicture}}}
\end{equation}

Combined with the changes in measurement bases that implement an SH gate on each data qubit as it is swapped with an auxiliary qubit, this protocol only requires one full syndrome extraction period (the measurement of four stabilizer generators) to implement SH transversally, and hence implement a logical SH gate. We note that after an SH gate has been performed on a data qubit, this must be taken into account for the subsequent stabilizer measurements, which must be performed with a corresponding basis change as well.

\section{Decoding and performance estimates for the 5+3 code}\label{Performance}

To evaluate the performance of the 5+3 code, we estimate the logical failure rate in the presence of circuit level noise, both during logical idle and during a logical two-qubit measurement. 
In the circuit noise model, each one- or two-qubit mesurement fails independently with probability $p_{physical}$. A measurement that fails acts as an ideal measurement followed by an error drawn uniformly from a set of non-trivial errors.
For a one-qubit measurement, the error is drawn from $\{ (P_1 , F) \} - \{ (I, 0) \}$, while, for a two-qubit measurement, the error is drawn from $\{ (P_1\otimes  P_2 , F) \} - \{ (I\otimes I , 0) \}$, where $P_1,P_2 \in \{ I,X,Y,Z\}$ are Pauli errors acting on the support of the measurement and $F\in \{0,1\}$ denotes a bit flip associated with a readout error of the measurement.
We will evaluate code performance with and without idling errors.
In the case where we include idling errors, the model includes each qubit failing with probability $p_{physical}$ in a step where it is idling, with the error drawn uniformly from the set $\{X,Y,Z\}$.
Setting the idle error to $p_{physical}$ is likely to overestimate the relative error rate of idling compared to the measurement error rate, but provides a sense of how idle errors may impact performance.

We use the square lattice syndrome extraction circuit shown in Appendix \ref{SquareCircuit},  referring to the 28 steps of this circuit as one \emph{round}, and consider two setups for performance evaluation. In the first setup, following Ref.~\onlinecite{Chao2020}, 
we run the code with a time boundary consisting of one noiseless round of syndrome extraction at the beginning and one at the end, together with $T$ noisy rounds
on which we apply noise according to the circuit noise model. 
\begin{center}
\begin{tikzpicture} 
 \draw [pattern=north west lines, pattern color=black,  opacity= 1] (0,0) rectangle (1,1);
 \draw [] (1,0) rectangle (2,1);
 \draw [] (2,0) rectangle (3,1);
 \draw [] (3,0) rectangle (4,1);
 \draw [] (4,0) rectangle (5,1);
 \draw [pattern=north west lines, pattern color=black,  opacity= 1] (5,0) rectangle (6,1);
\node at (0.5,0) [below]{$\;$ prior $\;$};
\node at (3,0) [below]{noisy};
\node at(5.5,0) [below]{posterior};
 \end{tikzpicture}
\end{center}
In the second setup, which we use when evaluating the effect of post-selection, we consider a closed circuit consisting of logical state preparation, syndrome extraction and final measurement of the logical qubit, with noise on the entire circuit. Files containing computer parsable descriptions of the circuits used in our simulations are provided in the supplementary material.

The logical failure rate depends on the decoder. We consider simple decoder constructions that do not guarantee optimal (maximum likelihood) decoding -- as such, the performance estimates may underestimate the performance that could be obtained with a better decoder. When evaluating performance in the setup of ideal prior and posterior rounds we use sliding-window decoding, as described below, while we use a lookup decoder based on degree one faults in the closed circuit setup. 
An example of an optimal decoder would be a lookup decoder that stores the unique syndromes for all configurations of faults with  degrees $t=1,2,...$ up to some highest correctable degree $t_{max}(T)$. In practice, such a large lookup decoder becomes prohibitively costly to construct already for small values of $t_{max}$.

\subsection{Sliding-window decoding of faults}

Instead of constructing a lookup decoder for the entire circuit, we construct a lookup decoder based on part of the circuit, and apply it iteratively. We refer to this as sliding-window decoding, following Ref.~\onlinecite{PRXQuantum.4.040344}. The idea of overlapping recovery, on which sliding-window decoding is based, was introduced in Ref.~\onlinecite{Dennis2002}.

The decoder we construct is able to correct faults that are sufficiently spaced out in time. As an example of what counts as sufficient, consider the following picture, where each red mark represents a degree one circuit fault somewhere in the corresponding noisy round of syndrome extraction:
\begin{center}$\vcenter{\hbox{\scalebox{0.5}{
\begin{tikzpicture}
 \draw [pattern=north west lines, pattern color=black,  opacity= 1] (-1,0) rectangle (0,1);
 \draw [] (0,0) rectangle (1,1);
 \draw [] (1,0) rectangle (2,1);
 \draw [] (2,0) rectangle (3,1);
 \draw [] (3,0) rectangle (4,1);
 \draw [] (4,0) rectangle (5,1);
 \draw [] (5,0) rectangle (6,1);
 \draw [] (6,0) rectangle (7,1);
 \draw [pattern=north west lines, pattern color=black,  opacity= 1] (7,0) rectangle (8,1);
 \node at (0.5,0.5) []{\color{red}$\times$};
 \node at (2.5,0.5) []{\color{red}$\times$};
 \node at (4.5,0.5) []{\color{red}$\times$};
 \node at (6.5,0.5) []{\color{red}$\times$};
 \end{tikzpicture}}
 }}$
\end{center}
As the detectors of the 5+3 code (in a suitable basis) stretch across at most two rounds, the faults in the above picture can be corrected by iteratively applying a lookup decoder based on only three rounds of syndrome extraction, such that it can uniquely identify any  
degree one fault on the middle round out of the three. This decoder is first applied to rounds 1-3 (the first of which by assumption is noiseless). After correcting the fault on round 2, the decoder is applied to rounds 2-4, 3-5, and so forth, until all faults are corrected. With this method, sufficiently spaced out means that any two faults are separated by a round containing no faults. We note that care must be taken to not correct faults `too early''. Consider the following picture:
\begin{center}$\vcenter{\hbox{\scalebox{0.5}{
\begin{tikzpicture}
 \draw [pattern=north west lines, pattern color=black,  opacity= 1] (-1,0) rectangle (0,1);
 \draw [] (0,0) rectangle (1,1);
 \draw [] (1,0) rectangle (2,1);
 \draw [] (2,0) rectangle (3,1);
 \draw [] (3,0) rectangle (4,1);
 \draw [] (4,0) rectangle (5,1);
 \draw [] (5,0) rectangle (6,1);
 \draw [] (6,0) rectangle (7,1);
 \draw [pattern=north west lines, pattern color=black,  opacity= 1] (7,0) rectangle (8,1);
 \node at (1.5,0.5) []{\color{red}$\times$};
 \node at (4.5,0.5) []{\color{red}$\times$};
 \node at (6.5,0.5) []{\color{red}$\times$};
 \end{tikzpicture}}
 }}$
\end{center}
If we naively apply the lookup decoder described above to rounds 1-3, when there is an error in round 3, but not 1 and 2, it may detect a non-trivial syndrome from the detectors stretching across rounds 2 and 3. However, since the  decoder is only built to distinguish between faults on the \emph{middle} round out of three, it might not match that syndrome to the correct fault.
We must therefore 
construct the sliding-window decoder so that it does not attempt a correction before the next step in the iteration in such a scenario. This is accomplished by the use of a lookup decoder based on four rounds rather than three, together with consideration of which detectors are triggered, which rounds they stretch across, and whether the detectors have already been triggered in the previous step of the iteration.

\subsection{Performance during logical idle}

We define $\tilde{p}_{logical}(T)$ to be the logical failure rate during $T$ noisy rounds of syndrome extraction, for a given physical error rate $p_{physical}$. 
Using the sliding-window decoder, errors and correcting operators have short time correlations. As discussed in Ref.~\onlinecite{Chao2020}, the total logical failure rate is therefore expected to grow linearly with $T$, $\tilde{p}_{logical}(T) \sim \alpha T + \beta$, with the constant $\beta$ depending on the time boundary. To account for the time boundary effect, we estimate the logical failure rate for increasingly large values of $T$, and consider the normalized value 
\begin{equation}
    p_{logical}(T) \eqdef \frac{1}{T}\tilde{p}_{logical}(2T)\,.
\end{equation}
The normalization is chosen to match the number of rounds of syndrome extraction needed to perform logical two-qubit measurement; we define a logical idle to consist of two rounds of syndrome extraction. We note that other choices of normalization could be considered, and that the linear dependence of the logical error rates on the number of rounds makes conversion straightforward.

Degree two circuit faults within two adjacent rounds are (generally) not sufficiently spaced out in time, as they could give rise to the same syndrome as a degree one fault, leading to an incorrect correction that induces a logical error. From this, we can make a rough combinatorial estimate of the logical failure rate, by counting the number of possible degree two faults within two rounds of syndrome extraction. In the circuit noise model with no noise on idle qubits, each measurement can give rise to a degree one fault with probability $p_{physical}$, and there are 56 measurements per round of syndrome extraction, leading to the estimate ${p}_{logical}(1) = \tilde{p}_{logical}(2) \sim \binom{2\times 56}{2}  p_{physical}^2$. This would yield a pseudo-threshold of 
$p_{intersect} \sim 1 /\binom{2\times 56}{2} \approx 1.6\times 10^{-4}$. 
In the circuit noise model with noise on idle qubits included, the logical failure rate is expected to be significantly higher, as the data qubits spend most of their time idle. This noise model yields the rough combinatorial estimate $p_{intersect} \sim 1/\binom{2\times 224}{2} \approx 1\times 10^{-5}$.

The above estimates are rather imprecise, as there can be degree two circuit circuit faults that do not lead to an error (decreasing the logical failure rate) and higher degree faults are not taken into account but can also cause logical errors (increasing the logical failure rate). To obtain better estimates, we turn to numerical simulations. In order to efficiently sample at low error rates, we use the importance sampling technique outlined in Ref.~\onlinecite{Paetznick2023}, section VI. 
This technique amounts to estimating the coefficients of the polynomial
\begin{equation}\label{failure_polynomial}
    \tilde{p}_{logical}(T) = \sum_{w\geq 0} f_w \binom{n}{w} p^w (1-p)^{n-w} \,,
\end{equation}
where $n$ is the number of locations in the circuit where faults can occur (in the circuit noise model: the number of measurements, optionally plus the number of idle steps) and $w$ is the degree of a fault configuration. The fraction $f_w$ of degree $w$ fault configurations that induce a logical failure after decoding is estimated numerically, by sampling uniformly over degree $w$ faults and evaluating the logical effect after decoding. We see that if $f_2 = 1$, the leading term of eq.~\eqref{failure_polynomial} would correspond to the rough estimate of $\tilde{p}_{logical}(2)$ given above.

\begin{figure}
    \centering
    \includegraphics[scale=0.3]{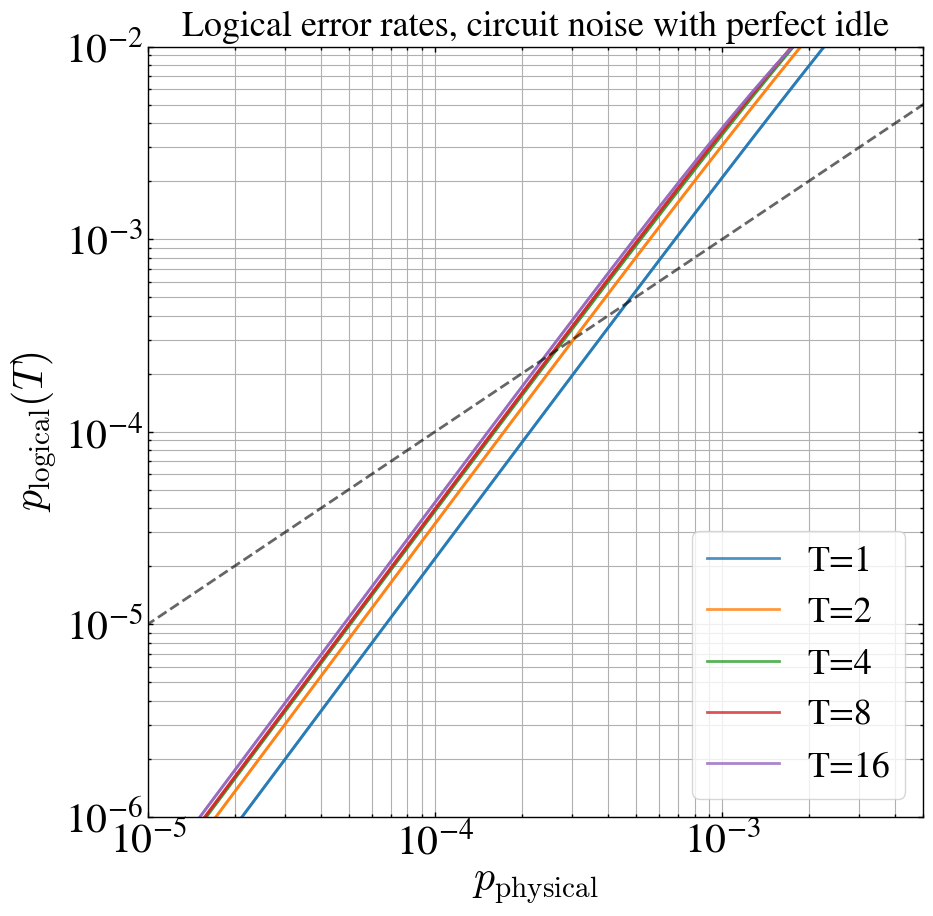}\includegraphics[scale=0.3]{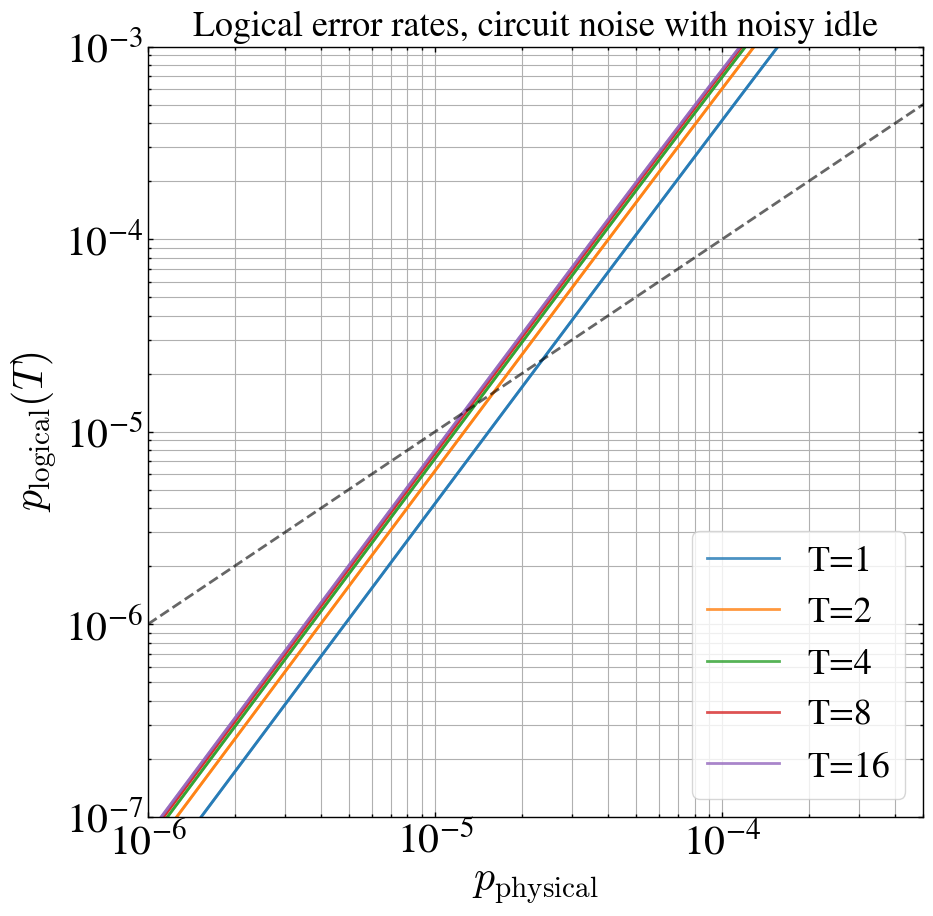}
    \caption{Logical failure rate during logical idle.
    The logical failure rate $p_{logical}(T)$ is shown as a function of the physical failure rate $p_{physical}$, for $2T$ noisy rounds of syndrome extraction. The estimate is based on importance sampling to relative error of 0.01, under the circuit noise model with perfect idle fidelity (left) and with noise on idle qubits (right).
    }
    \label{fig:failure_rate_circuit}
\end{figure}

Denoting by $\mu_{logical} $ and $\sigma_{logical}$ the estimated mean and standard deviation of $\tilde{p}_{logical}(T)$, sampling is performed until $\sigma_{logical} \leq \epsilon \mu_{logical}$ within a chosen range of $p_{physical}$,  for a chosen relative error $\epsilon$. Statistical uncertainty is estimated as in Ref.~\onlinecite{Paetznick2023}, and we sample to relative error $\epsilon = 0.01$. As in Ref.~\onlinecite{Paetznick2023} we truncate $w \leq w_{\rm max}$, excluding faults of high degree from the sampling. We take $w_{\rm max}$ between 16 and 32, depending on the size of the circuit.  
We plot the estimated mean of ${p}_{logical}(T)$ as a function of $p_{physical}$ for different $T$ in Fig. \ref{fig:failure_rate_circuit}, and observe that the results converge for large $T$, with pseudo-threshold estimates in line with the above rough combinatorial estimates. Together with the estimated mean, we also plot an upper bound on the logical failure rate based on the probability mass of the degrees left unsampled by the truncation. At the truncation levels and error ranges considered, this upper bound is indistinguishable from the estimated mean.

The performance estimates in Fig. \ref{fig:failure_rate_circuit}  highlight the importance of time boundary effects in the setting of ideal prior and posterior rounds of syndrome extraction. In particular, the results at $T=2$ (four noisy rounds of syndrome extraction) are clearly distinguishable from those at $T=4$, showing that convergence may take more than $d=3$ noisy rounds.

\subsection{Performance of a logical two-qubit measurement}

In addition to estimating performance during logical idle, we also simulate a logical two-qubit measurement. 
We consider two patches of the 5+3 code, and perform a logical $\bar{X}\bar{X}$ measurement by making inter-patch measurements as shown: 
\begin{equation}
\vcenter{\hbox{
\rotatebox{90}{
\begin{tikzpicture}[scale = 0.5]
\newcommand{\circlesize}{2}
 \draw [thick, red] (3, 1) -- (3, -1+3) ; 
 \draw [thick, red] (1+3, 1) -- (1+3, -1+3) ;
 \draw [thick, red] (2+3, 1) -- (2+3, -1+3) ; 
%
%
 \draw (1+3, 0+3) -- (1+3, -1+3) ;
 \draw (2+3, 0+3) -- (2+3, 1+3) ;
 \draw (0+3,0+3) -- (0+3,1+3);
  \draw (0+3,0+3) -- (2+3,0+3);
 \draw (0+3, 0+3) -- (0+3, -1+3) ;
 \draw (2+3, 0+3) -- (2+3, -1+3) ;
  \draw (0+3, 0+3) node[circle, draw, fill=black, minimum size = \circlesize mm, inner sep = 0]{};
  \draw (1+3, 0+3) node[circle, draw, fill=black, minimum size = \circlesize mm, inner sep = 0]{};
  \draw (2+3, 0+3) node[circle, draw, fill=black, minimum size = \circlesize mm, inner sep = 0]{};
 \draw (1+3, -1+3) node[fill, circle, draw , fill=white, minimum size = \circlesize mm, inner sep = 0]{};
  \draw (0+3, 1+3) node[fill, circle, draw , fill=white, minimum size = \circlesize mm, inner sep = 0]{};
  \draw (2+3, 1+3) node[fill, circle, draw , fill=white, minimum size = \circlesize mm, inner sep = 0]{};
  \draw (0+3, -1+3) node[fill, circle, draw , fill=white, minimum size = \circlesize mm, inner sep = 0]{};
  \draw (2+3, -1+3) node[fill, circle, draw , fill=white, minimum size = \circlesize mm, inner sep = 0]{};
%
%
 \draw (1+3, 0) -- (1+3, 1) ;
 \draw (2+3, 0) -- (2+3, 1) ;
 \draw (0+3,0) -- (0+3,1);
  \draw (0+3,0) -- (2+3,0);
 \draw (0+3, 0) -- (0+3, -1) ;
 \draw (2+3, 0) -- (2+3, -1) ;
  \draw (0+3, 0) node[circle, draw, fill=black, minimum size = \circlesize mm, inner sep = 0]{};
  \draw (1+3, 0) node[circle, draw, fill=black, minimum size = \circlesize mm, inner sep = 0]{};
  \draw (2+3, 0) node[circle, draw, fill=black, minimum size = \circlesize mm, inner sep = 0]{};
  %
  \draw (0+3, 1) node[fill, circle, draw , fill=white, minimum size = \circlesize mm, inner sep = 0]{};
   \draw (1+3, 1) node[fill, circle, draw , fill=white, minimum size = \circlesize mm, inner sep = 0]{};
  \draw (2+3, 1) node[fill, circle, draw , fill=white, minimum size = \circlesize mm, inner sep = 0]{};
  \draw (0+3, -1) node[fill, circle, draw , fill=white, minimum size = \circlesize mm, inner sep = 0]{};
   \draw (2+3, -1) node[fill, circle, draw , fill=white, minimum size = \circlesize mm, inner sep = 0]{};
%
\node [red] at (3.3,1.5) {\rotatebox{-90}{\scriptsize ZZ}};
\node [red] at (4.3,1.5) {\rotatebox{-90}{\scriptsize XX}};
\node [red] at (5.3,1.5) {\rotatebox{-90}{\scriptsize ZZ}};
 \end{tikzpicture}
 }
 }}
\end{equation}
A logical failure occurs if a set of faults, together with the correction given by the decoder, either induces a logical error on the single effective logical qubit that remains after the measurement, or flips the logical $\bar{X}\bar{X}$ measurement outcome.

The setup for performance simulation includes an ideal round of syndrome extraction for each patch of the 5+3 code in the very beginning and end. The noisy part of the circuit consists of repeated logical $\bar{X}\bar{X}$ measurements. More specifically, with $M_{zxz}$ being the above three inter-patch measurements, and $M_{s_1,s_2}$ being a round of syndrome extraction performed on each patch, we measure the sequence $M_{zxz} M_{zxz} M_{s_1,s_2} M_{zxz} M_{zxz} M_{s_1,s_2}$. 
Note that this sequence differs slightly from the protocol described in Subsection \ref{LogicalTwoQubit}, which only involves three inter-patch measurements. Performing four inter-patch measurements is done to simplify the sliding-window decoding. To simplify the handling of faults close to the time boundary, where the time translation symmetry is broken, the decoding procedure for the logical $\bar{X}\bar{X}$ measurement also includes an application of a degree one lookup decoder based on the whole circuit, before applying sliding-window decoding. This may affect the convergence of $p_{logical}(T)$ as $T$ is increased.

\begin{figure}
    \centering
    \includegraphics[scale=0.3]{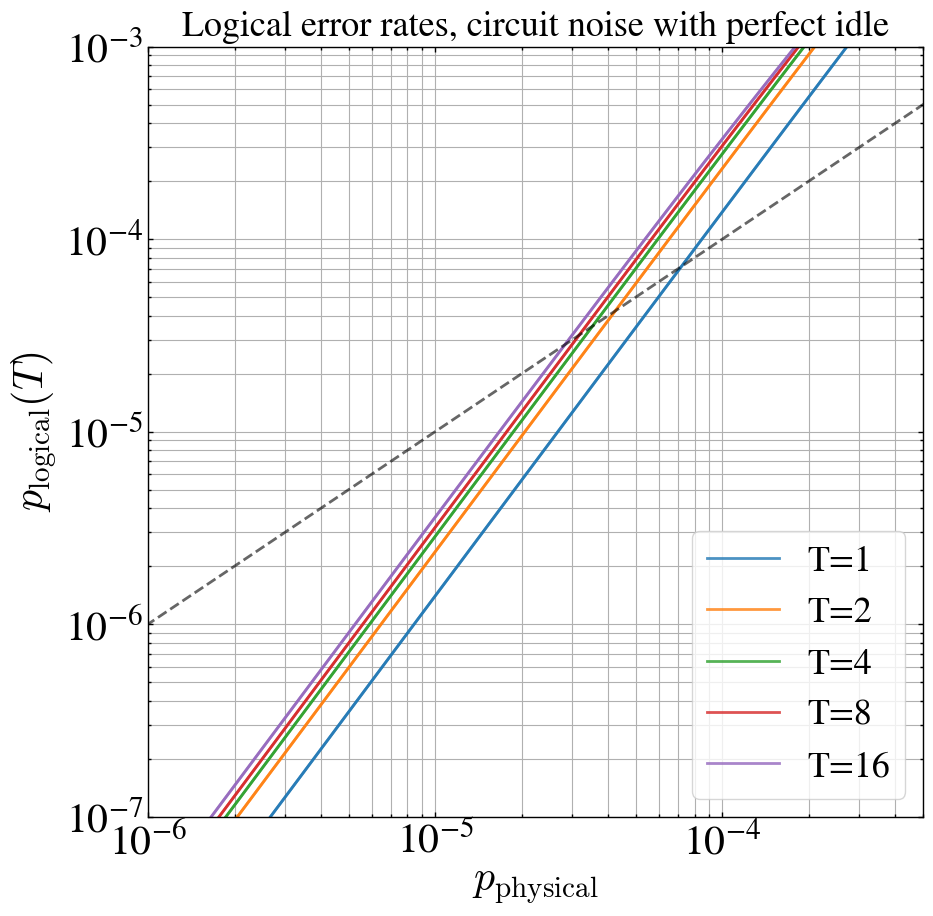}\includegraphics[scale=0.3]{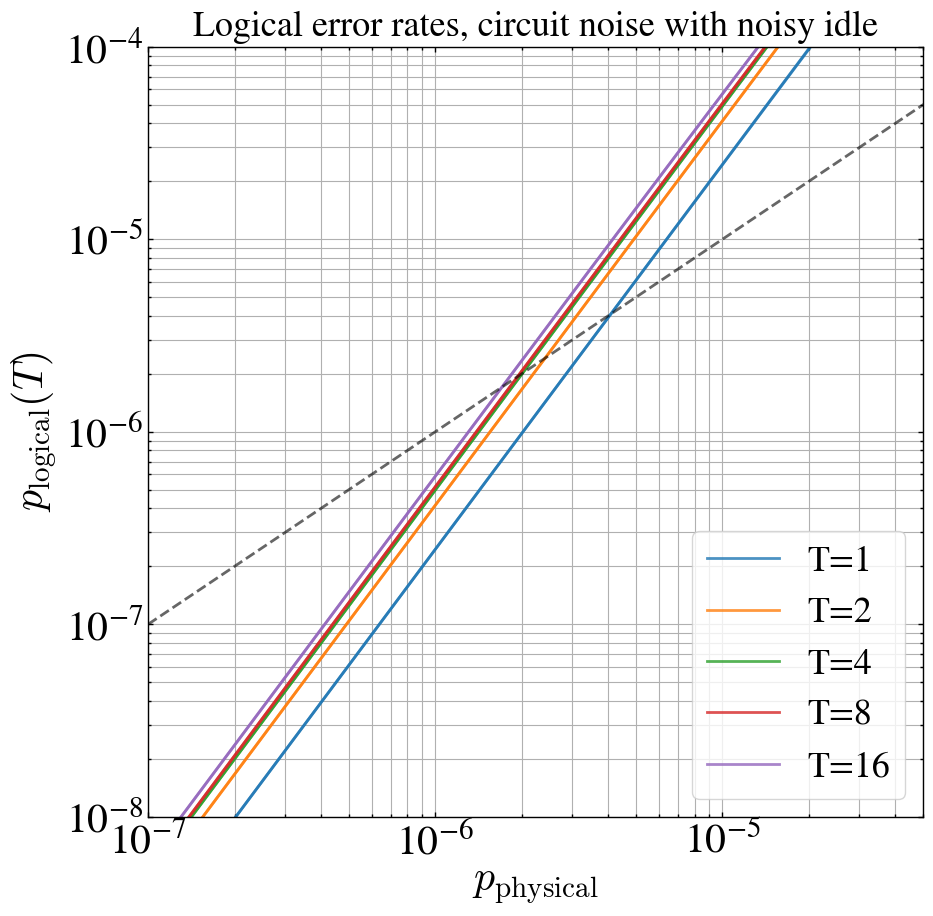}
    \caption{Logical failure rate per logical $\bar{X}\bar{X}$  measurement sequence $M_{zxz} M_{zxz} M_{s_1,s_2} M_{zxz} M_{zxz} M_{s_1,s_2}$.
    The logical failure rate $p_{logical}(T)$ is shown as a function of the physical failure rate $p_{physical}$, for $2T$ noisy rounds of the sequence $M_{zxz} M_{zxz} M_{s_1,s_2}$. The estimate is based on stratified Monte Carlo sampling to relative error of 0.01, under the circuit noise model with perfect idle fidelity (left) and with noise on idle qubits (right).
    }
    \label{fig:failure_rate_surgery}
\end{figure}

Making the combinatorial estimate that two degree one faults within two rounds of the sequence $M_{zxz} M_{zxz}$ $  M_{s_1,s_2}M_{zxz} M_{zxz} M_{s_1,s_2}$ cause a logical error, we obtain $p_{intersect} \sim 1/\binom{2\times 236}{2} \approx 9\times10^{-6}$ with no noise on idle qubits and $p_{intersect} \sim 1/\binom{2\times 948}{2}\approx 5.6\times10^{-7}$ with noise on idle qubits. However, while we need the full sequence to be able to correct all degree one faults, we expect that \emph{most} degree one faults can be corrected based on only half the sequence -- and this is, in practice, how we build the sliding-window decoder. (This choice may also affect the convergence of $p_{logical}(T)$.) 
If we instead base the combinatorial estimate on two rounds of $M_{zxz} M_{zxz} M_{s_1,s_2}$ we obtain $p_{intersect} \sim 1/\binom{2\times 118}{2} \approx 3.6\times10^{-5}$ with no noise on idle qubits and $p_{intersect} \sim 1/\binom{2\times 474}{2}\approx 2.2\times10^{-6}$ with noise on idle qubits.
The numerical estimates for the logical failure rates per round of $M_{zxz} M_{zxz} M_{s_1,s_2} M_{zxz} M_{zxz} M_{s_1,s_2}$ are shown in Fig. \ref{fig:failure_rate_surgery}, and are roughly in line with the latter combinatorial estimates.

\FloatBarrier

\subsection{The effect of adding post-selection on top of error correction}

In the 5+3 code, the introduction of additional detectors means that there are more possible syndromes than those corresponding to degree one faults. It is possible to use post-selection to improve the logical failure rates, rejecting shots whenever the observed syndrome does not correspond to a degree one fault.  
All consequential degree two faults can be detected, although since it is not a distance 4 code, some are indistinguishable from degree one faults by their syndromes. (The 5+3 code could be run as a purely error detecting code, in which case the logical failure rate would be proportional to $p_{physical}^3$.) To leading order, we expect the improvement in the logical error rate, when adding post-selection on top of error correction, to be given by a constant factor: the ratio between degree two faults that cause a logical error with and without post-selection. This ratio depends on the noise model, and may in particular be different between the circuit noise model with and without noise on idle qubits. We emphasize that the name post-selection should not be taken to mean that only syndrome information from the end of the circuit is used -- in the sliding-window decoding strategy, either both correction and rejection could be performed in each ``window'', or correction only could be performed in each window, after which the remaining syndrome of the entire circuit would determine whether to reject. 

We consider a closed circuit setup when evaluating the effect of post-selection, in order to avoid any possible complications from the artificial time boundary in the ideal prior and posterior setup. More specifically, we fault-tolerantly prepare the logical qubit in the $|\bar{0}\rangle$ state, perform two rounds of syndrome extraction (i.e. one logical idle), and finally measure the logical qubit in the $Z$ basis. Pictorially, drawing lines only for the data qubits, the closed circuit is given by
\begin{equation}\label{closedcircuit}
\vcenter{\hbox{
\begin{tikzpicture}[scale=2]
\foreach \y in {1,2,3,4,5}{
  \node[align=center] at (-0.35,\y*0.2) {$|0\rangle$};
}
\foreach \y in {1,2,3,4,5}{
  \draw (-0.2,\y*0.2)--(1.2,\y*0.2);
}
\filldraw[fill=blue!15!white, rounded corners] (0,0) rectangle (1,5*0.2 + 0.2);
\node[align=center] at (0.5,3*0.2) {XZZXI,\\ IXXZX, \\XIXZZ,\\ ZXIXZ};
\node[align=center] at (0.5,-0.2) {(1)};
\begin{scope}[shift={(1.2,0)}]
\foreach \y in {1,2,3,4,5}{
  \draw (-0.2,\y*0.2)--(1.2,\y*0.2);
}
\filldraw[fill=blue!15!white, rounded corners] (0,0) rectangle (1,5*0.2 + 0.2);
\node[align=center] at (0.5,3*0.2) {YZYII,\\ YIIYZ, \\IIYZY};
\node[align=center] at (0.5,-0.2) {(2)};
\end{scope}
\begin{scope}[shift={(1.2*2,0)}]
\foreach \y in {1,2,3,4,5}{
  \draw (-0.2,\y*0.2)--(1.2,\y*0.2);
}
\filldraw[fill=orange!20!white, rounded corners] (0,0) rectangle (1,5*0.2 + 0.2);
\node[align=center] at (0.5,3*0.2) {XZZXI,\\ IXXZX, \\XIXZZ,\\ ZXIXZ};
\node[align=center] at (0.5,-0.2) {(3)};
\end{scope}
\begin{scope}[shift={(1.2*3,0)}]
\foreach \y in {1,2,3,4,5}{
  \draw (-0.2,\y*0.2)--(1.2,\y*0.2);
}
\filldraw[fill=orange!20!white, rounded corners] (0,0) rectangle (1,5*0.2 + 0.2);
\node[align=center] at (0.5,3*0.2) {XZZXI,\\ IXXZX, \\XIXZZ,\\ ZXIXZ};
\node[align=center] at (0.5,-0.2) {(4)};
\end{scope}
\begin{scope}[shift={(1.2*4,0)}]
\foreach \y in {1,2,3,4,5}{
  \draw (-0.2,\y*0.2)--(1.2,\y*0.2);
}
\filldraw[fill=green!15!white, rounded corners] (0,0) rectangle (1,5*0.2 + 0.2);
\node[align=center] at (0.5,3*0.2) {YZYII,\\ YIIYZ, \\IIYZY};
\node[align=center] at (0.5,-0.2) {(5)};
\end{scope}
\begin{scope}[shift={(1.2*5,0)}]
\foreach \y in {1,2,3,4,5}{
  \draw (-0.2,\y*0.2)--(1.2,\y*0.2);
}
\filldraw[fill=green!15!white, rounded corners] (0,0) rectangle (1,5*0.2 + 0.2);
\node[align=center] at (0.5,3*0.2) {XZZXI,\\ IXXZX, \\XIXZZ,\\ ZXIXZ};
\node[align=center] at (0.5,-0.2) {(6)};
\end{scope}
\end{tikzpicture}}}
\end{equation}

Within each step, the stabilizers and logical $\bar{Z}$ representatives are measured in the order given, using the circuits detailed in Appendix \ref{SquareCircuit}. Steps (1) and (2), in blue, correspond to the fault-tolerant logical state preparation. Measuring the four stabilizer group generators in step (1) projects onto the code space, while the three measurements of $\bar{Z}$ in step (2) act as fault-tolerance checks for pre-selection: if any of these three checks results in a nontrivial measurement outcome, we reject. This is done in order to avoid correlated errors from step (1). Steps (3) and (4), in orange, correspond to a logical idle. Finally in step (5) the logical qubit is measured using three different representatives of $\bar{Z}$, with the result determined by a majority vote, and in step (6) a final round of syndrome extraction provides additional syndrome information needed for error correction. This circuit can be seen as a measurement-based version of the circuit in Ref.~\onlinecite{Chao_2018}, Figure 10 (arXiv version).  

When processing the measurement outcomes, we consider steps (1-2) and steps (3-6) of the circuit separately. We use measurement outcomes from only steps (1-2) when preparing the state, inserting Pauli corrections based on the outcomes in step (1) to ensure that we are in the code space and pre-selecting based on the outcomes in step (2). Error correction and (optionally) post-selection is performed using only measurement outcomes from steps (3-6), and the state of the logical qubit is determined based on the (corrected) outcomes in step (5). Error correction is performed using a lookup decoder based on degree one circuit faults. When post-selecting, we reject whenever the observed syndrome is not found in the lookup table, meaning it does not correspond to the syndrome of any degree one circuit fault.

\begin{figure}
    \centering
    \includegraphics[scale=0.3]{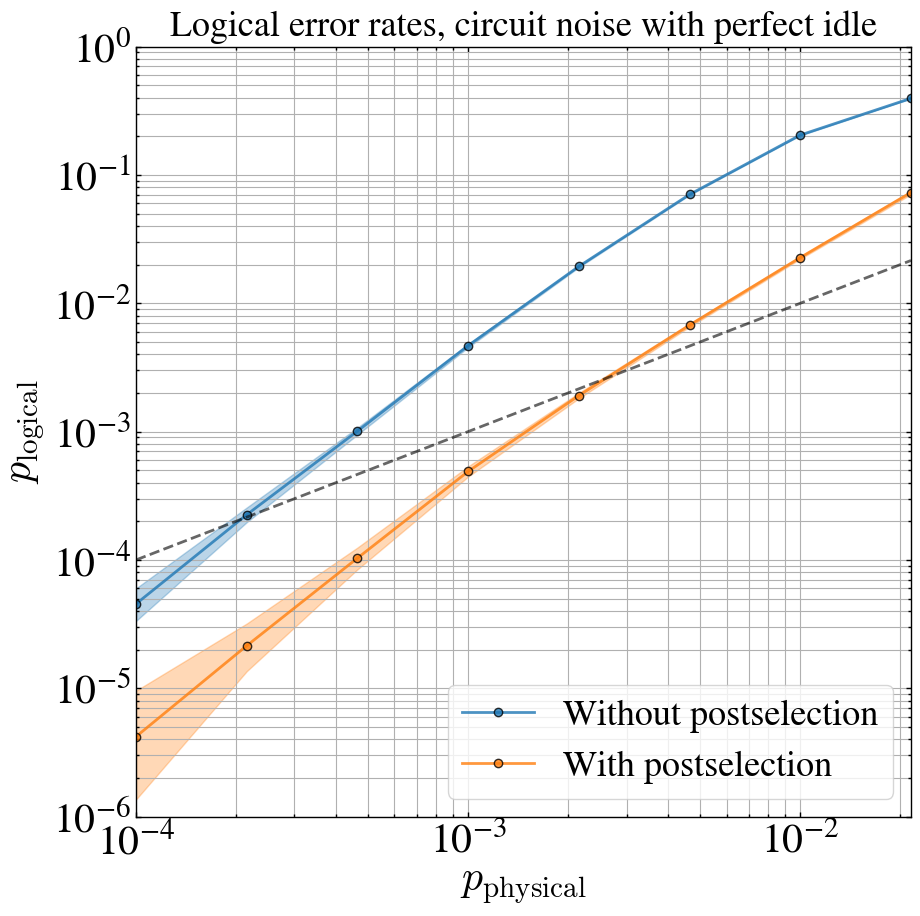}\includegraphics[scale=0.3]{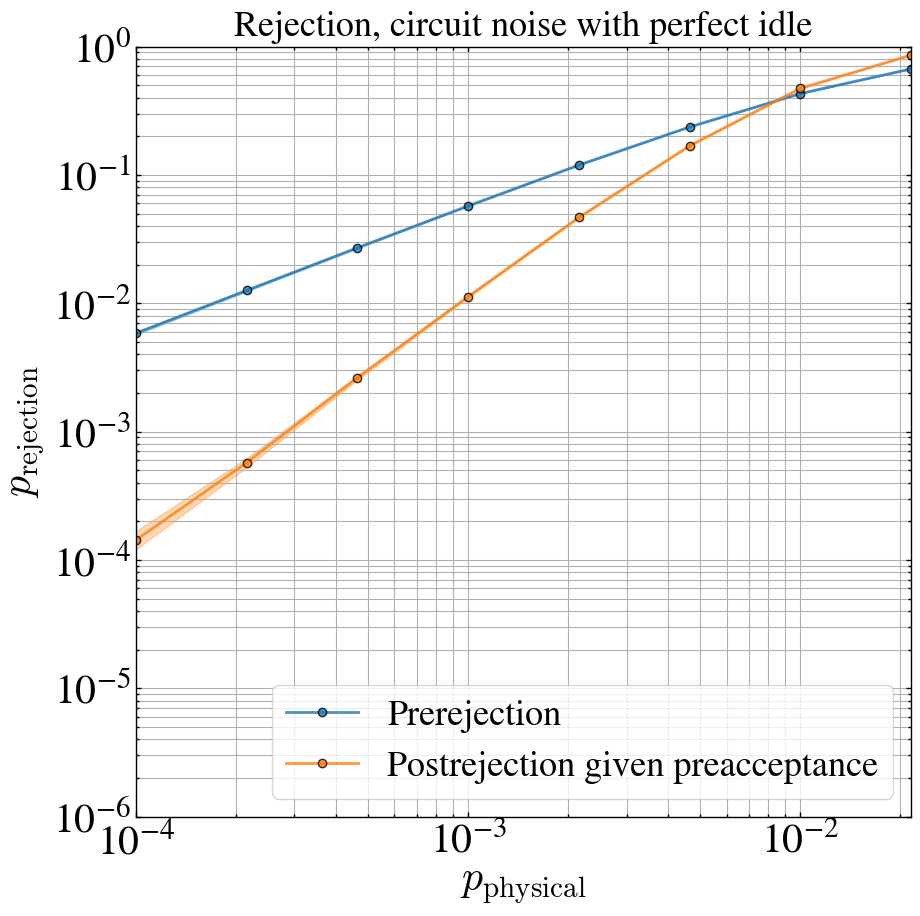}
    \caption{To the left: Estimates of the logical failure rate during a circuit of fault-tolerant state preparation, idle, measurement. A logical idle corresponds to two rounds of syndrome extraction. The estimate is based on Monte Carlo sampling under the circuit noise model with perfect idle fidelity, $10^6$ shots. The logical qubit is prepared and measured in the $Z$ basis, with a logical idle step in between, as described in and below \eqref{closedcircuit}. To the right: Rejection rates during pre- and post-selection.
    }
    \label{fig:postselect_noidle}
\end{figure}

\begin{figure}
    \centering
    \includegraphics[scale=0.3]{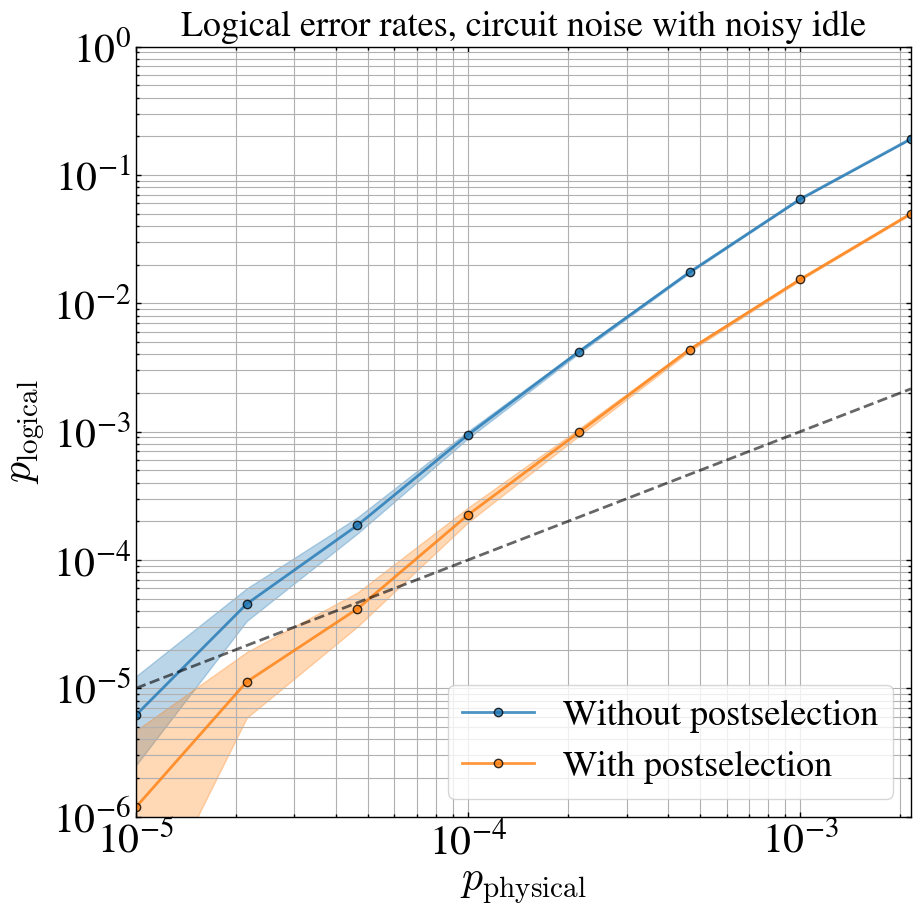}\includegraphics[scale=0.3]{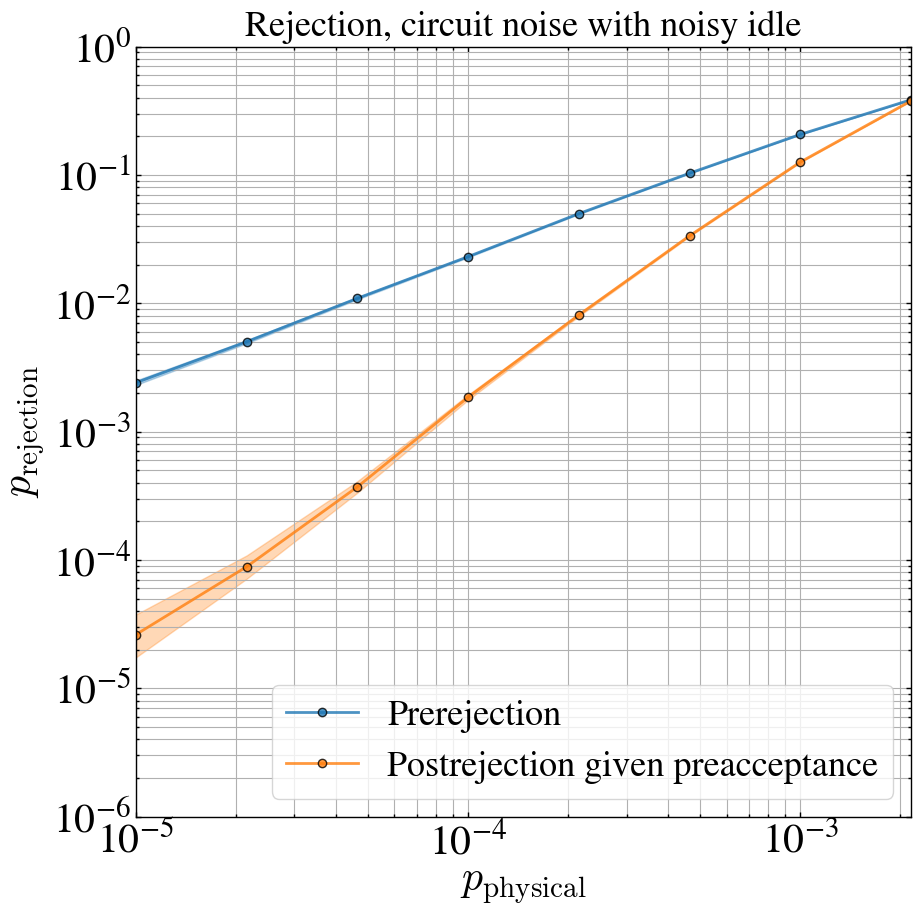}

    \caption{To the left: Estimates of the logical failure rate during a circuit of fault-tolerant state preparation, idle, measurement. A logical idle corresponds to two rounds of syndrome extraction. The estimate is based on Monte Carlo sampling under the circuit noise model with noise on idle qubits, $10^6$ shots. The logical qubit is prepared and measured in the $Z$ basis, with a logical idle step in between, as described in and below \eqref{closedcircuit}. To the right: Rejection rates during pre- and post-selection.
    }
    \label{fig:postselect_idle}
\end{figure}

We show the estimated logical failure rates and rejection rates with and without post-selection in Figs. \ref{fig:postselect_noidle} (circuit noise with perfect idle) and \ref{fig:postselect_idle} (circuit noise with noise on idle qubits). A logical failure occurs if a set of faults, together with the correction from the decoder, flips more than one of the logical measurements in step (5), so that the wrong outcome is obtained from the majority vote. We note that the pseudo-threshold estimates in the absence of post-selection are in line with the estimates in Fig. \ref{fig:failure_rate_circuit}.\footnote{Another possible way of estimating the pseudothreshold in a closed circuit would be to subtract the logical error rates of a circuit with only logical state preparation and measurement from the above results. We find that this error rate is far lower, and that subtracting it does not significantly change the results.} 
The results in Figs. \ref{fig:postselect_noidle} and \ref{fig:postselect_idle} are obtained from direct Monte Carlo sampling (as compared to importance sampling), with the points and the shaded regions representing the median and the 95\% credible intervals, respectively, of a posterior beta distribution. We take $\alpha=\beta=0.5$ in the prior distribution. We see that the improvement in performance is roughly constant across the range of $p_{physical}$, as expected, with the constant factor depending on the noise model: in the circuit noise model with perfect idle, the logical error rate is decreased by roughly a factor of ten, whereas in the circuit noise model with noisy idle qubits it is decreased by roughly a factor of four.

\begin{acknowledgments}

We thank A. Paetznick for many helpful discussions about decoding strategy and help with the numerical performance estimation. We also thank R. Mishmash and M. Beverland for helpful discussions about performance estimation, and M. Steinberg for helpful discussions about fault-tolerance. We thank  A. Paz for providing the parsable circuit format we use to share the simulated circuits.

\end{acknowledgments}

\clearpage

\newpage 

\appendix

\section{Detectors in the 5+2 code and the 5+3 code}\label{Detectors}

In the 5+2 code, the detector corresponding to repeated XZZXI stabilizer measurements is given by the set of measurements highlighted in yellow:
\begin{equation}
\vcenter{\hbox{
\scalebox{0.67}{

}}}
\end{equation}
The detectors corresponding to the other stabilizer generators follow by cyclic permutation of the data qubits.

In the 5+3 code we show the detectors in the less compact, depth 32 syndrome measurement circuit that is not adapted to the square lattice layout, for simplicity of presentation. The detector corresponding to repeated XZZXI stabilizer measurements is given by the set of measurements highlighted in yellow:
\begin{multline}
\vcenter{\hbox{
\scalebox{0.67}{
%
}}}
\end{multline}
Cyclic permutation of the data qubits gives the detectors corresponding to the other stabilizer generators. In the 5+3 code there are also smaller detectors (in terms of the number of measurements involved), corresponding to repeated auxiliary qubit measurements. These are shown for the first part of the depth 32 syndrome extraction circuit, and take the same form for the rest of the circuit. First, repeated single auxiliary qubit measurements introduce additional detectors that make the code robust to readout errors on these measurements. Three such detectors are shown highlighted in yellow, green and blue:
\begin{equation}
\vcenter{\hbox{
\scalebox{0.67}{
\begin{tikzpicture}
   \newcommand{\scaling}{0.55}
   \node at (-2*\scaling, 0.0000*\scaling)  [anchor=center] () {1};
   \draw[line width = 0.6mm, gray] (-1.5*\scaling, 0.0000*\scaling)   -- (19.5000*\scaling, 0.0000*\scaling);
   \node at (-2*\scaling, -1.0000*\scaling)  [anchor=center] () {2};
   \draw[line width = 0.6mm, gray] (-1.5*\scaling, -1.0000*\scaling)   -- (19.5000*\scaling, -1.0000*\scaling);
   \node at (-2*\scaling, -2.0000*\scaling)  [anchor=center] () {3};
   \draw[line width = 0.6mm, gray] (-1.5*\scaling, -2.0000*\scaling)   -- (19.5000*\scaling, -2.0000*\scaling);
   \node at (-2*\scaling, -3.0000*\scaling)  [anchor=center] () {4};
   \draw[line width = 0.6mm, gray] (-1.5*\scaling, -3.0000*\scaling)   -- (19.5000*\scaling, -3.0000*\scaling);
   \node at (-2*\scaling, -4.0000*\scaling)  [anchor=center] () {5};
   \draw[line width = 0.6mm, gray] (-1.5*\scaling, -4.0000*\scaling)   -- (19.5000*\scaling, -4.0000*\scaling);
   \node at (-2*\scaling, -5.0000*\scaling)  [anchor=center] () {A};
   \draw[line width = 0.6mm] (-1.5*\scaling, -5.0000*\scaling)   -- (19.5000*\scaling, -5.0000*\scaling);
   \node at (-2*\scaling, -6.0000*\scaling)  [anchor=center] () {B};
   \draw[line width = 0.6mm] (-1.5*\scaling, -6.0000*\scaling)   -- (19.5000*\scaling, -6.0000*\scaling);
   \node at (-2*\scaling, -7.0000*\scaling)  [anchor=center] () {C};
   \draw[line width = 0.6mm] (-1.5*\scaling, -7.0000*\scaling)   -- (19.5000*\scaling, -7.0000*\scaling);
   \draw[dotted] (-1.000000*\scaling, 0.000000*\scaling +1*\scaling)  -- (-1.000000*\scaling, -7.000000*\scaling -1*\scaling);
   \draw[dotted] (1.000000*\scaling, 0.000000*\scaling +1*\scaling)  -- (1.000000*\scaling, -7.000000*\scaling -1*\scaling);
   \draw[dotted] (3.000000*\scaling, 0.000000*\scaling +1*\scaling)  -- (3.000000*\scaling, -7.000000*\scaling -1*\scaling);
   \draw[dotted] (5.000000*\scaling, 0.000000*\scaling +1*\scaling)  -- (5.000000*\scaling, -7.000000*\scaling -1*\scaling);
   \draw[dotted] (7.000000*\scaling, 0.000000*\scaling +1*\scaling)  -- (7.000000*\scaling, -7.000000*\scaling -1*\scaling);
   \draw[dotted] (9.000000*\scaling, 0.000000*\scaling +1*\scaling)  -- (9.000000*\scaling, -7.000000*\scaling -1*\scaling);
   \draw[dotted] (11.000000*\scaling, 0.000000*\scaling +1*\scaling)  -- (11.000000*\scaling, -7.000000*\scaling -1*\scaling);
   \draw[dotted] (13.000000*\scaling, 0.000000*\scaling +1*\scaling)  -- (13.000000*\scaling, -7.000000*\scaling -1*\scaling);
   \draw[dotted] (15.000000*\scaling, 0.000000*\scaling +1*\scaling)  -- (15.000000*\scaling, -7.000000*\scaling -1*\scaling);
   \draw[dotted] (17.000000*\scaling, 0.000000*\scaling +1*\scaling)  -- (17.000000*\scaling, -7.000000*\scaling -1*\scaling);
   \draw[dotted] (19.000000*\scaling, 0.000000*\scaling +1*\scaling)  -- (19.000000*\scaling, -7.000000*\scaling -1*\scaling);
   \node at (0.0000*\scaling, -7.0000*\scaling -1*\scaling) {1};
   \node at (2.0000*\scaling, -7.0000*\scaling -1*\scaling) {2};
   \node at (4.0000*\scaling, -7.0000*\scaling -1*\scaling) {3};
   \node at (6.0000*\scaling, -7.0000*\scaling -1*\scaling) {4};
   \node at (8.0000*\scaling, -7.0000*\scaling -1*\scaling) {5};
   \node at (10.0000*\scaling, -7.0000*\scaling -1*\scaling) {6};
   \node at (12.0000*\scaling, -7.0000*\scaling -1*\scaling) {7};
   \node at (14.0000*\scaling, -7.0000*\scaling -1*\scaling) {8};
   \node at (16.0000*\scaling, -7.0000*\scaling -1*\scaling) {9};
   \node at (18.0000*\scaling, -7.0000*\scaling -1*\scaling) {10};
   \node at (0.000000*\scaling, -5.000000*\scaling)  [rectangle,fill=white,draw, minimum width = 12pt, minimum height = 12pt,text width=, anchor=center] () {\tiny Z};
   \node at (0.000000*\scaling, -6.000000*\scaling)  [rectangle,fill=white,draw, minimum width = 12pt, minimum height = 12pt,text width=, anchor=center] () {\tiny Z};
   \node at (0.000000*\scaling, -7.000000*\scaling)  [rectangle,fill=white,draw, minimum width = 12pt, minimum height = 12pt,text width=, anchor=center] () {\tiny X};
   \draw[line width=1mm, white] (1.6600*\scaling,  -5.000000*\scaling)   -- (1.6600*\scaling, 0.000000*\scaling);
   \draw (1.6600*\scaling,  -5.000000*\scaling)   -- (1.6600*\scaling, 0.000000*\scaling);
   \node at (1.6600*\scaling, -5.000000*\scaling)  [rectangle,fill=white,draw, minimum width = 12pt, minimum height = 12pt,text width=, anchor=center] () {\tiny X};
   \node at (1.6600*\scaling, 0.000000*\scaling)  [rectangle,fill=white,draw, minimum width = 12pt, minimum height = 12pt,text width=, anchor=center] () {\tiny X};
   \draw[line width=1mm, white] (2.3400*\scaling,  -6.000000*\scaling)   -- (2.3400*\scaling, -3.000000*\scaling);
   \draw (2.3400*\scaling,  -6.000000*\scaling)   -- (2.3400*\scaling, -3.000000*\scaling);
   \node at (2.3400*\scaling, -6.000000*\scaling)  [rectangle,fill=white,draw, minimum width = 12pt, minimum height = 12pt,text width=, anchor=center] () {\tiny X};
   \node at (2.3400*\scaling, -3.000000*\scaling)  [rectangle,fill=white,draw, minimum width = 12pt, minimum height = 12pt,text width=, anchor=center] () {\tiny X};
   \draw[line width=1mm, white] (4.000000*\scaling,  -7.000000*\scaling)   -- (4.000000*\scaling, -5.000000*\scaling);
   \draw (4.000000*\scaling,  -7.000000*\scaling)   -- (4.000000*\scaling, -5.000000*\scaling);
   \node at (4.000000*\scaling, -7.000000*\scaling)  [rectangle,fill=white,draw, minimum width = 12pt, minimum height = 12pt,text width=, anchor=center] () {\tiny Z};
   \node at (4.000000*\scaling, -5.000000*\scaling)  [rectangle,fill=white,draw, minimum width = 12pt, minimum height = 12pt,text width=, anchor=center] () {\tiny Z};
   \draw[line width=1mm, white] (6.000000*\scaling,  -7.000000*\scaling)   -- (6.000000*\scaling, -6.000000*\scaling);
   \draw (6.000000*\scaling,  -7.000000*\scaling)   -- (6.000000*\scaling, -6.000000*\scaling);
   \node at (6.000000*\scaling, -7.000000*\scaling)  [rectangle,fill=white,draw, minimum width = 12pt, minimum height = 12pt,text width=, anchor=center] () {\tiny Z};
   \node at (6.000000*\scaling, -6.000000*\scaling)  [rectangle,fill=white,draw, minimum width = 12pt, minimum height = 12pt,text width=, anchor=center] () {\tiny Z};
   \draw[line width=1mm, white] (8.000000*\scaling,  -7.000000*\scaling)   -- (8.000000*\scaling, -5.000000*\scaling);
   \draw (8.000000*\scaling,  -7.000000*\scaling)   -- (8.000000*\scaling, -5.000000*\scaling);
   \node at (8.000000*\scaling, -7.000000*\scaling)  [rectangle,fill=white,draw, minimum width = 12pt, minimum height = 12pt,text width=, anchor=center] () {\tiny Z};
   \node at (8.000000*\scaling, -5.000000*\scaling)  [rectangle,fill=white,draw, minimum width = 12pt, minimum height = 12pt,text width=, anchor=center] () {\tiny Z};
   \draw[line width=1mm, white] (10.000000*\scaling,  -7.000000*\scaling)   -- (10.000000*\scaling, -6.000000*\scaling);
   \draw (10.000000*\scaling,  -7.000000*\scaling)   -- (10.000000*\scaling, -6.000000*\scaling);
   \node at (10.000000*\scaling, -7.000000*\scaling)  [rectangle,fill=white,draw, minimum width = 12pt, minimum height = 12pt,text width=, anchor=center] () {\tiny Z};
   \node at (10.000000*\scaling, -6.000000*\scaling)  [rectangle,fill=white,draw, minimum width = 12pt, minimum height = 12pt,text width=, anchor=center] () {\tiny Z};
   \draw[line width=1mm, white] (11.6600*\scaling,  -5.000000*\scaling)   -- (11.6600*\scaling, -1.000000*\scaling);
   \draw (11.6600*\scaling,  -5.000000*\scaling)   -- (11.6600*\scaling, -1.000000*\scaling);
   \node at (11.6600*\scaling, -5.000000*\scaling)  [rectangle,fill=white,draw, minimum width = 12pt, minimum height = 12pt,text width=, anchor=center] () {\tiny X};
   \node at (11.6600*\scaling, -1.000000*\scaling)  [rectangle,fill=white,draw, minimum width = 12pt, minimum height = 12pt,text width=, anchor=center] () {\tiny Z};
   \draw[line width=1mm, white] (12.3400*\scaling,  -6.000000*\scaling)   -- (12.3400*\scaling, -2.000000*\scaling);
   \draw (12.3400*\scaling,  -6.000000*\scaling)   -- (12.3400*\scaling, -2.000000*\scaling);
   \node at (12.3400*\scaling, -6.000000*\scaling)  [rectangle,fill=white,draw, minimum width = 12pt, minimum height = 12pt,text width=, anchor=center] () {\tiny X};
   \node at (12.3400*\scaling, -2.000000*\scaling)  [rectangle,fill=white,draw, minimum width = 12pt, minimum height = 12pt,text width=, anchor=center] () {\tiny Z};
   \node at (14.000000*\scaling, -5.000000*\scaling)  [rectangle,fill=white,draw, minimum width = 12pt, minimum height = 12pt,text width=, anchor=center] () {\tiny Z};
   \node at (14.000000*\scaling, -5.000000*\scaling)  [rectangle,fill=yellow,draw, minimum width = 12pt, minimum height = 12pt,text width=, anchor=center] () {\tiny Z};
   \node at (14.000000*\scaling, -6.000000*\scaling)  [rectangle,fill=white,draw, minimum width = 12pt, minimum height = 12pt,text width=, anchor=center] () {\tiny Z};
   \node at (14.000000*\scaling, -6.000000*\scaling)  [rectangle,fill=green,draw, minimum width = 12pt, minimum height = 12pt,text width=, anchor=center] () {\tiny Z};
   \node at (14.000000*\scaling, -7.000000*\scaling)  [rectangle,fill=white,draw, minimum width = 12pt, minimum height = 12pt,text width=, anchor=center] () {\tiny X};
   \node at (14.000000*\scaling, -7.000000*\scaling)  [rectangle,fill=blue!30!white,draw, minimum width = 12pt, minimum height = 12pt,text width=, anchor=center] () {\tiny X};
   \node at (16.000000*\scaling, -5.000000*\scaling)  [rectangle,fill=white,draw, minimum width = 12pt, minimum height = 12pt,text width=, anchor=center] () {\tiny Z};
   \node at (16.000000*\scaling, -5.000000*\scaling)  [rectangle,fill=yellow,draw, minimum width = 12pt, minimum height = 12pt,text width=, anchor=center] () {\tiny Z};
   \node at (16.000000*\scaling, -6.000000*\scaling)  [rectangle,fill=white,draw, minimum width = 12pt, minimum height = 12pt,text width=, anchor=center] () {\tiny Z};
   \node at (16.000000*\scaling, -6.000000*\scaling)  [rectangle,fill=green,draw, minimum width = 12pt, minimum height = 12pt,text width=, anchor=center] () {\tiny Z};
   \node at (16.000000*\scaling, -7.000000*\scaling)  [rectangle,fill=white,draw, minimum width = 12pt, minimum height = 12pt,text width=, anchor=center] () {\tiny X};
   \node at (16.000000*\scaling, -7.000000*\scaling)  [rectangle,fill=blue!30!white,draw, minimum width = 12pt, minimum height = 12pt,text width=, anchor=center] () {\tiny X};
   \draw[line width=1mm, white] (17.6600*\scaling,  -5.000000*\scaling)   -- (17.6600*\scaling, -1.000000*\scaling);
   \draw (17.6600*\scaling,  -5.000000*\scaling)   -- (17.6600*\scaling, -1.000000*\scaling);
   \node at (17.6600*\scaling, -5.000000*\scaling)  [rectangle,fill=white,draw, minimum width = 12pt, minimum height = 12pt,text width=, anchor=center] () {\tiny X};
   \node at (17.6600*\scaling, -1.000000*\scaling)  [rectangle,fill=white,draw, minimum width = 12pt, minimum height = 12pt,text width=, anchor=center] () {\tiny X};
   \draw[line width=1mm, white] (18.3400*\scaling,  -6.000000*\scaling)   -- (18.3400*\scaling, -4.000000*\scaling);
   \draw (18.3400*\scaling,  -6.000000*\scaling)   -- (18.3400*\scaling, -4.000000*\scaling);
   \node at (18.3400*\scaling, -6.000000*\scaling)  [rectangle,fill=white,draw, minimum width = 12pt, minimum height = 12pt,text width=, anchor=center] () {\tiny X};
   \node at (18.3400*\scaling, -4.000000*\scaling)  [rectangle,fill=white,draw, minimum width = 12pt, minimum height = 12pt,text width=, anchor=center] () {\tiny X};

\end{tikzpicture}
}}}
\end{equation}
Second, repeated pairwise auxiliary qubit measurements introduce additional detectors that make the code robust to Pauli errors on the mediating auxiliary qubit $C$, and to readout errors on the pairwise auxiliary qubit measurements. Two such detectors are shown highlighted in yellow and green:
\begin{equation}\label{hook_detectors}
\vcenter{\hbox{
\scalebox{0.67}{
\begin{tikzpicture}
   \newcommand{\scaling}{0.55}
   \node at (-2*\scaling, 0.0000*\scaling)  [anchor=center] () {1};
   \draw[line width = 0.6mm, gray] (-1.5*\scaling, 0.0000*\scaling)   -- (19.5000*\scaling, 0.0000*\scaling);
   \node at (-2*\scaling, -1.0000*\scaling)  [anchor=center] () {2};
   \draw[line width = 0.6mm, gray] (-1.5*\scaling, -1.0000*\scaling)   -- (19.5000*\scaling, -1.0000*\scaling);
   \node at (-2*\scaling, -2.0000*\scaling)  [anchor=center] () {3};
   \draw[line width = 0.6mm, gray] (-1.5*\scaling, -2.0000*\scaling)   -- (19.5000*\scaling, -2.0000*\scaling);
   \node at (-2*\scaling, -3.0000*\scaling)  [anchor=center] () {4};
   \draw[line width = 0.6mm, gray] (-1.5*\scaling, -3.0000*\scaling)   -- (19.5000*\scaling, -3.0000*\scaling);
   \node at (-2*\scaling, -4.0000*\scaling)  [anchor=center] () {5};
   \draw[line width = 0.6mm, gray] (-1.5*\scaling, -4.0000*\scaling)   -- (19.5000*\scaling, -4.0000*\scaling);
   \node at (-2*\scaling, -5.0000*\scaling)  [anchor=center] () {A};
   \draw[line width = 0.6mm] (-1.5*\scaling, -5.0000*\scaling)   -- (19.5000*\scaling, -5.0000*\scaling);
   \node at (-2*\scaling, -6.0000*\scaling)  [anchor=center] () {B};
   \draw[line width = 0.6mm] (-1.5*\scaling, -6.0000*\scaling)   -- (19.5000*\scaling, -6.0000*\scaling);
   \node at (-2*\scaling, -7.0000*\scaling)  [anchor=center] () {C};
   \draw[line width = 0.6mm] (-1.5*\scaling, -7.0000*\scaling)   -- (19.5000*\scaling, -7.0000*\scaling);
   \draw[dotted] (-1.000000*\scaling, 0.000000*\scaling +1*\scaling)  -- (-1.000000*\scaling, -7.000000*\scaling -1*\scaling);
   \draw[dotted] (1.000000*\scaling, 0.000000*\scaling +1*\scaling)  -- (1.000000*\scaling, -7.000000*\scaling -1*\scaling);
   \draw[dotted] (3.000000*\scaling, 0.000000*\scaling +1*\scaling)  -- (3.000000*\scaling, -7.000000*\scaling -1*\scaling);
   \draw[dotted] (5.000000*\scaling, 0.000000*\scaling +1*\scaling)  -- (5.000000*\scaling, -7.000000*\scaling -1*\scaling);
   \draw[dotted] (7.000000*\scaling, 0.000000*\scaling +1*\scaling)  -- (7.000000*\scaling, -7.000000*\scaling -1*\scaling);
   \draw[dotted] (9.000000*\scaling, 0.000000*\scaling +1*\scaling)  -- (9.000000*\scaling, -7.000000*\scaling -1*\scaling);
   \draw[dotted] (11.000000*\scaling, 0.000000*\scaling +1*\scaling)  -- (11.000000*\scaling, -7.000000*\scaling -1*\scaling);
   \draw[dotted] (13.000000*\scaling, 0.000000*\scaling +1*\scaling)  -- (13.000000*\scaling, -7.000000*\scaling -1*\scaling);
   \draw[dotted] (15.000000*\scaling, 0.000000*\scaling +1*\scaling)  -- (15.000000*\scaling, -7.000000*\scaling -1*\scaling);
   \draw[dotted] (17.000000*\scaling, 0.000000*\scaling +1*\scaling)  -- (17.000000*\scaling, -7.000000*\scaling -1*\scaling);
   \draw[dotted] (19.000000*\scaling, 0.000000*\scaling +1*\scaling)  -- (19.000000*\scaling, -7.000000*\scaling -1*\scaling);
   \node at (0.0000*\scaling, -7.0000*\scaling -1*\scaling) {1};
   \node at (2.0000*\scaling, -7.0000*\scaling -1*\scaling) {2};
   \node at (4.0000*\scaling, -7.0000*\scaling -1*\scaling) {3};
   \node at (6.0000*\scaling, -7.0000*\scaling -1*\scaling) {4};
   \node at (8.0000*\scaling, -7.0000*\scaling -1*\scaling) {5};
   \node at (10.0000*\scaling, -7.0000*\scaling -1*\scaling) {6};
   \node at (12.0000*\scaling, -7.0000*\scaling -1*\scaling) {7};
   \node at (14.0000*\scaling, -7.0000*\scaling -1*\scaling) {8};
   \node at (16.0000*\scaling, -7.0000*\scaling -1*\scaling) {9};
   \node at (18.0000*\scaling, -7.0000*\scaling -1*\scaling) {10};
   \node at (0.000000*\scaling, -5.000000*\scaling)  [rectangle,fill=white,draw, minimum width = 12pt, minimum height = 12pt,text width=, anchor=center] () {\tiny Z};
   \node at (0.000000*\scaling, -6.000000*\scaling)  [rectangle,fill=white,draw, minimum width = 12pt, minimum height = 12pt,text width=, anchor=center] () {\tiny Z};
   \node at (0.000000*\scaling, -7.000000*\scaling)  [rectangle,fill=white,draw, minimum width = 12pt, minimum height = 12pt,text width=, anchor=center] () {\tiny X};
   \draw[line width=1mm, white] (1.6600*\scaling,  -5.000000*\scaling)   -- (1.6600*\scaling, 0.000000*\scaling);
   \draw (1.6600*\scaling,  -5.000000*\scaling)   -- (1.6600*\scaling, 0.000000*\scaling);
   \node at (1.6600*\scaling, -5.000000*\scaling)  [rectangle,fill=white,draw, minimum width = 12pt, minimum height = 12pt,text width=, anchor=center] () {\tiny X};
   \node at (1.6600*\scaling, 0.000000*\scaling)  [rectangle,fill=white,draw, minimum width = 12pt, minimum height = 12pt,text width=, anchor=center] () {\tiny X};
   \draw[line width=1mm, white] (2.3400*\scaling,  -6.000000*\scaling)   -- (2.3400*\scaling, -3.000000*\scaling);
   \draw (2.3400*\scaling,  -6.000000*\scaling)   -- (2.3400*\scaling, -3.000000*\scaling);
   \node at (2.3400*\scaling, -6.000000*\scaling)  [rectangle,fill=white,draw, minimum width = 12pt, minimum height = 12pt,text width=, anchor=center] () {\tiny X};
   \node at (2.3400*\scaling, -3.000000*\scaling)  [rectangle,fill=white,draw, minimum width = 12pt, minimum height = 12pt,text width=, anchor=center] () {\tiny X};
   \draw[line width=1mm, white] (4.000000*\scaling,  -7.000000*\scaling)   -- (4.000000*\scaling, -5.000000*\scaling);
   \draw (4.000000*\scaling,  -7.000000*\scaling)   -- (4.000000*\scaling, -5.000000*\scaling);
   \node at (4.000000*\scaling, -7.000000*\scaling)  [rectangle,fill=white,draw, minimum width = 12pt, minimum height = 12pt,text width=, anchor=center] () {\tiny Z};
   \node at (4.000000*\scaling, -5.000000*\scaling)  [rectangle,fill=white,draw, minimum width = 12pt, minimum height = 12pt,text width=, anchor=center] () {\tiny Z};
   \node at (4.000000*\scaling, -7.000000*\scaling)  [rectangle,fill=yellow,draw, minimum width = 12pt, minimum height = 12pt,text width=, anchor=center] () {\tiny Z};
   \node at (4.000000*\scaling, -5.000000*\scaling)  [rectangle,fill=yellow,draw, minimum width = 12pt, minimum height = 12pt,text width=, anchor=center] () {\tiny Z};
   \draw[line width=1mm, white] (6.000000*\scaling,  -7.000000*\scaling)   -- (6.000000*\scaling, -6.000000*\scaling);
   \draw (6.000000*\scaling,  -7.000000*\scaling)   -- (6.000000*\scaling, -6.000000*\scaling);
   \node at (6.000000*\scaling, -7.000000*\scaling)  [rectangle,fill=white,draw, minimum width = 12pt, minimum height = 12pt,text width=, anchor=center] () {\tiny Z};
   \node at (6.000000*\scaling, -6.000000*\scaling)  [rectangle,fill=white,draw, minimum width = 12pt, minimum height = 12pt,text width=, anchor=center] () {\tiny Z};
   \node at (6.000000*\scaling, -7.000000*\scaling)  [rectangle,fill=green,draw, minimum width = 12pt, minimum height = 12pt,text width=, anchor=center] () {\tiny Z};
   \node at (6.000000*\scaling, -6.000000*\scaling)  [rectangle,fill=green,draw, minimum width = 12pt, minimum height = 12pt,text width=, anchor=center] () {\tiny Z};
   \draw[line width=1mm, white] (8.000000*\scaling,  -7.000000*\scaling)   -- (8.000000*\scaling, -5.000000*\scaling);
   \draw (8.000000*\scaling,  -7.000000*\scaling)   -- (8.000000*\scaling, -5.000000*\scaling);
   \node at (8.000000*\scaling, -7.000000*\scaling)  [rectangle,fill=white,draw, minimum width = 12pt, minimum height = 12pt,text width=, anchor=center] () {\tiny Z};
   \node at (8.000000*\scaling, -5.000000*\scaling)  [rectangle,fill=white,draw, minimum width = 12pt, minimum height = 12pt,text width=, anchor=center] () {\tiny Z};
   \node at (8.000000*\scaling, -7.000000*\scaling)  [rectangle,fill=yellow,draw, minimum width = 12pt, minimum height = 12pt,text width=, anchor=center] () {\tiny Z};
   \node at (8.000000*\scaling, -5.000000*\scaling)  [rectangle,fill=yellow,draw, minimum width = 12pt, minimum height = 12pt,text width=, anchor=center] () {\tiny Z};
   \draw[line width=1mm, white] (10.000000*\scaling,  -7.000000*\scaling)   -- (10.000000*\scaling, -6.000000*\scaling);
   \draw (10.000000*\scaling,  -7.000000*\scaling)   -- (10.000000*\scaling, -6.000000*\scaling);
   \node at (10.000000*\scaling, -7.000000*\scaling)  [rectangle,fill=white,draw, minimum width = 12pt, minimum height = 12pt,text width=, anchor=center] () {\tiny Z};
   \node at (10.000000*\scaling, -6.000000*\scaling)  [rectangle,fill=white,draw, minimum width = 12pt, minimum height = 12pt,text width=, anchor=center] () {\tiny Z};
   \node at (10.000000*\scaling, -7.000000*\scaling)  [rectangle,fill=green,draw, minimum width = 12pt, minimum height = 12pt,text width=, anchor=center] () {\tiny Z};
   \node at (10.000000*\scaling, -6.000000*\scaling)  [rectangle,fill=green,draw, minimum width = 12pt, minimum height = 12pt,text width=, anchor=center] () {\tiny Z};
   \draw[line width=1mm, white] (11.6600*\scaling,  -5.000000*\scaling)   -- (11.6600*\scaling, -1.000000*\scaling);
   \draw (11.6600*\scaling,  -5.000000*\scaling)   -- (11.6600*\scaling, -1.000000*\scaling);
   \node at (11.6600*\scaling, -5.000000*\scaling)  [rectangle,fill=white,draw, minimum width = 12pt, minimum height = 12pt,text width=, anchor=center] () {\tiny X};
   \node at (11.6600*\scaling, -1.000000*\scaling)  [rectangle,fill=white,draw, minimum width = 12pt, minimum height = 12pt,text width=, anchor=center] () {\tiny Z};
   \draw[line width=1mm, white] (12.3400*\scaling,  -6.000000*\scaling)   -- (12.3400*\scaling, -2.000000*\scaling);
   \draw (12.3400*\scaling,  -6.000000*\scaling)   -- (12.3400*\scaling, -2.000000*\scaling);
   \node at (12.3400*\scaling, -6.000000*\scaling)  [rectangle,fill=white,draw, minimum width = 12pt, minimum height = 12pt,text width=, anchor=center] () {\tiny X};
   \node at (12.3400*\scaling, -2.000000*\scaling)  [rectangle,fill=white,draw, minimum width = 12pt, minimum height = 12pt,text width=, anchor=center] () {\tiny Z};
   \node at (14.000000*\scaling, -5.000000*\scaling)  [rectangle,fill=white,draw, minimum width = 12pt, minimum height = 12pt,text width=, anchor=center] () {\tiny Z};
   \node at (14.000000*\scaling, -6.000000*\scaling)  [rectangle,fill=white,draw, minimum width = 12pt, minimum height = 12pt,text width=, anchor=center] () {\tiny Z};
   \node at (14.000000*\scaling, -7.000000*\scaling)  [rectangle,fill=white,draw, minimum width = 12pt, minimum height = 12pt,text width=, anchor=center] () {\tiny X};
   \node at (16.000000*\scaling, -5.000000*\scaling)  [rectangle,fill=white,draw, minimum width = 12pt, minimum height = 12pt,text width=, anchor=center] () {\tiny Z};
   \node at (16.000000*\scaling, -6.000000*\scaling)  [rectangle,fill=white,draw, minimum width = 12pt, minimum height = 12pt,text width=, anchor=center] () {\tiny Z};
   \node at (16.000000*\scaling, -7.000000*\scaling)  [rectangle,fill=white,draw, minimum width = 12pt, minimum height = 12pt,text width=, anchor=center] () {\tiny X};
   \draw[line width=1mm, white] (17.6600*\scaling,  -5.000000*\scaling)   -- (17.6600*\scaling, -1.000000*\scaling);
   \draw (17.6600*\scaling,  -5.000000*\scaling)   -- (17.6600*\scaling, -1.000000*\scaling);
   \node at (17.6600*\scaling, -5.000000*\scaling)  [rectangle,fill=white,draw, minimum width = 12pt, minimum height = 12pt,text width=, anchor=center] () {\tiny X};
   \node at (17.6600*\scaling, -1.000000*\scaling)  [rectangle,fill=white,draw, minimum width = 12pt, minimum height = 12pt,text width=, anchor=center] () {\tiny X};
   \draw[line width=1mm, white] (18.3400*\scaling,  -6.000000*\scaling)   -- (18.3400*\scaling, -4.000000*\scaling);
   \draw (18.3400*\scaling,  -6.000000*\scaling)   -- (18.3400*\scaling, -4.000000*\scaling);
   \node at (18.3400*\scaling, -6.000000*\scaling)  [rectangle,fill=white,draw, minimum width = 12pt, minimum height = 12pt,text width=, anchor=center] () {\tiny X};
   \node at (18.3400*\scaling, -4.000000*\scaling)  [rectangle,fill=white,draw, minimum width = 12pt, minimum height = 12pt,text width=, anchor=center] () {\tiny X};

\end{tikzpicture}
}}}
\end{equation}


\section{Square lattice measurement circuits for the 5+3 code}\label{SquareCircuit}

In this Appendix, we show the syndrome extraction circuit for the 5+3 code in the square lattice layout, as well as an example circuit for measuring logical operators. The data qubits are labelled 1-5 and the auxiliary qubits $A,B,C$ as follows:
\begin{center}
\begin{tikzpicture}
\newcommand{\halfscaling}{0.8}
\newcommand{\nodesize}{1.5}
\draw [fill=white,opacity=0] (-0.5*\halfscaling,-0.5*\halfscaling) rectangle (2.5*\halfscaling,2.5*\halfscaling); 
%
\foreach \i in {0,...,2}{\foreach \j in {1}{
\draw (\i*\halfscaling, \j*\halfscaling) node[circle, draw, fill=black, minimum size = \nodesize mm, inner sep = 0]{}; }; };
\foreach \i in {0,2}{\foreach \j in {0}{
\draw (\i*\halfscaling, \j*\halfscaling) node[circle, draw, fill=white, minimum size = \nodesize mm, inner sep = 0]{};};};
\foreach \i in {0,1,2}{\foreach \j in {2}{
\draw (\i*\halfscaling, \j*\halfscaling) node[circle, draw, fill=white, minimum size = \nodesize mm, inner sep = 0]{};};};
\node at (0*\halfscaling, 1*\halfscaling)[below] {\small A};
\node at (1*\halfscaling, 1*\halfscaling)[below] {\small C};
\node at (2*\halfscaling, 1*\halfscaling)[below] {\small B};
\node at (0*\halfscaling, 2*\halfscaling)[below] {\small 1};
\node at (1*\halfscaling, 2*\halfscaling)[below] {\small 2};
\node at (2*\halfscaling, 2*\halfscaling)[below] {\small 3};
\node at (0*\halfscaling, 0*\halfscaling)[below] {\small 5};
\node at (2*\halfscaling, 0*\halfscaling)[below] {\small 4};
\end{tikzpicture}
\end{center}
\noindent
We show the circuit for each stabilizer generator in turn, and recall that the last step $8$ of a stabilizer measurement is performed at the same time as the first step $1$ of the subsequent stabilizer measurement, so that each stabilizer measurement is effectively depth 7. In order to avoid collisions during the simultaneous measurement of the last step and the first subsequent step, we measure the stabilizers in the order $XZZXI$, $IXZZX$, $ZXIXZ$, $XIXZZ$.

\begin{minipage}{0.15\linewidth}
${XZZXI}$

\vspace{1.5cm}
\end{minipage}
\begin{minipage}{0.8\linewidth}
\begin{tikzpicture}
   \newcommand{\scaling}{0.6}
   \node at (-2*\scaling, 0.0000*\scaling)  [anchor=center] () {1};
   \draw[line width = 0.6mm, gray] (-1.5*\scaling, 0.0000*\scaling)   -- (19.7000*\scaling, 0.0000*\scaling);
   \node at (-2*\scaling, -1.0000*\scaling)  [anchor=center] () {2};
   \draw[line width = 0.6mm, gray] (-1.5*\scaling, -1.0000*\scaling)   -- (19.7000*\scaling, -1.0000*\scaling);
   \node at (-2*\scaling, -2.0000*\scaling)  [anchor=center] () {3};
   \draw[line width = 0.6mm, gray] (-1.5*\scaling, -2.0000*\scaling)   -- (19.7000*\scaling, -2.0000*\scaling);
   \node at (-2*\scaling, -3.0000*\scaling)  [anchor=center] () {4};
   \draw[line width = 0.6mm, gray] (-1.5*\scaling, -3.0000*\scaling)   -- (19.7000*\scaling, -3.0000*\scaling);
   \node at (-2*\scaling, -4.0000*\scaling)  [anchor=center] () {5};
   \draw[line width = 0.6mm, gray] (-1.5*\scaling, -4.0000*\scaling)   -- (19.7000*\scaling, -4.0000*\scaling);
   \node at (-2*\scaling, -5.0000*\scaling)  [anchor=center] () {A};
   \draw[line width = 0.6mm] (-1.5*\scaling, -5.0000*\scaling)   -- (19.7000*\scaling, -5.0000*\scaling);
   \node at (-2*\scaling, -6.0000*\scaling)  [anchor=center] () {B};
   \draw[line width = 0.6mm] (-1.5*\scaling, -6.0000*\scaling)   -- (19.7000*\scaling, -6.0000*\scaling);
   \node at (-2*\scaling, -7.0000*\scaling)  [anchor=center] () {C};
   \draw[line width = 0.6mm] (-1.5*\scaling, -7.0000*\scaling)   -- (19.7000*\scaling, -7.0000*\scaling);
   \draw[dotted] (-1.300000*\scaling, 0.000000*\scaling +1*\scaling)  -- (-1.300000*\scaling, -7.000000*\scaling -1*\scaling);
   \draw[dotted] (1.300000*\scaling, 0.000000*\scaling +1*\scaling)  -- (1.300000*\scaling, -7.000000*\scaling -1*\scaling);
   \draw[dotted] (3.900000*\scaling, 0.000000*\scaling +1*\scaling)  -- (3.900000*\scaling, -7.000000*\scaling -1*\scaling);
   \draw[dotted] (6.500000*\scaling, 0.000000*\scaling +1*\scaling)  -- (6.500000*\scaling, -7.000000*\scaling -1*\scaling);
   \draw[dotted] (9.100000*\scaling, 0.000000*\scaling +1*\scaling)  -- (9.100000*\scaling, -7.000000*\scaling -1*\scaling);
   \draw[dotted] (11.700000*\scaling, 0.000000*\scaling +1*\scaling)  -- (11.700000*\scaling, -7.000000*\scaling -1*\scaling);
   \draw[dotted] (14.300000*\scaling, 0.000000*\scaling +1*\scaling)  -- (14.300000*\scaling, -7.000000*\scaling -1*\scaling);
   \draw[dotted] (16.900000*\scaling, 0.000000*\scaling +1*\scaling)  -- (16.900000*\scaling, -7.000000*\scaling -1*\scaling);
   \draw[dotted] (19.500000*\scaling, 0.000000*\scaling +1*\scaling)  -- (19.500000*\scaling, -7.000000*\scaling -1*\scaling);
   \node at (0.0000*\scaling, -7.0000*\scaling -1*\scaling) {1};
   \node at (2.6000*\scaling, -7.0000*\scaling -1*\scaling) {2};
   \node at (5.2000*\scaling, -7.0000*\scaling -1*\scaling) {3};
   \node at (7.8000*\scaling, -7.0000*\scaling -1*\scaling) {4};
   \node at (10.4000*\scaling, -7.0000*\scaling -1*\scaling) {5};
   \node at (13.0000*\scaling, -7.0000*\scaling -1*\scaling) {6};
   \node at (15.6000*\scaling, -7.0000*\scaling -1*\scaling) {7};
   \node at (18.2000*\scaling, -7.0000*\scaling -1*\scaling) {8};
   \node at (0.000000*\scaling, -6.000000*\scaling)  [rectangle,fill=white,draw, minimum width = 12pt, minimum height = 12pt,text width=, anchor=center] () {\tiny Z};
   \node at (2.158000*\scaling, -5.000000*\scaling)  [rectangle,fill=white,draw, minimum width = 12pt, minimum height = 12pt,text width=, anchor=center] () {\tiny Z};
   \node at (2.158000*\scaling, -7.000000*\scaling)  [rectangle,fill=white,draw, minimum width = 12pt, minimum height = 12pt,text width=, anchor=center] () {\tiny X};
   \draw[line width=1mm, white] (3.0420*\scaling,  -6.000000*\scaling)   -- (3.0420*\scaling, -3.000000*\scaling);
   \draw (3.0420*\scaling,  -6.000000*\scaling)   -- (3.0420*\scaling, -3.000000*\scaling);
   \node at (3.0420*\scaling, -6.000000*\scaling)  [rectangle,fill=white,draw, minimum width = 12pt, minimum height = 12pt,text width=, anchor=center] () {\tiny X};
   \node at (3.0420*\scaling, -3.000000*\scaling)  [rectangle,fill=white,draw, minimum width = 12pt, minimum height = 12pt,text width=, anchor=center] () {\tiny X};
   \draw[line width=1mm, white] (5.2000*\scaling,  -5.000000*\scaling)   -- (5.2000*\scaling, 0.000000*\scaling);
   \draw (5.2000*\scaling,  -5.000000*\scaling)   -- (5.2000*\scaling, 0.000000*\scaling);
   \node at (5.2000*\scaling, -5.000000*\scaling)  [rectangle,fill=white,draw, minimum width = 12pt, minimum height = 12pt,text width=, anchor=center] () {\tiny X};
   \node at (5.2000*\scaling, 0.000000*\scaling)  [rectangle,fill=white,draw, minimum width = 12pt, minimum height = 12pt,text width=, anchor=center] () {\tiny X};
   \draw[line width=1mm, white] (5.2000*\scaling,  -7.000000*\scaling)   -- (5.2000*\scaling, -6.000000*\scaling);
   \draw (5.2000*\scaling,  -7.000000*\scaling)   -- (5.2000*\scaling, -6.000000*\scaling);
   \node at (5.2000*\scaling, -7.000000*\scaling)  [rectangle,fill=white,draw, minimum width = 12pt, minimum height = 12pt,text width=, anchor=center] () {\tiny Z};
   \node at (5.2000*\scaling, -6.000000*\scaling)  [rectangle,fill=white,draw, minimum width = 12pt, minimum height = 12pt,text width=, anchor=center] () {\tiny Z};
   \draw[line width=1mm, white] (7.8000*\scaling,  -7.000000*\scaling)   -- (7.8000*\scaling, -5.000000*\scaling);
   \draw (7.8000*\scaling,  -7.000000*\scaling)   -- (7.8000*\scaling, -5.000000*\scaling);
   \node at (7.8000*\scaling, -7.000000*\scaling)  [rectangle,fill=white,draw, minimum width = 12pt, minimum height = 12pt,text width=, anchor=center] () {\tiny Z};
   \node at (7.8000*\scaling, -5.000000*\scaling)  [rectangle,fill=white,draw, minimum width = 12pt, minimum height = 12pt,text width=, anchor=center] () {\tiny Z};
   \draw[line width=1mm, white] (10.4000*\scaling,  -7.000000*\scaling)   -- (10.4000*\scaling, -6.000000*\scaling);
   \draw (10.4000*\scaling,  -7.000000*\scaling)   -- (10.4000*\scaling, -6.000000*\scaling);
   \node at (10.4000*\scaling, -7.000000*\scaling)  [rectangle,fill=white,draw, minimum width = 12pt, minimum height = 12pt,text width=, anchor=center] () {\tiny Z};
   \node at (10.4000*\scaling, -6.000000*\scaling)  [rectangle,fill=white,draw, minimum width = 12pt, minimum height = 12pt,text width=, anchor=center] () {\tiny Z};
   \draw[line width=1mm, white] (12.5580*\scaling,  -7.000000*\scaling)   -- (12.5580*\scaling, -5.000000*\scaling);
   \draw (12.5580*\scaling,  -7.000000*\scaling)   -- (12.5580*\scaling, -5.000000*\scaling);
   \node at (12.5580*\scaling, -7.000000*\scaling)  [rectangle,fill=white,draw, minimum width = 12pt, minimum height = 12pt,text width=, anchor=center] () {\tiny Z};
   \node at (12.5580*\scaling, -5.000000*\scaling)  [rectangle,fill=white,draw, minimum width = 12pt, minimum height = 12pt,text width=, anchor=center] () {\tiny Z};
   \draw[line width=1mm, white] (13.4420*\scaling,  -6.000000*\scaling)   -- (13.4420*\scaling, -2.000000*\scaling);
   \draw (13.4420*\scaling,  -6.000000*\scaling)   -- (13.4420*\scaling, -2.000000*\scaling);
   \node at (13.4420*\scaling, -6.000000*\scaling)  [rectangle,fill=white,draw, minimum width = 12pt, minimum height = 12pt,text width=, anchor=center] () {\tiny X};
   \node at (13.4420*\scaling, -2.000000*\scaling)  [rectangle,fill=white,draw, minimum width = 12pt, minimum height = 12pt,text width=, anchor=center] () {\tiny Z};
   \draw[line width=1mm, white] (15.1580*\scaling,  -7.000000*\scaling)   -- (15.1580*\scaling, -1.000000*\scaling);
   \draw (15.1580*\scaling,  -7.000000*\scaling)   -- (15.1580*\scaling, -1.000000*\scaling);
   \node at (15.1580*\scaling, -7.000000*\scaling)  [rectangle,fill=white,draw, minimum width = 12pt, minimum height = 12pt,text width=, anchor=center] () {\tiny X};
   \node at (15.1580*\scaling, -1.000000*\scaling)  [rectangle,fill=white,draw, minimum width = 12pt, minimum height = 12pt,text width=, anchor=center] () {\tiny Z};
   \node at (16.042000*\scaling, -5.000000*\scaling)  [rectangle,fill=white,draw, minimum width = 12pt, minimum height = 12pt,text width=, anchor=center] () {\tiny X};
   \node at (16.042000*\scaling, -6.000000*\scaling)  [rectangle,fill=white,draw, minimum width = 12pt, minimum height = 12pt,text width=, anchor=center] () {\tiny Z};
   \node at (18.200000*\scaling, -7.000000*\scaling)  [rectangle,fill=white,draw, minimum width = 12pt, minimum height = 12pt,text width=, anchor=center] () {\tiny Z};
\end{tikzpicture}
\vspace{-0.5cm}

\hspace{0.64cm}
\scalebox{0.715}{
\input{XZZXI_connections_7step}}
\end{minipage}

\vspace{0.1cm}

\begin{minipage}{0.15\linewidth}
${IXZZX}$

\vspace{1.5cm}
\end{minipage}
\begin{minipage}{0.8\linewidth}
\begin{tikzpicture}
   \newcommand{\scaling}{0.6}
   \node at (-2*\scaling, 0.0000*\scaling)  [anchor=center] () {1};
   \draw[line width = 0.6mm, gray] (-1.5*\scaling, 0.0000*\scaling)   -- (19.7000*\scaling, 0.0000*\scaling);
   \node at (-2*\scaling, -1.0000*\scaling)  [anchor=center] () {2};
   \draw[line width = 0.6mm, gray] (-1.5*\scaling, -1.0000*\scaling)   -- (19.7000*\scaling, -1.0000*\scaling);
   \node at (-2*\scaling, -2.0000*\scaling)  [anchor=center] () {3};
   \draw[line width = 0.6mm, gray] (-1.5*\scaling, -2.0000*\scaling)   -- (19.7000*\scaling, -2.0000*\scaling);
   \node at (-2*\scaling, -3.0000*\scaling)  [anchor=center] () {4};
   \draw[line width = 0.6mm, gray] (-1.5*\scaling, -3.0000*\scaling)   -- (19.7000*\scaling, -3.0000*\scaling);
   \node at (-2*\scaling, -4.0000*\scaling)  [anchor=center] () {5};
   \draw[line width = 0.6mm, gray] (-1.5*\scaling, -4.0000*\scaling)   -- (19.7000*\scaling, -4.0000*\scaling);
   \node at (-2*\scaling, -5.0000*\scaling)  [anchor=center] () {A};
   \draw[line width = 0.6mm] (-1.5*\scaling, -5.0000*\scaling)   -- (19.7000*\scaling, -5.0000*\scaling);
   \node at (-2*\scaling, -6.0000*\scaling)  [anchor=center] () {B};
   \draw[line width = 0.6mm] (-1.5*\scaling, -6.0000*\scaling)   -- (19.7000*\scaling, -6.0000*\scaling);
   \node at (-2*\scaling, -7.0000*\scaling)  [anchor=center] () {C};
   \draw[line width = 0.6mm] (-1.5*\scaling, -7.0000*\scaling)   -- (19.7000*\scaling, -7.0000*\scaling);
   \draw[dotted] (-1.300000*\scaling, 0.000000*\scaling +1*\scaling)  -- (-1.300000*\scaling, -7.000000*\scaling -1*\scaling);
   \draw[dotted] (1.300000*\scaling, 0.000000*\scaling +1*\scaling)  -- (1.300000*\scaling, -7.000000*\scaling -1*\scaling);
   \draw[dotted] (3.900000*\scaling, 0.000000*\scaling +1*\scaling)  -- (3.900000*\scaling, -7.000000*\scaling -1*\scaling);
   \draw[dotted] (6.500000*\scaling, 0.000000*\scaling +1*\scaling)  -- (6.500000*\scaling, -7.000000*\scaling -1*\scaling);
   \draw[dotted] (9.100000*\scaling, 0.000000*\scaling +1*\scaling)  -- (9.100000*\scaling, -7.000000*\scaling -1*\scaling);
   \draw[dotted] (11.700000*\scaling, 0.000000*\scaling +1*\scaling)  -- (11.700000*\scaling, -7.000000*\scaling -1*\scaling);
   \draw[dotted] (14.300000*\scaling, 0.000000*\scaling +1*\scaling)  -- (14.300000*\scaling, -7.000000*\scaling -1*\scaling);
   \draw[dotted] (16.900000*\scaling, 0.000000*\scaling +1*\scaling)  -- (16.900000*\scaling, -7.000000*\scaling -1*\scaling);
   \draw[dotted] (19.500000*\scaling, 0.000000*\scaling +1*\scaling)  -- (19.500000*\scaling, -7.000000*\scaling -1*\scaling);
   \node at (0.0000*\scaling, -7.0000*\scaling -1*\scaling) {1};
   \node at (2.6000*\scaling, -7.0000*\scaling -1*\scaling) {2};
   \node at (5.2000*\scaling, -7.0000*\scaling -1*\scaling) {3};
   \node at (7.8000*\scaling, -7.0000*\scaling -1*\scaling) {4};
   \node at (10.4000*\scaling, -7.0000*\scaling -1*\scaling) {5};
   \node at (13.0000*\scaling, -7.0000*\scaling -1*\scaling) {6};
   \node at (15.6000*\scaling, -7.0000*\scaling -1*\scaling) {7};
   \node at (18.2000*\scaling, -7.0000*\scaling -1*\scaling) {8};
   \node at (0.000000*\scaling, -6.000000*\scaling)  [rectangle,fill=white,draw, minimum width = 12pt, minimum height = 12pt,text width=, anchor=center] () {\tiny Z};
   \node at (2.158000*\scaling, -5.000000*\scaling)  [rectangle,fill=white,draw, minimum width = 12pt, minimum height = 12pt,text width=, anchor=center] () {\tiny Z};
   \node at (2.158000*\scaling, -7.000000*\scaling)  [rectangle,fill=white,draw, minimum width = 12pt, minimum height = 12pt,text width=, anchor=center] () {\tiny X};
   \draw[line width=1mm, white] (3.0420*\scaling,  -6.000000*\scaling)   -- (3.0420*\scaling, -3.000000*\scaling);
   \draw (3.0420*\scaling,  -6.000000*\scaling)   -- (3.0420*\scaling, -3.000000*\scaling);
   \node at (3.0420*\scaling, -6.000000*\scaling)  [rectangle,fill=white,draw, minimum width = 12pt, minimum height = 12pt,text width=, anchor=center] () {\tiny X};
   \node at (3.0420*\scaling, -3.000000*\scaling)  [rectangle,fill=white,draw, minimum width = 12pt, minimum height = 12pt,text width=, anchor=center] () {\tiny Z};
   \draw[line width=1mm, white] (5.2000*\scaling,  -5.000000*\scaling)   -- (5.2000*\scaling, -4.000000*\scaling);
   \draw (5.2000*\scaling,  -5.000000*\scaling)   -- (5.2000*\scaling, -4.000000*\scaling);
   \node at (5.2000*\scaling, -5.000000*\scaling)  [rectangle,fill=white,draw, minimum width = 12pt, minimum height = 12pt,text width=, anchor=center] () {\tiny X};
   \node at (5.2000*\scaling, -4.000000*\scaling)  [rectangle,fill=white,draw, minimum width = 12pt, minimum height = 12pt,text width=, anchor=center] () {\tiny X};
   \draw[line width=1mm, white] (5.2000*\scaling,  -7.000000*\scaling)   -- (5.2000*\scaling, -6.000000*\scaling);
   \draw (5.2000*\scaling,  -7.000000*\scaling)   -- (5.2000*\scaling, -6.000000*\scaling);
   \node at (5.2000*\scaling, -7.000000*\scaling)  [rectangle,fill=white,draw, minimum width = 12pt, minimum height = 12pt,text width=, anchor=center] () {\tiny Z};
   \node at (5.2000*\scaling, -6.000000*\scaling)  [rectangle,fill=white,draw, minimum width = 12pt, minimum height = 12pt,text width=, anchor=center] () {\tiny Z};
   \draw[line width=1mm, white] (7.8000*\scaling,  -7.000000*\scaling)   -- (7.8000*\scaling, -5.000000*\scaling);
   \draw (7.8000*\scaling,  -7.000000*\scaling)   -- (7.8000*\scaling, -5.000000*\scaling);
   \node at (7.8000*\scaling, -7.000000*\scaling)  [rectangle,fill=white,draw, minimum width = 12pt, minimum height = 12pt,text width=, anchor=center] () {\tiny Z};
   \node at (7.8000*\scaling, -5.000000*\scaling)  [rectangle,fill=white,draw, minimum width = 12pt, minimum height = 12pt,text width=, anchor=center] () {\tiny Z};
   \draw[line width=1mm, white] (10.4000*\scaling,  -7.000000*\scaling)   -- (10.4000*\scaling, -6.000000*\scaling);
   \draw (10.4000*\scaling,  -7.000000*\scaling)   -- (10.4000*\scaling, -6.000000*\scaling);
   \node at (10.4000*\scaling, -7.000000*\scaling)  [rectangle,fill=white,draw, minimum width = 12pt, minimum height = 12pt,text width=, anchor=center] () {\tiny Z};
   \node at (10.4000*\scaling, -6.000000*\scaling)  [rectangle,fill=white,draw, minimum width = 12pt, minimum height = 12pt,text width=, anchor=center] () {\tiny Z};
   \draw[line width=1mm, white] (12.5580*\scaling,  -7.000000*\scaling)   -- (12.5580*\scaling, -5.000000*\scaling);
   \draw (12.5580*\scaling,  -7.000000*\scaling)   -- (12.5580*\scaling, -5.000000*\scaling);
   \node at (12.5580*\scaling, -7.000000*\scaling)  [rectangle,fill=white,draw, minimum width = 12pt, minimum height = 12pt,text width=, anchor=center] () {\tiny Z};
   \node at (12.5580*\scaling, -5.000000*\scaling)  [rectangle,fill=white,draw, minimum width = 12pt, minimum height = 12pt,text width=, anchor=center] () {\tiny Z};
   \draw[line width=1mm, white] (13.4420*\scaling,  -6.000000*\scaling)   -- (13.4420*\scaling, -2.000000*\scaling);
   \draw (13.4420*\scaling,  -6.000000*\scaling)   -- (13.4420*\scaling, -2.000000*\scaling);
   \node at (13.4420*\scaling, -6.000000*\scaling)  [rectangle,fill=white,draw, minimum width = 12pt, minimum height = 12pt,text width=, anchor=center] () {\tiny X};
   \node at (13.4420*\scaling, -2.000000*\scaling)  [rectangle,fill=white,draw, minimum width = 12pt, minimum height = 12pt,text width=, anchor=center] () {\tiny Z};
   \draw[line width=1mm, white] (15.1580*\scaling,  -7.000000*\scaling)   -- (15.1580*\scaling, -1.000000*\scaling);
   \draw (15.1580*\scaling,  -7.000000*\scaling)   -- (15.1580*\scaling, -1.000000*\scaling);
   \node at (15.1580*\scaling, -7.000000*\scaling)  [rectangle,fill=white,draw, minimum width = 12pt, minimum height = 12pt,text width=, anchor=center] () {\tiny X};
   \node at (15.1580*\scaling, -1.000000*\scaling)  [rectangle,fill=white,draw, minimum width = 12pt, minimum height = 12pt,text width=, anchor=center] () {\tiny X};
   \node at (16.042000*\scaling, -5.000000*\scaling)  [rectangle,fill=white,draw, minimum width = 12pt, minimum height = 12pt,text width=, anchor=center] () {\tiny X};
   \node at (16.042000*\scaling, -6.000000*\scaling)  [rectangle,fill=white,draw, minimum width = 12pt, minimum height = 12pt,text width=, anchor=center] () {\tiny Z};
   \node at (18.200000*\scaling, -7.000000*\scaling)  [rectangle,fill=white,draw, minimum width = 12pt, minimum height = 12pt,text width=, anchor=center] () {\tiny Z};
\end{tikzpicture}
\vspace{-0.5cm}

\hspace{0.64cm}
\scalebox{0.715}{
\input{IXZZX_connections_7step}}
\end{minipage}

\vspace{0.1cm}

\begin{minipage}{0.15\linewidth}
${ZXIXZ}$

\vspace{1.5cm}
\end{minipage}
\begin{minipage}{0.8\linewidth}
\begin{tikzpicture}
   \newcommand{\scaling}{0.6}
   \node at (-2*\scaling, 0.0000*\scaling)  [anchor=center] () {1};
   \draw[line width = 0.6mm, gray] (-1.5*\scaling, 0.0000*\scaling)   -- (19.7000*\scaling, 0.0000*\scaling);
   \node at (-2*\scaling, -1.0000*\scaling)  [anchor=center] () {2};
   \draw[line width = 0.6mm, gray] (-1.5*\scaling, -1.0000*\scaling)   -- (19.7000*\scaling, -1.0000*\scaling);
   \node at (-2*\scaling, -2.0000*\scaling)  [anchor=center] () {3};
   \draw[line width = 0.6mm, gray] (-1.5*\scaling, -2.0000*\scaling)   -- (19.7000*\scaling, -2.0000*\scaling);
   \node at (-2*\scaling, -3.0000*\scaling)  [anchor=center] () {4};
   \draw[line width = 0.6mm, gray] (-1.5*\scaling, -3.0000*\scaling)   -- (19.7000*\scaling, -3.0000*\scaling);
   \node at (-2*\scaling, -4.0000*\scaling)  [anchor=center] () {5};
   \draw[line width = 0.6mm, gray] (-1.5*\scaling, -4.0000*\scaling)   -- (19.7000*\scaling, -4.0000*\scaling);
   \node at (-2*\scaling, -5.0000*\scaling)  [anchor=center] () {A};
   \draw[line width = 0.6mm] (-1.5*\scaling, -5.0000*\scaling)   -- (19.7000*\scaling, -5.0000*\scaling);
   \node at (-2*\scaling, -6.0000*\scaling)  [anchor=center] () {B};
   \draw[line width = 0.6mm] (-1.5*\scaling, -6.0000*\scaling)   -- (19.7000*\scaling, -6.0000*\scaling);
   \node at (-2*\scaling, -7.0000*\scaling)  [anchor=center] () {C};
   \draw[line width = 0.6mm] (-1.5*\scaling, -7.0000*\scaling)   -- (19.7000*\scaling, -7.0000*\scaling);
   \draw[dotted] (-1.300000*\scaling, 0.000000*\scaling +1*\scaling)  -- (-1.300000*\scaling, -7.000000*\scaling -1*\scaling);
   \draw[dotted] (1.300000*\scaling, 0.000000*\scaling +1*\scaling)  -- (1.300000*\scaling, -7.000000*\scaling -1*\scaling);
   \draw[dotted] (3.900000*\scaling, 0.000000*\scaling +1*\scaling)  -- (3.900000*\scaling, -7.000000*\scaling -1*\scaling);
   \draw[dotted] (6.500000*\scaling, 0.000000*\scaling +1*\scaling)  -- (6.500000*\scaling, -7.000000*\scaling -1*\scaling);
   \draw[dotted] (9.100000*\scaling, 0.000000*\scaling +1*\scaling)  -- (9.100000*\scaling, -7.000000*\scaling -1*\scaling);
   \draw[dotted] (11.700000*\scaling, 0.000000*\scaling +1*\scaling)  -- (11.700000*\scaling, -7.000000*\scaling -1*\scaling);
   \draw[dotted] (14.300000*\scaling, 0.000000*\scaling +1*\scaling)  -- (14.300000*\scaling, -7.000000*\scaling -1*\scaling);
   \draw[dotted] (16.900000*\scaling, 0.000000*\scaling +1*\scaling)  -- (16.900000*\scaling, -7.000000*\scaling -1*\scaling);
   \draw[dotted] (19.500000*\scaling, 0.000000*\scaling +1*\scaling)  -- (19.500000*\scaling, -7.000000*\scaling -1*\scaling);
   \node at (0.0000*\scaling, -7.0000*\scaling -1*\scaling) {1};
   \node at (2.6000*\scaling, -7.0000*\scaling -1*\scaling) {2};
   \node at (5.2000*\scaling, -7.0000*\scaling -1*\scaling) {3};
   \node at (7.8000*\scaling, -7.0000*\scaling -1*\scaling) {4};
   \node at (10.4000*\scaling, -7.0000*\scaling -1*\scaling) {5};
   \node at (13.0000*\scaling, -7.0000*\scaling -1*\scaling) {6};
   \node at (15.6000*\scaling, -7.0000*\scaling -1*\scaling) {7};
   \node at (18.2000*\scaling, -7.0000*\scaling -1*\scaling) {8};
   \node at (0.000000*\scaling, -5.000000*\scaling)  [rectangle,fill=white,draw, minimum width = 12pt, minimum height = 12pt,text width=, anchor=center] () {\tiny Z};
   \node at (2.600000*\scaling, -6.000000*\scaling)  [rectangle,fill=white,draw, minimum width = 12pt, minimum height = 12pt,text width=, anchor=center] () {\tiny Z};
   \node at (2.600000*\scaling, -7.000000*\scaling)  [rectangle,fill=white,draw, minimum width = 12pt, minimum height = 12pt,text width=, anchor=center] () {\tiny X};
   \draw[line width=1mm, white] (2.6000*\scaling,  -5.000000*\scaling)   -- (2.6000*\scaling, 0.000000*\scaling);
   \draw (2.6000*\scaling,  -5.000000*\scaling)   -- (2.6000*\scaling, 0.000000*\scaling);
   \node at (2.6000*\scaling, -5.000000*\scaling)  [rectangle,fill=white,draw, minimum width = 12pt, minimum height = 12pt,text width=, anchor=center] () {\tiny X};
   \node at (2.6000*\scaling, 0.000000*\scaling)  [rectangle,fill=white,draw, minimum width = 12pt, minimum height = 12pt,text width=, anchor=center] () {\tiny Z};
   \draw[line width=1mm, white] (4.7580*\scaling,  -6.000000*\scaling)   -- (4.7580*\scaling, -3.000000*\scaling);
   \draw (4.7580*\scaling,  -6.000000*\scaling)   -- (4.7580*\scaling, -3.000000*\scaling);
   \node at (4.7580*\scaling, -6.000000*\scaling)  [rectangle,fill=white,draw, minimum width = 12pt, minimum height = 12pt,text width=, anchor=center] () {\tiny X};
   \node at (4.7580*\scaling, -3.000000*\scaling)  [rectangle,fill=white,draw, minimum width = 12pt, minimum height = 12pt,text width=, anchor=center] () {\tiny X};
   \draw[line width=1mm, white] (5.6420*\scaling,  -7.000000*\scaling)   -- (5.6420*\scaling, -5.000000*\scaling);
   \draw (5.6420*\scaling,  -7.000000*\scaling)   -- (5.6420*\scaling, -5.000000*\scaling);
   \node at (5.6420*\scaling, -7.000000*\scaling)  [rectangle,fill=white,draw, minimum width = 12pt, minimum height = 12pt,text width=, anchor=center] () {\tiny Z};
   \node at (5.6420*\scaling, -5.000000*\scaling)  [rectangle,fill=white,draw, minimum width = 12pt, minimum height = 12pt,text width=, anchor=center] () {\tiny Z};
   \draw[line width=1mm, white] (7.8000*\scaling,  -7.000000*\scaling)   -- (7.8000*\scaling, -6.000000*\scaling);
   \draw (7.8000*\scaling,  -7.000000*\scaling)   -- (7.8000*\scaling, -6.000000*\scaling);
   \node at (7.8000*\scaling, -7.000000*\scaling)  [rectangle,fill=white,draw, minimum width = 12pt, minimum height = 12pt,text width=, anchor=center] () {\tiny Z};
   \node at (7.8000*\scaling, -6.000000*\scaling)  [rectangle,fill=white,draw, minimum width = 12pt, minimum height = 12pt,text width=, anchor=center] () {\tiny Z};
   \draw[line width=1mm, white] (10.4000*\scaling,  -7.000000*\scaling)   -- (10.4000*\scaling, -5.000000*\scaling);
   \draw (10.4000*\scaling,  -7.000000*\scaling)   -- (10.4000*\scaling, -5.000000*\scaling);
   \node at (10.4000*\scaling, -7.000000*\scaling)  [rectangle,fill=white,draw, minimum width = 12pt, minimum height = 12pt,text width=, anchor=center] () {\tiny Z};
   \node at (10.4000*\scaling, -5.000000*\scaling)  [rectangle,fill=white,draw, minimum width = 12pt, minimum height = 12pt,text width=, anchor=center] () {\tiny Z};
   \draw[line width=1mm, white] (13.0000*\scaling,  -7.000000*\scaling)   -- (13.0000*\scaling, -6.000000*\scaling);
   \draw (13.0000*\scaling,  -7.000000*\scaling)   -- (13.0000*\scaling, -6.000000*\scaling);
   \node at (13.0000*\scaling, -7.000000*\scaling)  [rectangle,fill=white,draw, minimum width = 12pt, minimum height = 12pt,text width=, anchor=center] () {\tiny Z};
   \node at (13.0000*\scaling, -6.000000*\scaling)  [rectangle,fill=white,draw, minimum width = 12pt, minimum height = 12pt,text width=, anchor=center] () {\tiny Z};
   \draw[line width=1mm, white] (13.0000*\scaling,  -5.000000*\scaling)   -- (13.0000*\scaling, -4.000000*\scaling);
   \draw (13.0000*\scaling,  -5.000000*\scaling)   -- (13.0000*\scaling, -4.000000*\scaling);
   \node at (13.0000*\scaling, -5.000000*\scaling)  [rectangle,fill=white,draw, minimum width = 12pt, minimum height = 12pt,text width=, anchor=center] () {\tiny X};
   \node at (13.0000*\scaling, -4.000000*\scaling)  [rectangle,fill=white,draw, minimum width = 12pt, minimum height = 12pt,text width=, anchor=center] () {\tiny Z};
   \draw[line width=1mm, white] (15.1580*\scaling,  -7.000000*\scaling)   -- (15.1580*\scaling, -1.000000*\scaling);
   \draw (15.1580*\scaling,  -7.000000*\scaling)   -- (15.1580*\scaling, -1.000000*\scaling);
   \node at (15.1580*\scaling, -7.000000*\scaling)  [rectangle,fill=white,draw, minimum width = 12pt, minimum height = 12pt,text width=, anchor=center] () {\tiny X};
   \node at (15.1580*\scaling, -1.000000*\scaling)  [rectangle,fill=white,draw, minimum width = 12pt, minimum height = 12pt,text width=, anchor=center] () {\tiny X};
   \node at (16.042000*\scaling, -5.000000*\scaling)  [rectangle,fill=white,draw, minimum width = 12pt, minimum height = 12pt,text width=, anchor=center] () {\tiny Z};
   \node at (16.042000*\scaling, -6.000000*\scaling)  [rectangle,fill=white,draw, minimum width = 12pt, minimum height = 12pt,text width=, anchor=center] () {\tiny X};
   \node at (18.200000*\scaling, -7.000000*\scaling)  [rectangle,fill=white,draw, minimum width = 12pt, minimum height = 12pt,text width=, anchor=center] () {\tiny Z};
\end{tikzpicture}
\vspace{-0.5cm}

\hspace{0.64cm}
\scalebox{0.715}{
\input{ZXIXZ_connections_7step}}
\end{minipage}

\vspace{0.1cm}

\begin{minipage}{0.15\linewidth}
${XIXZZ}$

\vspace{1.5cm}
\end{minipage}
\begin{minipage}{0.8\linewidth}
\begin{tikzpicture}
   \newcommand{\scaling}{0.6}
   \node at (-2*\scaling, 0.0000*\scaling)  [anchor=center] () {1};
   \draw[line width = 0.6mm, gray] (-1.5*\scaling, 0.0000*\scaling)   -- (19.7000*\scaling, 0.0000*\scaling);
   \node at (-2*\scaling, -1.0000*\scaling)  [anchor=center] () {2};
   \draw[line width = 0.6mm, gray] (-1.5*\scaling, -1.0000*\scaling)   -- (19.7000*\scaling, -1.0000*\scaling);
   \node at (-2*\scaling, -2.0000*\scaling)  [anchor=center] () {3};
   \draw[line width = 0.6mm, gray] (-1.5*\scaling, -2.0000*\scaling)   -- (19.7000*\scaling, -2.0000*\scaling);
   \node at (-2*\scaling, -3.0000*\scaling)  [anchor=center] () {4};
   \draw[line width = 0.6mm, gray] (-1.5*\scaling, -3.0000*\scaling)   -- (19.7000*\scaling, -3.0000*\scaling);
   \node at (-2*\scaling, -4.0000*\scaling)  [anchor=center] () {5};
   \draw[line width = 0.6mm, gray] (-1.5*\scaling, -4.0000*\scaling)   -- (19.7000*\scaling, -4.0000*\scaling);
   \node at (-2*\scaling, -5.0000*\scaling)  [anchor=center] () {A};
   \draw[line width = 0.6mm] (-1.5*\scaling, -5.0000*\scaling)   -- (19.7000*\scaling, -5.0000*\scaling);
   \node at (-2*\scaling, -6.0000*\scaling)  [anchor=center] () {B};
   \draw[line width = 0.6mm] (-1.5*\scaling, -6.0000*\scaling)   -- (19.7000*\scaling, -6.0000*\scaling);
   \node at (-2*\scaling, -7.0000*\scaling)  [anchor=center] () {C};
   \draw[line width = 0.6mm] (-1.5*\scaling, -7.0000*\scaling)   -- (19.7000*\scaling, -7.0000*\scaling);
   \draw[dotted] (-1.300000*\scaling, 0.000000*\scaling +1*\scaling)  -- (-1.300000*\scaling, -7.000000*\scaling -1*\scaling);
   \draw[dotted] (1.300000*\scaling, 0.000000*\scaling +1*\scaling)  -- (1.300000*\scaling, -7.000000*\scaling -1*\scaling);
   \draw[dotted] (3.900000*\scaling, 0.000000*\scaling +1*\scaling)  -- (3.900000*\scaling, -7.000000*\scaling -1*\scaling);
   \draw[dotted] (6.500000*\scaling, 0.000000*\scaling +1*\scaling)  -- (6.500000*\scaling, -7.000000*\scaling -1*\scaling);
   \draw[dotted] (9.100000*\scaling, 0.000000*\scaling +1*\scaling)  -- (9.100000*\scaling, -7.000000*\scaling -1*\scaling);
   \draw[dotted] (11.700000*\scaling, 0.000000*\scaling +1*\scaling)  -- (11.700000*\scaling, -7.000000*\scaling -1*\scaling);
   \draw[dotted] (14.300000*\scaling, 0.000000*\scaling +1*\scaling)  -- (14.300000*\scaling, -7.000000*\scaling -1*\scaling);
   \draw[dotted] (16.900000*\scaling, 0.000000*\scaling +1*\scaling)  -- (16.900000*\scaling, -7.000000*\scaling -1*\scaling);
   \draw[dotted] (19.500000*\scaling, 0.000000*\scaling +1*\scaling)  -- (19.500000*\scaling, -7.000000*\scaling -1*\scaling);
   \node at (0.0000*\scaling, -7.0000*\scaling -1*\scaling) {1};
   \node at (2.6000*\scaling, -7.0000*\scaling -1*\scaling) {2};
   \node at (5.2000*\scaling, -7.0000*\scaling -1*\scaling) {3};
   \node at (7.8000*\scaling, -7.0000*\scaling -1*\scaling) {4};
   \node at (10.4000*\scaling, -7.0000*\scaling -1*\scaling) {5};
   \node at (13.0000*\scaling, -7.0000*\scaling -1*\scaling) {6};
   \node at (15.6000*\scaling, -7.0000*\scaling -1*\scaling) {7};
   \node at (18.2000*\scaling, -7.0000*\scaling -1*\scaling) {8};
   \node at (0.000000*\scaling, -6.000000*\scaling)  [rectangle,fill=white,draw, minimum width = 12pt, minimum height = 12pt,text width=, anchor=center] () {\tiny Z};
   \node at (2.158000*\scaling, -5.000000*\scaling)  [rectangle,fill=white,draw, minimum width = 12pt, minimum height = 12pt,text width=, anchor=center] () {\tiny Z};
   \node at (2.158000*\scaling, -7.000000*\scaling)  [rectangle,fill=white,draw, minimum width = 12pt, minimum height = 12pt,text width=, anchor=center] () {\tiny X};
   \draw[line width=1mm, white] (3.0420*\scaling,  -6.000000*\scaling)   -- (3.0420*\scaling, -3.000000*\scaling);
   \draw (3.0420*\scaling,  -6.000000*\scaling)   -- (3.0420*\scaling, -3.000000*\scaling);
   \node at (3.0420*\scaling, -6.000000*\scaling)  [rectangle,fill=white,draw, minimum width = 12pt, minimum height = 12pt,text width=, anchor=center] () {\tiny X};
   \node at (3.0420*\scaling, -3.000000*\scaling)  [rectangle,fill=white,draw, minimum width = 12pt, minimum height = 12pt,text width=, anchor=center] () {\tiny Z};
   \draw[line width=1mm, white] (5.2000*\scaling,  -5.000000*\scaling)   -- (5.2000*\scaling, -4.000000*\scaling);
   \draw (5.2000*\scaling,  -5.000000*\scaling)   -- (5.2000*\scaling, -4.000000*\scaling);
   \node at (5.2000*\scaling, -5.000000*\scaling)  [rectangle,fill=white,draw, minimum width = 12pt, minimum height = 12pt,text width=, anchor=center] () {\tiny X};
   \node at (5.2000*\scaling, -4.000000*\scaling)  [rectangle,fill=white,draw, minimum width = 12pt, minimum height = 12pt,text width=, anchor=center] () {\tiny Z};
   \draw[line width=1mm, white] (5.2000*\scaling,  -7.000000*\scaling)   -- (5.2000*\scaling, -6.000000*\scaling);
   \draw (5.2000*\scaling,  -7.000000*\scaling)   -- (5.2000*\scaling, -6.000000*\scaling);
   \node at (5.2000*\scaling, -7.000000*\scaling)  [rectangle,fill=white,draw, minimum width = 12pt, minimum height = 12pt,text width=, anchor=center] () {\tiny Z};
   \node at (5.2000*\scaling, -6.000000*\scaling)  [rectangle,fill=white,draw, minimum width = 12pt, minimum height = 12pt,text width=, anchor=center] () {\tiny Z};
   \draw[line width=1mm, white] (7.8000*\scaling,  -7.000000*\scaling)   -- (7.8000*\scaling, -5.000000*\scaling);
   \draw (7.8000*\scaling,  -7.000000*\scaling)   -- (7.8000*\scaling, -5.000000*\scaling);
   \node at (7.8000*\scaling, -7.000000*\scaling)  [rectangle,fill=white,draw, minimum width = 12pt, minimum height = 12pt,text width=, anchor=center] () {\tiny Z};
   \node at (7.8000*\scaling, -5.000000*\scaling)  [rectangle,fill=white,draw, minimum width = 12pt, minimum height = 12pt,text width=, anchor=center] () {\tiny Z};
   \draw[line width=1mm, white] (10.4000*\scaling,  -7.000000*\scaling)   -- (10.4000*\scaling, -6.000000*\scaling);
   \draw (10.4000*\scaling,  -7.000000*\scaling)   -- (10.4000*\scaling, -6.000000*\scaling);
   \node at (10.4000*\scaling, -7.000000*\scaling)  [rectangle,fill=white,draw, minimum width = 12pt, minimum height = 12pt,text width=, anchor=center] () {\tiny Z};
   \node at (10.4000*\scaling, -6.000000*\scaling)  [rectangle,fill=white,draw, minimum width = 12pt, minimum height = 12pt,text width=, anchor=center] () {\tiny Z};
   \draw[line width=1mm, white] (12.5580*\scaling,  -7.000000*\scaling)   -- (12.5580*\scaling, -5.000000*\scaling);
   \draw (12.5580*\scaling,  -7.000000*\scaling)   -- (12.5580*\scaling, -5.000000*\scaling);
   \node at (12.5580*\scaling, -7.000000*\scaling)  [rectangle,fill=white,draw, minimum width = 12pt, minimum height = 12pt,text width=, anchor=center] () {\tiny Z};
   \node at (12.5580*\scaling, -5.000000*\scaling)  [rectangle,fill=white,draw, minimum width = 12pt, minimum height = 12pt,text width=, anchor=center] () {\tiny Z};
   \draw[line width=1mm, white] (13.4420*\scaling,  -6.000000*\scaling)   -- (13.4420*\scaling, -2.000000*\scaling);
   \draw (13.4420*\scaling,  -6.000000*\scaling)   -- (13.4420*\scaling, -2.000000*\scaling);
   \node at (13.4420*\scaling, -6.000000*\scaling)  [rectangle,fill=white,draw, minimum width = 12pt, minimum height = 12pt,text width=, anchor=center] () {\tiny X};
   \node at (13.4420*\scaling, -2.000000*\scaling)  [rectangle,fill=white,draw, minimum width = 12pt, minimum height = 12pt,text width=, anchor=center] () {\tiny X};
   \draw[line width=1mm, white] (15.6000*\scaling,  -5.000000*\scaling)   -- (15.6000*\scaling, 0.000000*\scaling);
   \draw (15.6000*\scaling,  -5.000000*\scaling)   -- (15.6000*\scaling, 0.000000*\scaling);
   \node at (15.6000*\scaling, -5.000000*\scaling)  [rectangle,fill=white,draw, minimum width = 12pt, minimum height = 12pt,text width=, anchor=center] () {\tiny X};
   \node at (15.6000*\scaling, 0.000000*\scaling)  [rectangle,fill=white,draw, minimum width = 12pt, minimum height = 12pt,text width=, anchor=center] () {\tiny X};
   \node at (15.600000*\scaling, -6.000000*\scaling)  [rectangle,fill=white,draw, minimum width = 12pt, minimum height = 12pt,text width=, anchor=center] () {\tiny Z};
   \node at (15.600000*\scaling, -7.000000*\scaling)  [rectangle,fill=white,draw, minimum width = 12pt, minimum height = 12pt,text width=, anchor=center] () {\tiny X};
   \node at (18.200000*\scaling, -5.000000*\scaling)  [rectangle,fill=white,draw, minimum width = 12pt, minimum height = 12pt,text width=, anchor=center] () {\tiny Z};
\end{tikzpicture}
\vspace{-0.5cm}

\hspace{0.64cm}
\scalebox{0.715}{
\input{XIXZZ_connections_7step}}
\end{minipage}


Three-qubit representatives of the single qubit logical operators $\bar{X},\bar{Y},\bar{Z}$ are measured through a small modification of the stabilizer measurement circuits, in which one pairwise measurement has been removed and one single qubit measurement is performed in a different basis. The modified circuit involves only the three data qubits in the support of the representative. As an example, the following circuit measures one of the representatives of $\bar{X}$, $ZXZII$:

\begin{minipage}{0.15\linewidth}
${ZXZII}$

\vspace{0cm}
\end{minipage}
\begin{minipage}{0.8\linewidth}
\begin{tikzpicture}
   \newcommand{\scaling}{0.6}
   \node at (-2*\scaling, 0.0000*\scaling)  [anchor=center] () {1};
   \draw[line width = 0.6mm, gray] (-1.5*\scaling, 0.0000*\scaling)   -- (17.1000*\scaling, 0.0000*\scaling);
   \node at (-2*\scaling, -1.0000*\scaling)  [anchor=center] () {2};
   \draw[line width = 0.6mm, gray] (-1.5*\scaling, -1.0000*\scaling)   -- (17.1000*\scaling, -1.0000*\scaling);
   \node at (-2*\scaling, -2.0000*\scaling)  [anchor=center] () {3};
   \draw[line width = 0.6mm, gray] (-1.5*\scaling, -2.0000*\scaling)   -- (17.1000*\scaling, -2.0000*\scaling);
   \node at (-2*\scaling, -3.0000*\scaling)  [anchor=center] () {4};
   \draw[line width = 0.6mm, gray] (-1.5*\scaling, -3.0000*\scaling)   -- (17.1000*\scaling, -3.0000*\scaling);
   \node at (-2*\scaling, -4.0000*\scaling)  [anchor=center] () {5};
   \draw[line width = 0.6mm, gray] (-1.5*\scaling, -4.0000*\scaling)   -- (17.1000*\scaling, -4.0000*\scaling);
   \node at (-2*\scaling, -5.0000*\scaling)  [anchor=center] () {A};
   \draw[line width = 0.6mm] (-1.5*\scaling, -5.0000*\scaling)   -- (17.1000*\scaling, -5.0000*\scaling);
   \node at (-2*\scaling, -6.0000*\scaling)  [anchor=center] () {B};
   \draw[line width = 0.6mm] (-1.5*\scaling, -6.0000*\scaling)   -- (17.1000*\scaling, -6.0000*\scaling);
   \node at (-2*\scaling, -7.0000*\scaling)  [anchor=center] () {C};
   \draw[line width = 0.6mm] (-1.5*\scaling, -7.0000*\scaling)   -- (17.1000*\scaling, -7.0000*\scaling);
   \draw[dotted] (-1.300000*\scaling, 0.000000*\scaling +1*\scaling)  -- (-1.300000*\scaling, -7.000000*\scaling -1*\scaling);
   \draw[dotted] (1.300000*\scaling, 0.000000*\scaling +1*\scaling)  -- (1.300000*\scaling, -7.000000*\scaling -1*\scaling);
   \draw[dotted] (3.900000*\scaling, 0.000000*\scaling +1*\scaling)  -- (3.900000*\scaling, -7.000000*\scaling -1*\scaling);
   \draw[dotted] (6.500000*\scaling, 0.000000*\scaling +1*\scaling)  -- (6.500000*\scaling, -7.000000*\scaling -1*\scaling);
   \draw[dotted] (9.100000*\scaling, 0.000000*\scaling +1*\scaling)  -- (9.100000*\scaling, -7.000000*\scaling -1*\scaling);
   \draw[dotted] (11.700000*\scaling, 0.000000*\scaling +1*\scaling)  -- (11.700000*\scaling, -7.000000*\scaling -1*\scaling);
   \draw[dotted] (14.300000*\scaling, 0.000000*\scaling +1*\scaling)  -- (14.300000*\scaling, -7.000000*\scaling -1*\scaling);
   \draw[dotted] (16.900000*\scaling, 0.000000*\scaling +1*\scaling)  -- (16.900000*\scaling, -7.000000*\scaling -1*\scaling);
   \node at (0.0000*\scaling, -7.0000*\scaling -1*\scaling) {1};
   \node at (2.6000*\scaling, -7.0000*\scaling -1*\scaling) {2};
   \node at (5.2000*\scaling, -7.0000*\scaling -1*\scaling) {3};
   \node at (7.8000*\scaling, -7.0000*\scaling -1*\scaling) {4};
   \node at (10.4000*\scaling, -7.0000*\scaling -1*\scaling) {5};
   \node at (13.0000*\scaling, -7.0000*\scaling -1*\scaling) {6};
   \node at (15.6000*\scaling, -7.0000*\scaling -1*\scaling) {7};
   \node at (0.000000*\scaling, -5.000000*\scaling)  [rectangle,fill=white,draw, minimum width = 12pt, minimum height = 12pt,text width=, anchor=center] () {\tiny Z};
   \node at (0.000000*\scaling, -6.000000*\scaling)  [rectangle,fill=white,draw, minimum width = 12pt, minimum height = 12pt,text width=, anchor=center] () {\tiny X};
   \node at (0.000000*\scaling, -7.000000*\scaling)  [rectangle,fill=white,draw, minimum width = 12pt, minimum height = 12pt,text width=, anchor=center] () {\tiny X};
   \draw[line width=1mm, white] (2.6000*\scaling,  0.000000*\scaling)   -- (2.6000*\scaling, -5.000000*\scaling);
   \draw (2.6000*\scaling,  0.000000*\scaling)   -- (2.6000*\scaling, -5.000000*\scaling);
   \node at (2.6000*\scaling, 0.000000*\scaling)  [rectangle,fill=white,draw, minimum width = 12pt, minimum height = 12pt,text width=, anchor=center] () {\tiny Z};
   \node at (2.6000*\scaling, -5.000000*\scaling)  [rectangle,fill=white,draw, minimum width = 12pt, minimum height = 12pt,text width=, anchor=center] () {\tiny X};
   \draw[line width=1mm, white] (2.6000*\scaling,  -7.000000*\scaling)   -- (2.6000*\scaling, -6.000000*\scaling);
   \draw (2.6000*\scaling,  -7.000000*\scaling)   -- (2.6000*\scaling, -6.000000*\scaling);
   \node at (2.6000*\scaling, -7.000000*\scaling)  [rectangle,fill=white,draw, minimum width = 12pt, minimum height = 12pt,text width=, anchor=center] () {\tiny Z};
   \node at (2.6000*\scaling, -6.000000*\scaling)  [rectangle,fill=white,draw, minimum width = 12pt, minimum height = 12pt,text width=, anchor=center] () {\tiny Z};
   \draw[line width=1mm, white] (5.2000*\scaling,  -7.000000*\scaling)   -- (5.2000*\scaling, -5.000000*\scaling);
   \draw (5.2000*\scaling,  -7.000000*\scaling)   -- (5.2000*\scaling, -5.000000*\scaling);
   \node at (5.2000*\scaling, -7.000000*\scaling)  [rectangle,fill=white,draw, minimum width = 12pt, minimum height = 12pt,text width=, anchor=center] () {\tiny Z};
   \node at (5.2000*\scaling, -5.000000*\scaling)  [rectangle,fill=white,draw, minimum width = 12pt, minimum height = 12pt,text width=, anchor=center] () {\tiny Z};
   \draw[line width=1mm, white] (7.8000*\scaling,  -7.000000*\scaling)   -- (7.8000*\scaling, -6.000000*\scaling);
   \draw (7.8000*\scaling,  -7.000000*\scaling)   -- (7.8000*\scaling, -6.000000*\scaling);
   \node at (7.8000*\scaling, -7.000000*\scaling)  [rectangle,fill=white,draw, minimum width = 12pt, minimum height = 12pt,text width=, anchor=center] () {\tiny Z};
   \node at (7.8000*\scaling, -6.000000*\scaling)  [rectangle,fill=white,draw, minimum width = 12pt, minimum height = 12pt,text width=, anchor=center] () {\tiny Z};
   \draw[line width=1mm, white] (10.4000*\scaling,  -7.000000*\scaling)   -- (10.4000*\scaling, -5.000000*\scaling);
   \draw (10.4000*\scaling,  -7.000000*\scaling)   -- (10.4000*\scaling, -5.000000*\scaling);
   \node at (10.4000*\scaling, -7.000000*\scaling)  [rectangle,fill=white,draw, minimum width = 12pt, minimum height = 12pt,text width=, anchor=center] () {\tiny Z};
   \node at (10.4000*\scaling, -5.000000*\scaling)  [rectangle,fill=white,draw, minimum width = 12pt, minimum height = 12pt,text width=, anchor=center] () {\tiny Z};
   \draw[line width=1mm, white] (12.5580*\scaling,  -7.000000*\scaling)   -- (12.5580*\scaling, -1.000000*\scaling);
   \draw (12.5580*\scaling,  -7.000000*\scaling)   -- (12.5580*\scaling, -1.000000*\scaling);
   \node at (12.5580*\scaling, -7.000000*\scaling)  [rectangle,fill=white,draw, minimum width = 12pt, minimum height = 12pt,text width=, anchor=center] () {\tiny X};
   \node at (12.5580*\scaling, -1.000000*\scaling)  [rectangle,fill=white,draw, minimum width = 12pt, minimum height = 12pt,text width=, anchor=center] () {\tiny X};
   \draw[line width=1mm, white] (13.4420*\scaling,  -6.000000*\scaling)   -- (13.4420*\scaling, -2.000000*\scaling);
   \draw (13.4420*\scaling,  -6.000000*\scaling)   -- (13.4420*\scaling, -2.000000*\scaling);
   \node at (13.4420*\scaling, -6.000000*\scaling)  [rectangle,fill=white,draw, minimum width = 12pt, minimum height = 12pt,text width=, anchor=center] () {\tiny X};
   \node at (13.4420*\scaling, -2.000000*\scaling)  [rectangle,fill=white,draw, minimum width = 12pt, minimum height = 12pt,text width=, anchor=center] () {\tiny Z};
   \node at (15.600000*\scaling, -5.000000*\scaling)  [rectangle,fill=white,draw, minimum width = 12pt, minimum height = 12pt,text width=, anchor=center] () {\tiny X};
   \node at (15.600000*\scaling, -6.000000*\scaling)  [rectangle,fill=white,draw, minimum width = 12pt, minimum height = 12pt,text width=, anchor=center] () {\tiny Z};
   \node at (15.600000*\scaling, -7.000000*\scaling)  [rectangle,fill=white,draw, minimum width = 12pt, minimum height = 12pt,text width=, anchor=center] () {\tiny Z};
\end{tikzpicture}

\hspace{0.64cm}
\scalebox{0.715}{
\input{logical_connections}}
\end{minipage}

\vspace{0cm}

\noindent
On the auxiliary qubit that now lacks a pairwise measurement connecting to a fourth data qubit, the single qubit measurement that would have preceded or followed that removed pairwise measurement (depending on whether the pairwise measurement was removed from the first or second half of the circuit) is performed in the $X$ basis rather than the $Z$ basis. This basis change can be understood from the fusion rules of ZX-calculus. In particular, in the above example both auxiliary qubits $B$ and $C$ are measured in the $X$ basis in step 1, rather than one being measured in the $Z$ basis.


\section{Pairwise measurement-based syndrome extraction in the [[8,3,2]] color code}\label{ColorCode}

The stabilizer group of the $[[8,3,2]]$ color code is generated by $X^{\otimes 8}$, $ZZZZIIII$, $IIIIZZZZ$, $ZZIIZZII$ and $ZIZIZIZI$. The four weight 4 stabilizers can be measured fault-tolerantly through circuits of the form used in the 5+3 code. Meanwhile, the following circuit measures the weight 8 stabilizer:
\begin{equation}
\vcenter{\hbox{
\scalebox{0.67}{
\input{colorcode_diagram}
}}}
\end{equation}
This circuit is \emph{not} fault-tolerant, in the sense that there are degree one circuit faults that are equivalent to faults on multiple data qubits. However, through the ordering of the measurements above, which connects to the data qubits in the order $8,1,2,3,4,5,6,7$ in order to ``spread'' faults strategically, any such circuit fault is still detected by the weight 4 stabilizers.

\bibliographystyle{apsrev4-2}
\bibliography{references}

\end{document}